\documentclass{article}

\usepackage{amsthm,amsmath,amsfonts,amssymb}
\usepackage[margin=1in]{geometry}
\usepackage[authoryear,sort,round]{natbib}
\usepackage[colorlinks,citecolor=blue,urlcolor=blue]{hyperref}
\usepackage{graphicx}
\usepackage{algorithm}
\usepackage{algorithmic}
\usepackage{subcaption}
\usepackage{enumitem}
\usepackage{tikz}
\usepackage{authblk}
\usetikzlibrary{decorations.pathreplacing}

\theoremstyle{plain}
\newtheorem{theorem}{Theorem}[section]
\newtheorem{lemma}[theorem]{Lemma}
\newtheorem{corollary}[theorem]{Corollary}

\theoremstyle{remark}
\newtheorem{definition}[theorem]{Definition}

\newtheorem{remark}[theorem]{Remark}

\title{Mitigating Included- and Omitted-Variable Bias\\in Estimates of Disparate
  Impact\thanks{%
    We thank Josh Grossman, Julian Nyarko, and João M. Souto-Maior for helpful
    conversations. All data and replication code are available at:
    \url{https://github.com/jgaeb/rar-repro}. The sensitivity analysis is
    implemented in the
    \href{https://cran.r-project.org/package=rar}{\texttt{rar}} \texttt{R}
    package available on CRAN.
  }
}
\author[ ]{Jongbin Jung\kern-3.5pt\relax}
\author[ ]{Sam Corbett-Davies\kern-3.5pt\relax}
\author[1]{Johann D. Gaebler}
\author[2]{Ravi Shroff}
\author[3]{Sharad Goel}

\affil[1]{Department of Statistics, Harvard University}
\affil[2]{%
  Department of Applied Statistics, Social Science, and Humanities, New York
  University
}
\affil[3]{Harvard Kennedy School, Harvard University}
\date{}

\begin{document}

\maketitle

\begin{abstract}
\noindent
Managers, employers, policymakers, and others often seek to understand whether
decisions are biased against certain groups. One popular analytic strategy is to
estimate disparities after adjusting for observed covariates, typically with a
regression model. This approach, however, suffers from two key statistical
challenges. First, omitted-variable bias can skew results if the model does not
adjust for all relevant factors; second, and conversely, included-variable
bias---a lesser-known phenomenon---can skew results if the set of covariates
includes irrelevant factors. Here we introduce a new, three-step statistical
method, which we call risk-adjusted regression, to address both concerns in
settings where decision makers have clearly measurable objectives. In the first
step, we use all available covariates to estimate the value, or inversely, the
\emph{risk}, of taking a certain action, such as approving a loan application or
hiring a job candidate. Second, we measure disparities in decisions after
adjusting for these risk estimates alone, mitigating the problem of
included-variable bias. Finally, in the third step, we assess the sensitivity of
results to potential mismeasurement of risk, addressing concerns about
omitted-variable bias. To do so, we develop a novel, non-parametric sensitivity
analysis that yields tight bounds on the true disparity in terms of the average
gap between true and estimated risk---a single interpretable parameter that
facilitates credible estimates. We demonstrate this approach on a detailed
dataset of 2.2 million police stops of pedestrians in New York City, and show
that traditional statistical tests of discrimination can substantially
underestimate the magnitude of disparities.
\end{abstract}

\section{Introduction}

Studies of discrimination generally start by assessing whether certain groups,
particularly those defined by race and gender, receive favorable decisions more
often than others. For example, one might examine whether loan applications from
white candidates are granted more often than those from racial minorities, or
whether male employees are promoted more often than women. Although observed
disparities may be the result of bias, it is also possible that they stem from
statistical differences between groups. In particular, if some groups contain
disproportionately many qualified members, then one would also expect those
groups to receive disproportionately many favorable decisions, even in the
absence of discrimination.

To tease apart these two possibilities---group differences versus
discrimination---the most popular statistical approach is ordinary linear or
logistic regression. In the banking context, for instance, one could examine
race-contingent lending rates after adjusting for relevant factors, such as
income and credit history. Disparities that persist after accounting for such
factors are often interpreted as evidence of discrimination. This basic
statistical strategy has been used in numerous studies to test for bias across
domains, including education~\citep{espenshade2004admission,
grossman2023disparate}, employment~\citep{polachek2008earnings}, criminal
justice~\citep{%
  gelman2007analysis, macdonald2019effect, gaebler2022causal,
  abrams2014law, rehavi2012racial%
}, and medicine~\citep{balsa2005testing}.

Despite the ubiquity of such regression-based tests for discrimination, the
approach suffers from two serious statistical limitations. First, the well-known
problem of omitted-variable bias arises when decisions are based in part on
relevant factors that correlate with group membership, but which are omitted
from the regression~\citep{angrist2008mostly}. For example, if lending officers
consider an applicant's payment history, and if payment history correlates with
race but is not recorded in the data (and thus cannot be included in the
regression), the results of the regression can suggest discrimination where
there is none, or \emph{vice versa}. Unfortunately, omitted-variable bias is the
rule rather than the exception. It is generally prohibitive to measure every
variable relevant to a decision, and it is likely that most unmeasured variables
are at least weakly correlated with demographic attributes, skewing results.

A second problem with regression-based tests is what
\cite{ayres2005three,ayres2010included} calls \emph{included-variable bias}, an
issue as important as omitted-variable bias in studies of discrimination but one
that receives far less attention. To take an extreme example, it is problematic
to include control variables in a regression that are obvious proxies for
legally protected attributes---such as vocal register as a proxy for
gender---when examining the extent to which observed disparities stem from group
differences in qualification. Including such proxies will typically lead one to
underestimate the true magnitude of discrimination in decisions. But what counts
as a ``proxy'' is not always clear. For example, given existing patterns of
residential segregation, one might argue that ZIP codes are a proxy for race,
and thus should be excluded when testing for racial bias. But one could also
argue that ZIP code provides legitimate information relevant to a decision, and
so excluding it would lead to omitted-variable bias. \cite{ayres2010included}
proposes a middle ground, suggesting that potential proxies should be included,
but their coefficients capped to a ``justifiable'' level; in practice, however,
it is difficult to determine and  defend specific constraints on regression
coefficients.\footnote{%
  Such problems have prompted a search for alternatives to regression-based
  approaches. Most prominently, \cite{becker1993} proposed the outcome test,
  which is based not on the rate at which decisions are made, but on the success
  rate of those decisions. In the context of banking, Becker argued that even
  if, hypothetically, racial minorities were less creditworthy than white
  applicants, minorities who were granted loans should still be found to repay
  their loans at the same rate as white applicants who were granted loans. If
  loans to minorities had a higher repayment rate than loans to white borrowers,
  it suggests that lenders applied a double standard (intentionally or not),
  granting loans only to exceptionally qualified minorities---potentially
  violating disparate impact laws. The outcome-based approach has been applied
  almost as broadly as simple regression to study discrimination in, for
  example, policing~\citep{goel_2016c,goel2016b,ayres2002,knowles2001},
  lending~\citep{berkovec1996mortgage}, and scientific
  publication~\citep{smart1996citation}. Outcome tests, however, have their own
  significant statistical shortcomings, most notably the problem of
  infra-marginality~\citep{ayres2002,simoiu2017problem,pierson2018fast}, which
  can lead the test to incorrectly suggest an absence of
  discrimination~\citep{pierson2018large}.
}

Here we present a statistically principled and logistically straightforward
method for measuring discrimination that addresses both omitted- and
included-variable bias. Our method, which we call risk-adjusted regression,
proceeds in three steps. In  the  first  step,  we  use  all available
information, including potential proxies of protected traits, to estimate the
value---or, equivalently, the risk---of taking a particular action. For example,
in the lending context, we might estimate an applicant's risk of default if
granted a loan, conditional on all available covariates. In the second step, we
assess disparities by regressing decisions (e.g., loan offers) on
individual-level risk estimates and protected traits alone, allowing us to
measure the extent to which similarly qualified individuals are treated
differently. This strategy can be seen as formalizing the coefficient-capping
procedure of~\cite{ayres2010included}---with covariates used only to the extent
that they are statistically justified by risk---and thus circumvents the problem
of included-variable bias. Finally, we assess the sensitivity of results to
potential mismeasurement of risk. In particular, we derive tight analytic bounds
on  risk-adjusted disparities as a function of the extent to which risk
estimates differ from true risk.

To demonstrate this approach, we examine 2.2 million stops of pedestrians
conducted by the New York City Police Department between 2008 and 2011. After
adjusting for a stopped individual's statistical risk of carrying a
weapon---based in part on detailed behavioral indicators recorded by
officers---we find that stopped Black and Hispanic pedestrians are searched for
weapons substantially more often than stopped white individuals. We find that
these risk-adjusted disparities are considerably larger than disparities
suggested by a standard regression that adjusts for all available covariates,
underscoring the importance of accounting for included-variable bias. Finally,
we show that our results are robust to potentially large errors in risk
estimates.

\section{Theories of Discrimination}%
\label{ssec:background}

There are two main legal doctrines of discrimination in the United States:
disparate treatment and disparate impact. Here we describe the conceptual
underpinnings of these theories. We further connect these ideas to standard
statistical tests of discrimination, highlighting the problem of
included-variable bias.

\subsection{The Jurisprudence of Discrimination}

Disparate treatment derives force from the Equal Protection Clause of the U.S.
Constitution's Fourteenth Amendment, and it prohibits government agents from
acting with ``discriminatory purpose''~\citep{davis1976}. Although equal
protection law bars policies undertaken with animus, it allows for the limited
use of protected attributes to further a compelling government interest. For
example, until recently, certain affirmative action programs for college
admissions were legally permissible to further the government's interest in
promoting diversity. In 2023, the U.S. Supreme Court overturned the legality of
such affirmative action programs, ruling that it was unlawful to explicitly
consider race in college admissions decisions~\citep{2023students}.

The most widespread statistical test of such intentional discrimination is
ordinary linear or logistic regression, in which one estimates the likelihood of
favorable (or unfavorable) decisions across groups defined by race, gender, or
other legally protected traits. In this approach, the investigator adjusts for
all potentially relevant risk factors, excluding only clear proxies for the
protected attributes. Barring omitted-variable bias, non-zero coefficients on
the protected traits suggest those factors influenced the decision maker's
actions; in the absence of a compelling justification, such evidence is
suggestive of a discriminatory purpose. It is difficult---and perhaps
impossible---to rigorously define the \emph{influence}, or causal effect, of
largely immutable traits like race on
decisions~\citep{vanderweele2014,greiner2011}, but a regression of this type is
nevertheless considered a reasonable first step to identify discriminatory
motive, both by criminologists and by legal
scholars~\citep{fagan2010report,gaebler2022causal}. For an equal protection
claim to succeed in court, however, one typically needs additional documentary
evidence (e.g., acknowledgement of an illegitimate motive) to bolster the
statistical evidence.

In contrast to disparate treatment, the disparate impact doctrine is concerned
with the effects of a policy, not a decision maker's intentions, and it is the
primary form of discrimination we study in this paper. Under the disparate
impact standard, a practice may be deemed discriminatory if it has an
unjustified adverse effect on protected groups, even in the absence of explicit
categorization or animus. The doctrine stems from statutory rules, rather than
constitutional law, and applies only in certain contexts, such as employment
(via Title VII of the 1964 Civil Rights Act) and housing (via the Fair Housing
Act of 1968). Apart from federal statutes, some states have passed more
expansive disparate impact laws, including Illinois and California.

The disparate impact doctrine was formalized in the landmark U.S. Supreme Court
case \emph{Griggs v. Duke Power Co.} (1971). In 1955, the Duke Power Company
instituted a policy that mandated employees have a high school diploma to be
considered for promotion, which had the effect of drastically limiting the
eligibility of Black employees. The Court found that this requirement had little
relation to job performance, and thus deemed it to have an unjustified disparate
impact. Importantly, the employer's motivation for instituting the policy was
irrelevant to the Court's decision; even if enacted without discriminatory
purpose, the policy was deemed discriminatory in its effects and hence illegal.

More specifically, the legal test of disparate impact developed over half a
century of case law has three principal elements~\citep{grossmanreconciling}:
\begin{enumerate}
    \item \textbf{Adverse impact:} The plaintiff first must establish that the
      policy disproportionately impacts the minority group.
    \item \textbf{No Justification:} Next, the defendant must establish that the
      adverse impact has a substantial justification rooted in a legitimate
      policy goal.
    \item \textbf{Less discriminatory alternative:} Even if the disparate impact
      is justified, the plaintiff can nevertheless prevail if they demonstrate
      that there is an alternative feasible policy with less adverse impact on
      the minority group.
\end{enumerate}
Our concern here is with the second element, namely, whether the adverse impact
has some legitimate policy justification.

As discussed above, the standard statistical test for disparate treatment is a
``kitchen sink'' regression, where one examines the residual explanatory power
of protected group status after including all other available covariates as
controls. That approach, however, is ill-suited to assess whether practices are
rationally justified, which is the relevant standard in disparate impact claims.
\cite{ayres2005three} makes the point persuasively in the context of the
original \emph{Griggs} decision:
\begin{quote}
  ``One could imagine running a regression to test whether an employer was less
  likely to hire African American applicants than white applicants. It would be
  possible to control in this regression for whether the applicant had received
  a high-school diploma. Under the facts of \emph{Griggs}, such a control would
  likely have reduced the racial disparity in the hiring rates. But including in
  the regression a variable controlling for applicants' education would be
  inappropriate. The central point of \emph{Griggs} was to determine whether the
  employer's diploma requirement had a disparate racial impact. The possibility
  that including a diploma variable would reduce the estimated race effect in
  the regression would in no way be inconsistent with a theory that the
  employer's diploma requirement disparately excluded African Americans from
  employment.''
\end{quote}
In short, by including educational status in the regression, one would mask the
policy's unjustified disparate impact.

To assess claims of \emph{unjustified} disparate impact---in \emph{Griggs} and
beyond---one would ideally compare decision rates for similarly \emph{qualified}
groups of applicants (e.g., similarly qualified white and Black candidates).
Unfortunately, if one does not, or cannot, adjust for sufficiently many
covariates, omitted-variable bias may skew results; conversely, if one does
adjust for a rich set of covariates, included-variable bias may corrupt
conclusions.\footnote{%
  A related body of work conceptualizes discrimination as race-specific
  differences in decision error rates~\citep{%
    arnold2020measuring, arnold2021measuring, bohren2022systemic,
    coston2020counterfactual, hardt2016equality%
  }. For example, one might consider differences in loan rejection rates between
  white and Black applicants who in reality would repay their loans (i.e., a
  counterfactual false negative rate). Recent work, however, has observed that
  by conditioning on \emph{ex post} (potential) outcomes---as opposed to
  \emph{ex ante} risk, as we do---such approaches suffer from infra-marginality
  and associated statistical issues, rendering them problematic measures of
  common legal and policy understandings of discrimination~\citep{%
    ayres2005three, mayson2018bias, nilforoshan2022causal, corbett2018measure%
  }.
}

\subsection{A formal illustration of included-variable bias}%
\label{sec:formalization}

We give a simple formal illustration of the statistical phenomenon at issue in
\emph{Griggs}. Consider the following data-generating process, depicted by the
DAG below.
\[
  \begin{tikzpicture}[xscale=3, line width = 0.8pt]
    \node (X) at (0, 0)
    {\vbox{%
        \hbox to 2cm {\hfil\(E\)\hfil}
        \hbox to 2cm {\hfil \footnotesize Experience\hfil}
    }};
    \node (A) at (1, 0)
    {\vbox{%
      \hbox to 2cm {\hfil\(A\)\hfil}
      \hbox to 2cm {\hfil \footnotesize Promotion\hfil}
    }};
    \node (E) at (2, 0)
    {\vbox{%
      \hbox to 2cm {\hfil\(H\)\hfil}
      \hbox to 2cm {\hfil \footnotesize H.S.\ Degree\hfil}
    }};
    \node (C) at (3, 0)
    {\vbox{%
      \hbox to 1.3cm {\hfil\(C\)\hfil}
      \hbox to 1.3cm {\hfil \footnotesize Race\hfil}
    }};

    \draw[->] (X) -- (A);
    \draw[->] (E) -- (A);
    \draw[dashed, <->] (C) --  node[midway, above] {\(\rho\)} (E);
  \end{tikzpicture}
\]
Here, \(C\) indicates the race of an employee, with \(C = 0\) representing a
White employee and \(C = 1\) a Black employee; and \(H\) indicates whether an
employee has a high-school degree. We assume that \(H\) and \(C\) have a joint
distribution given by
\[
  \Pr(H = 1, C = 1) = \Pr(H = 0, C = 0) = \frac {1 + \rho} 4, \qquad \Pr(H = 1,
  C = 0) = \Pr(H = 0, C = 1) = \frac {1 - \rho} 4.
\]
In particular, it follows that both \(H\) and \(C\) are marginally
\(\operatorname{\mathrm{Bern}}(\tfrac 1 2)\) with correlation \(\rho\).
Importantly, we need not assume that race \(C\) ``causes'' an employee to have a
high school degree but only that race and having a high school degree have some
correlation \(\rho\) (although, in this particular example, a causal mechanism
like discriminatory enrollment policies is plausible). 

Further, \(E \sim \operatorname{\mathrm{Unif}}(0, 1)\) denotes an employee's
level of experience, interpreted as the proportion of employees who have been
employed for less time than them. Finally, by \(A\), we denote whether the
employee was promoted. Recalling the details of \emph{Griggs}, we assume that
\[
    \Pr(A = 1 \mid E, H) = E \cdot H.
\]
That is, only employees with high school degrees are promoted, and their
probability of promotion depends on their experience.

A natural quantity to consider is the average difference in promotion rates for
``similarly situated'' Black and White employees---that is, the average
difference in promotion rates for employees with the same experience level and
high school degree. In this case, it is straightforward to show\footnote{%
  See Appendix~\ref{app:ex} for proof.
}
that
\begin{equation}
\label{eq:griggs-dt}
  \mathbb E[\Pr(A = 1 \mid E, H, C = 1) - \Pr(A = 1 \mid E, H, C = 0)] = 0.
\end{equation}
Since race does not factor into promotion decisions, the average
difference in hiring rates for similarly situated Black and White employees is
zero, suggesting an absence of disparate \emph{treatment}.
However, if, following the Court's reasoning in \emph{Griggs}, one were to
consider the average difference in promotion rates for employees with the same
\emph{experience level}---regardless of whether they had a high school
degree---one would obtain
\begin{equation}
\label{eq:griggs-di}
  \mathbb E[\Pr(A = 1 \mid E, C = 1) - \Pr(A = 1 \mid E, C = 0)] = \frac
  \rho 2,
\end{equation}
Following the facts of \emph{Griggs}, we expect \(\rho < 0\), indicating that
Black employees are less likely than White employees to have attained a
high-school degree. As a result, as shown in Eq.~\eqref{eq:griggs-di}, Black
employees at the same level of experience are less likely to be promoted. In
particular, this estimand captures the disparate \emph{impact} of the promotion
policy. This stylized example illustrates the importance of conditioning on the
appropriate variables to avoid included-variable bias in studies of
discrimination. Here, high-school degree attainment is inappropriate to
condition on because it has little relationship to job performance.

Non-parametric estimands like those shown in Eqs.~\eqref{eq:griggs-dt}
and~\eqref{eq:griggs-di} can be statistically challenging to estimate in
practice, so it is common instead to use linear regression or other parametric
models to approximate these estimands. Imagine if an analyst were to fit a
linear kitchen-sink regression model of \(A\) given experience level (\(E\)),
high school degree (\(H\)), the interaction between these two terms (\(E \cdot
H\)), and race (\(C\)),
\[
    \Pr(A = 1 \mid E, H, C) = \beta_0 + \beta_E \cdot E + \beta_H \cdot H +
    \beta_{E : H} \cdot E \cdot H + \beta_C \cdot C.
\]
Here, \(\beta_C\) is typically interpreted as the magnitude of
``discrimination.'' In this example, \(\hat\beta_C\) tends to zero in large
samples, suggesting a lack of disparate \emph{treatment}; see
Appendix~\ref{app:ex}. But, despite fitting a correctly specified model of
decisions, the analyst would fail to detect the disparate impact of the hiring
policy.

Imagine, instead, that the analyst regressed decisions only on years of
experience (\(E\)) and race (\(C\)), as the Court indicated they should, i.e.,
\[
    \Pr(A = 1 \mid E, C) = \beta_0' + \beta_E' \cdot E + \beta_C' \cdot C.
\]
In this case, as with the non-parametric estimand in Eq.~\eqref{eq:griggs-di},
\(\hat \beta_C'\) tends to \(\tfrac \rho 2\), reflecting the fact that it
correctly captures the disparate impact of the promotion policy. (In this
specific example, the parametric and non-parametric estimands are identical, but
this is not always the case.)

Following the Court's ruling, we have assumed that any use of education is
\emph{unjustified}, while use of experience is \emph{justified} when making
promotion decisions. As a result, the key expression of interest is that given
in Eq.~\eqref{eq:griggs-di}, which quantifies disparities in decision rates
after conditioning on experience (\(E\)) alone. In practice, however, all
covariates are usually at least weakly informative about one's qualifications.
In the context of \emph{Griggs}, high school degree attainment, even among
employees with the same level of experience, is likely at least somewhat
predictive of job performance---making it unclear how one should formally define
a measure of disparate impact. In the subsequent section, we introduce one
useful way of conceptualizing disparate impact in this more realistic setting.

\section{A Statistical Approach to Assessing Disparate Impact}%
\label{sec:setup}

We now formally describe our approach to measuring disparate impact---a
procedure we call risk-adjusted regression. The data generating process consists
of draws of tuples \((X, \tilde X, C, W, A)\) where
\[
  X \in \mathcal X, \qquad \tilde X \in \tilde {\mathcal X}, \qquad C \in \{1,
  \ldots, m\}, \qquad W \in \{0, 1\}, \qquad \text{and} \quad A \in \{0, 1\}.
\]
Here \(A\) is a binary decision, \(X\) is the set of covariates on which that
decision is based, and \(C\) represents membership in some protected class. (In
general, we make no assumption about whether \(C\) is encoded in \(X\).) The
binary variable \(W\) represents a latent property of interest to the decision
maker. However, we assume the decision is made based on \(X\) alone, i.e.,
\[
  A \mathrel{\rlap{\(\perp\)}\mkern2mu\perp} W \mid X.
\]
Finally, we assume that an analyst observes the tuple \((\tilde X, C, A, A \cdot
W)\), based on which they seek to estimate disparate impact in the decision
process. Here \(\tilde X\) represents an alternative set of covariates available
to the analyst that may differ from \(\tilde X\) in arbitrary ways.
Additionally, the term \(A \cdot W\) reflects the fact that \(W\) is only
observed by the analyst when action \(A = 1\) occurs.

As a concrete example, consider estimating risk-adjusted racial disparities in
police searches of pedestrians for weapons---the application we discuss in more
detail below. In this case, \(A_i\), \(X_i\), \(C_i\), and \(W_i\),
respectively, indicate whether the \(i\)-th stopped pedestrian was searched, the
information available to the officer when deciding whether to conduct a search,
the stopped individual's race, and whether the individual was in possession of a
weapon. Further, \(\tilde X_i\) denotes the information available to the analyst
from administrative records of the \(i\)-th stop. The fact that the analyst
observes \(A_i \cdot W_i\) means that they know whether searched individuals
were carrying a weapon, but not whether unsearched individuals were carrying a
weapon, which is generally the case. Although we do not consider them further
here, this general framing applies to a wide variety of settings, such as hiring
(where \(W\) represents a prospective employee's productivity and \(A\) whether
or not they were hired) and lending (where \(W\) represents the amount a
prospective borrower will repay and \(A\) represents the bank's lending
decision).

Given this setup, we define \emph{ex ante} \emph{risk} to be:
\begin{equation}
\label{eq:risk}
    R = \Pr(W = 1 \mid X).
\end{equation}
In our policing example, \(R_i\) is the probability that the \(i\)-th stopped
individual, with covariates \(X_i\), is carrying a weapon.\footnote{%
  To allow for the possibility of statistical discrimination~\citep{arrow1973,
  phelps1972, chaudhuri2008statistical} in the decision maker's course of
  action, risk may be conditioned on group membership as well as other
  observables---formally, by including \(C\) among the covariates \(X\).
  However, whether the inclusion of race or other protected attributes in risk
  models is legally appropriate is uncertain. Importantly, our theoretical and
  empirical results remain essentially unchanged whether or not one assumes
  \(X\) encodes \(C\); see Figures~\ref{fig:policy_rb}
  through~\ref{fig:modelchecks_rb} in Appendix~\ref{app:figures}.
}
To ensure that various quantities of interest are well-defined, we assume a
variant of the overlap condition
holds~\citep{rosenbaum_1983a,rosenbaum_1983b}:
\[
  0 < \Pr(C = j \mid R) < 1, \qquad \text{for all } j = 1, \ldots, m.
\]
We note that this version of the overlap condition is immediately implied by an
analogue of the stronger standard overlap condition: \(0 < \Pr(C = j \mid X) <
1\).

Our goal is to quantify whether decisions systematically differ for individuals
at the same level of risk. Many estimands summarizing such potential differences
are possible, but, for simplicity, we consider the following non-parametric
quantity:
\begin{equation}
\label{eq:nonpar}
    \mathbb{E}[\Pr(A = 1 \mid C = j, R) - \Pr(A = 1 \mid C = 1, R)].
\end{equation}
Eq.~\eqref{eq:nonpar} defines risk-adjusted disparities to be the difference in
the probability of taking an action for group \(C = j\) relative to the
reference group \(C = 1\), after accounting for potential differences in risk
across groups. Our overlap assumption ensures that this quantity is
well-defined.

To facilitate computation---and, in particular, the sensitivity analysis
presented below---we approximate the estimand in \eqref{eq:nonpar} by a linear
probability model. Specifically, we estimate \(\Pr(A = 1 \mid C, R)\) as
\begin{equation}
\label{eq:rar}
    \sum_{j=1}^m \hat\beta_j \cdot \mathbf 1(C = j) + \hat\beta_{R} \cdot R,
\end{equation}
where \(\mathbf 1(C = j)\) indicates membership in group \(j\) and
\(\hat\beta_j\) is its corresponding fitted coefficient. The difference between
fitted coefficients, \(\hat\beta_j - \hat{\beta}_1\), yields an estimate of
\eqref{eq:nonpar}---see Appendix~\ref{app:sim} for further details on the
statistical quality of this estimator.

Under this model, if \(\hat\beta_j\) is greater than \(\hat{\beta}_1\), this
indicates that members of group \(j\) are more likely to receive action \(A =
1\) than members of the base group with similar estimated risk. In our policing
example, this means that members of group \(j\) are searched more often than
members of the reference group who were equally likely to be carrying a weapon,
and we would say that such elevated search rates are \emph{unjustified} by risk.
We note, however, that \(\hat\beta_j > \hat\beta_1\) does not imply
\emph{intentional} discrimination---as in \emph{Griggs}, unjustified disparate
impact is possible even under a facially neutral policy undertaken without
animus.

\section{Sensitivity analysis}
\label{sec:sens}

Our definition of disparate impact depends on the true risks, \(R_i\). But, in
practice, analysts observe only partial information on \(X\) and \(W\), and so
at best can construct only imperfect estimates of risks, \(\hat R_i\). For
instance, in our running police example, officers might base search decisions in
part on subtle behavioral cues that are not documented in the data; and analysts
typically would not know whether individuals who were \emph{not} searched were
in fact carrying a weapon. In Section~\ref{sec:sqfny}, we discuss various
approaches to estimating risk in light of these challenges. Whatever approach
one adopts, the accuracy of one's conclusions depends critically on the accuracy
of one's estimates of risk. To address this issue, we develop a novel
sensitivity analysis (implemented in the
\href{https://cran.r-project.org/package=rar}{\texttt{rar}} \texttt{R} package
on CRAN) that draws inspiration from methods for sensitivity analysis popular in
the causal inference literature~\citep{rosenbaum_1983a,simplerules}---though we
emphasize that our framework does not itself involve estimating causal effects.

To start, we assume that there is a known constant \(\epsilon \geq 0\) such that
\begin{equation}
\label{eq:epsilon}
    \frac 1 n \sum_{i=1}^n |R_i - \hat R_i| \leq \epsilon;
\end{equation}
that is, that the true risks and the estimated risks differ on average by at
most \(\epsilon\).\footnote{%
  These differences could arise due to either measurement or modelling error.
  However, since measurement error tends to vanish as the sample size increases,
  it is less of a concern, and is better captured through bootstrapped bounds on
  the sensitivity analysis, as we discuss below.
}
Given this assumption, the goal of our sensitivity analysis
is to understand how different our estimate of disparate impact, \(\hat \beta_j
- \hat \beta_1\), could be if we had access to the true risk \(R_i\) instead of
the estimated risk \(\hat R_i\). In practice, an analyst would examine the
robustness of conclusions to different choices of \(\epsilon\).

We could attempt to bound our estimate of disparate impact by searching over all
possible choices of \(R_i\) satisfying the constraint in Eq.~\eqref{eq:epsilon}.
But such an approach is overly conservative, as the observed data themselves
rule out possible values of \(R_i\). In our policing example, for instance, we
expect the average true risk of searched individuals to approximately equal the
proportion of searched individuals carrying a weapon. Since, by assumption, the
observed data tell us about this latter quantity, they constrain the possible
risk distributions we must consider.

To formalize our approach, a key quantity to consider is the average (true) risk
on each stratum, 
\begin{equation}
  \frac 1 {n_{j,a}} \sum_{i \in \mathcal S_{j,a}} R_i,
\end{equation}
where
\[
  \mathcal S_{j,a} = \{\, i : C_i = j, A_i = a \, \},\qquad n_{j,a} = |\mathcal S_{j,a}|,
\]
i.e., \(\mathcal S_{j,a}\) denotes the stratum containing all those individuals
in group \(j\) for whom the decision maker took action \(a\), and \(n_{j,a}\) is
its size. By the law of large numbers, the average risk on stratum \(\mathcal
S_{j,a}\) is approximated by \(\mathbb E[R \mid C = j, A = a]\). Further,
\begin{align*}
  \mathbb{E}[R \mid C = j, A = a] 
    &= \mathbb{E}[\mathbb{E}[W \mid X] \mid C = j, A = a] \\
    &= \mathbb{E}[\mathbb{E}[W \mid X, U] \mid C = j, A = a] \\
    &= \mathbb{E}[W \mid C = j, A = a],
\end{align*}
where the first equality follows by definition, the second by the fact that \(W
\mathrel{\rlap{\(\perp\)}\mkern2mu\perp} U \mid X\), and the third by the law of
iterated expectations, since \(C\) is a function of \(X\) and \(A\) is a
function of \(X\) and \(U\). When \(A = 1\), by our assumption in
Section~\ref{sec:setup}, the above quantity is identified by data available to
the analyst. In particular, in our policing example, a consistent estimator of
\(\mathbb{E}[W \mid C = j, A = 1]\) is the proportion of searched individuals
possessing a weapon, among those individuals belonging to group \(j\). We denote
by \(\rho_j\) the analyst's estimate of \(\mathbb{E}[W \mid C = j, A = 1]\)
based on the observed data.\footnote{%
  One might alternatively estimate \(\mathbb{E}[W \mid C = j, A = 1]\) by the
  average \emph{estimated} risks on the stratum, namely \(\frac 1 {|\mathcal
  S_{j,1}|} \cdot \sum_{i \in \mathcal S_{j,1}} \hat R_i\). This alternative,
  which we adopt in our empirical analysis in Section~\ref{sec:sqfny}, has the
  interpretive advantage that our sensitivity analysis exactly recovers the
  correct answer when \(\epsilon = 0\).
} 

Finally, putting the pieces together, for each particular \(j^*\) we seek the
largest and smallest values of \(\hat \beta_{j^*} - \hat \beta_1\), as estimated
with \(R_i\), subject to two key constraints: (1) the average absolute deviation
between \(R_i\) and \(\hat R_i\) is less than \(\epsilon\); and (2) \(R_i\) is
consistent with the observed data on each stratum \(\mathcal S_{j,1}\). This is
succinctly expressed as an optimization problem (henceforth \textbf{``the base
problem''}):
\begin{equation}
\label{eq:basic-optim}
  \arraycolsep=1.4pt
  \begin{array}{rrclc}
    \mathop\mathrm{Optimize}\limits_{\mathbf R \in \mathbb R^n} &
      \multicolumn{3}{c}{\displaystyle\hat \beta_{j^*} - \hat \beta_1} \\
    \text{s.t.} & \frac 1 n \sum_{i=1}^n |R_i - \hat R_i| &\leq& \epsilon \\[4pt]
                & \frac 1 {n_{j,1}} \sum_{i \in \mathcal S_{j,1}} R_i
                      &=&     \rho_j, & (j = 1, \ldots, m) \\
                & R_i &\leq&  u_i,    & (i = 1, \ldots, n) \\
                & R_i &\geq&  \ell_i, & (i = 1, \ldots, n)
    \end{array}
\end{equation}
where, by ``\(\mathop\mathrm{Optimize}\),'' we simply mean both maximize and
minimize, and \(\hat R_i\) and \(\rho_j\) are fixed by the data as discussed
above.\footnote{%
  It may seem inappropriate to fix the average risks on the observed stratum
  \(\mathcal S_{j,1}\) to be exactly \(\rho_j\), since \(\rho_j\) is estimated
  with error. One could instead allow \(\rho_j\) to vary within some range
  (e.g., a 95\% confidence interval), solving the parameterized problem for
  various \(\rho_j\) as well as \(\tau_j\), as detailed below. However, doing so
  doubles the dimension of the search space, which is computationally costly,
  and does not fully take advantage of available information about the
  distribution of the estimation error. A more principled and computationally
  feasible approach to dealing with estimation error is to bootstrap confidence
  intervals for the bounds following \citet{zhao2019sensitivity}, as we do in
  Section~\ref{ssec:grounding} below, where \(\rho_j\) is re-estimated in each
  bootstrapped resample, and the base problem is solved with corresponding
  equality constraints.
}
The additional upper and lower bounds on \(R_i\) arise from the fact that the
risks must be probabilities, i.e., \(0 \leq R_i \leq 1\), though our approach
accommodates tighter, individual-specific bounds \(\ell_i \leq R_i \leq u_i\)
set by the analyst.

Below, we give a polynomial time approximation to the solution of the base
problem when the constraints are ``sortable.'' Sortability is a mild hypothesis
which holds in the typical case, \(\ell_i = 0\) and \(u_i = 1\) for all \(i\).
It also holds if, e.g., \(\ell_i\) and \(u_i\) differ from \(\hat R_i\) by the
additive constant \(\Gamma\) on the log odds scale, as is common in many kinds
of sensitivity analysis \citep[e.g.,][]{rosenbaum2002sensitivity}.

\begin{definition}[Sortability]
\label{defn:sortability}
  We say that the \emph{constraints are sortable} if there exists a permutation
  \(\pi\) of \(\{1, \ldots, n\}\) such that \(\hat R_{\pi(1)} \leq \ldots \leq
  \hat R_{\pi(n)}\), \(\ell_{\pi(1)} \leq \ldots \leq \ell_{\pi(n)}\), and
  \(u_{\pi(1)} \leq \ldots \leq u_{\pi(n)}\).
\end{definition}

There are two main obstacles to solving the base problem above. First, the
dimension of the search space is \(n - m\), where \(n\) is the number of
observations and \(m\) is the number of groups. Second, the regression
coefficients \(\hat \beta_j\) are non-convex functions of the true risk vector
\(\mathbf R = (R_1, \ldots, R_n)\). Addressing these twin challenges involves a
detailed analysis of the underlying geometry of the problem. Here we present an
outline of our approach that illustrates the key ideas, with the full exposition
in Appendix~\ref{app:sens}.

It is useful to think of the base problem as an optimization over subproblems
(henceforth \textbf{``the parameterized problem''}) where the average (true)
risk is assumed known and equal to some \(\tau_j\) for \emph{unobserved} strata
\(\mathcal S_{j,0}\) as well. Using the closed form of OLS regression, we can
show that solving the parameterized problem reduces to---although is not
equivalent to---solving the following optimization problem (henceforth
\textbf{``the simplified problem''}):
\begin{equation}
\label{eq:simp-optim}
  \arraycolsep=1.4pt
  \begin{array}{rrclc}
    \mathop\mathrm{Optimize}\limits_{\mathbf R \in \mathbb R^n}
      & \multicolumn{3}{c}{\displaystyle\frac 1 n \displaystyle\sum_{i=1}^n R_i^2} \\
    \mathrm{s.t.} & \frac 1 n \sum_{i=1}^n |R_i - \hat R_i| &\leq& \epsilon, \\
                  & \frac 1 {n_{j,1}} \cdot \sum_{i \in \mathcal S_{j,1}} R_i
                        &=&     \rho_j, & (j = 1, \ldots, m) \\[4pt]
                  & \frac 1 {n_{j,0}} \cdot \sum_{i \in \mathcal S_{j,0}} R_i
                        &=&     \tau_j, & (j = 1, \ldots, m) \\
                  & R_i &\leq&  u_i,    & (i = 1, \ldots, n) \\
                  & R_i &\geq&  \ell_i. & (i = 1, \ldots, n) \\
    \end{array}
\end{equation}
An efficient method for solving the simplified problem then yields an efficient
method for solving the base problem simply by searching over this much smaller
\(m\)-dimensional parameter space of \((\tau_1, \ldots, \tau_m)\). In practice,
\(m\) typically equals two or three, and so this final optimization step can be
solved either using a grid search, as we do in Section~\ref{sec:sqfny}, or using
other optimization techniques for low-dimensional non-convex problems.

The objective of the simplified problem, \(\tfrac 1 n \sum_{i=1}^n R_i^2\), is
convex, as are the other constraints on \(\mathbf R\), and so, in principle,
minimization in the simplified problem can be achieved efficiently and
accurately using interior-point methods~\citep{boyd2004convex}. However, by
carefully examining the KKT conditions~\citep{karush1939minima, MR0047303}, it
is possible to see that the minimizing risk vector must adhere to a special
form, which we term ``minimization normal form.'' Specifically, there exist
thresholds \(t^{\mathrm{lwr}}_{j,a}\) and \(t^{\mathrm{upr}}_{j,a}\) for each
stratum such that the minimizing \(R_i\) is obtained by ``pulling up'' low
estimated risks to \(t^{\mathrm{lwr}}_{j,a}\), ``pulling down'' high estimated
risks to \(t^{\mathrm{upr}}_{j,a}\), and leaving unchanged intermediate risks.
An example of a risk vector in minimization normal form is shown in
Figure~\ref{fig:min-norm}. 

This observation reduces the dimension of the search space from \(n-2m\) to
\(2m\); that is, by the above, one need only search over the \(2m\) thresholds.
One can, however, do better. A careful examination of the KKT conditions reveals
that while the thresholds \(t^{\mathrm{lwr}}_{j,a}\) and
\(t^{\mathrm{upr}}_{j,a}\) may vary by stratum, the gap between them \(\Delta =
t^{\mathrm{upr}}_{j,a} - t^{\mathrm{lwr}}_{j,a}\) must be the same across
strata. Moreover, once \(\Delta\) is fixed, the thresholds
\(t^{\mathrm{lwr}}_{j,a}\) and \(t^{\mathrm{upr}}_{j,a}\) are determined by the
strata-specific average risks \(\rho_j\) and \(\tau_j\). As a result, finding
the minimizing risk vector for the simplified problem reduces to optimizing over
a single parameter, \(\Delta\), which, in our setting, can be done very
efficiently. In particular, we give an \(O(n \cdot \log(m))\) algorithm that
finds the exact solution.

\begin{figure}[t]
    \begin{center}
        \begin{tikzpicture}
            \draw (-1,0) -- (11,0);
            \node[anchor = east] at (-1, 0) {\(R = 0\)};
            \draw (-1, 5) -- (11, 5);
            \node[anchor = east] at (-1, 5) {\(R = 1\)};
    
            \node[anchor = east] at (-1, 2) {\(t^{\mathrm{lwr}}_{a,j}\)};
            \node[anchor = east] at (-1, 4) {\(t^{\mathrm{upr}}_{a,j}\)};
            \draw[decoration={brace,raise=5pt},decorate] (-1.75,2) -- node[left=10pt] {\(\Delta\)} (-1.75,4);
            
            \filldraw[black]  (0,1/2) circle (2pt);
            \filldraw[black]  (1,2/3) circle (2pt);
            \filldraw[black]  (2, 1)  circle (2pt);
            \filldraw[black]  (3,6/5) circle (2pt);
            \filldraw[black]  (4,3/2) circle (2pt);
            \filldraw[black]  (5,7/3) circle (2pt);
            \filldraw[black]  (6,10/3) circle (2pt);
            \filldraw[black]  (7,7/2)  circle (2pt);
            \filldraw[black]  (8,9/2)  circle (2pt);
            \filldraw[black]  (9,14/3) circle (2pt);
            \filldraw[black]  (10,19/4) circle (2pt);
    
            \draw (0,0) -- (0,1); \draw (-0.1,1) -- (0.1,1);
            \draw (1,0) -- (1,2); \draw (0.9,2) -- (1.1,2);
            \draw (2,0) -- (2,2.2); \draw (1.9, 2.2) -- (2.1, 2.2);
            \draw (3,0.2) -- (3, 2.4); \draw (2.9, 0.2) -- (3.1, 0.2); \draw (2.9, 2.4) -- (3.1, 2.4);
            \draw (4,0.2) -- (4,3); \draw (3.9, 0.2) -- (4.1, 0.2); \draw (3.9, 3) -- (4.1, 3);
            \draw (5,1) -- (5, 3.3); \draw (4.9, 1) -- (5.1, 1); \draw (4.9, 3.3) -- (5.1, 3.3);
            \draw (6, 7/3) -- (6, 13/3); \draw (5.9, 7/3) -- (6.1, 7/3); \draw (5.9, 13/3) -- (6.1, 13/3);
            \draw (7, 5/2) -- (7, 19/4); \draw (6.9, 5/2) -- (7.1, 5/2); \draw (6.9, 19/4) -- (7.1, 19/4);
            \draw (8, 3) -- (8, 5); \draw (7.9, 3) -- (8.1, 3); \draw (7.9, 5) -- (8.1, 5);
            \draw (9, 4) -- (9, 5); \draw (8.9, 4) -- (9.1, 4); \draw (8.9, 5) -- (9.1, 5);
            \draw (10, 4.5) -- (10, 5); \draw (9.9, 4.5) -- (10.1, 4.5); \draw (9.9, 5) -- (10.1, 5);
    
            \draw[dashed] (-1, 2) -- (11, 2);
            \draw[dashed] (-1, 4) -- (11, 4);
    
            \filldraw[red] (0.2, 1) circle (2pt);
            \filldraw[red] (1.2, 2) circle (2pt);
            \filldraw[red] (2.2, 2) circle (2pt);
            \filldraw[red] (3.2, 2) circle (2pt);
            \filldraw[red] (4.2, 2) circle (2pt);
            \filldraw[red] (5.2, 7/3) circle (2pt);
            \filldraw[red] (6.2, 10/3) circle (2pt);
            \filldraw[red] (7.2, 7/2) circle (2pt);
            \filldraw[red] (8.2, 4) circle (2pt);
            \filldraw[red] (9.2, 4) circle (2pt);
            \filldraw[red] (10.2, 9/2) circle (2pt);
    
            \draw[->, color = red] (0.2, 1/2) -- (0.2, 0.9);
            \draw[->, color = red] (1.2, 2/3) -- (1.2, 1.9);
            \draw[->, color = red] (2.2, 1) -- (2.2, 1.9);
            \draw[->, color = red] (3.2, 6/5) -- (3.2, 1.9);
            \draw[->, color = red] (4.2, 3/2) -- (4.2, 1.9);
            \draw[->, color = red] (8.2, 9/2) -- (8.2, 4.1);
            \draw[->, color = red] (9.2, 14/3) -- (9.2, 4.1); 
            \draw[->, color = red] (10.2, 19/4) -- (10.2, 4.6);
        \end{tikzpicture}
    \end{center}
    \caption{\emph{%
      Minimization normal form. Black dots represent estimated risks \(\hat
      R_i\), the error bars represent the allowable range \([\ell_i, u_i]\), and
      red dots indicate the corresponding value of the normal form vector
      \(R_i\). Here \(\Delta\) denotes the gap between the upper and lower
      thresholds.
    }}
\label{fig:min-norm}%
\end{figure}

Maximization requires more care, since the maximization problem is non-convex,
and the general problem of maximizing a quadratic objective over a convex set is
NP-hard~\citep{sahni1974computationally}. Using a similar, but more delicate,
version of the techniques used to minimize the simplified problem, we give an
\(O(m \cdot (\epsilon / \gamma)^2 + n)\)-time algorithm for approximating the
maximum of the simplified problem. The key steps described above for optimizing
the base problem are given in Algorithm~\ref{algo:main}.

Putting these results together, we have the following theorem, the proof of
which is given in Appendix~\ref{app:sens}.

\begin{theorem}
\label{thm:main}
  Suppose the constraints are sortable. Consider Algorithm~\ref{algo:main} with
  step-size parameters \(\eta\) and \(\gamma\) and maximum average absolute
  deviation \(\epsilon\). Let \(\delta^*\) denote the true optimum of the
  base problem, and let \(\delta^\dagger\) denote the value of the objective
  returned by Algorithm~\ref{algo:main}. Then, there exists a constant \(c_0\)
  such that Algorithm~\ref{algo:main} runs in time at most
  \[
    c_0 \left( n \cdot \log(n) + m / \gamma^2 \right) \cdot \eta^{-m}.
  \]
  Moreover, there exists a constant \(c_1\) and a problem-specific constant
  \(V(\epsilon)\) (i.e., depending on \(\hat R_i\), \(\ell_i\), \(u_i\), and
  \(\epsilon\)) such that
  \[
    |\delta^* - \delta^\dagger| \leq \frac {c_1 \cdot m (\eta + \gamma)}
    {V(\epsilon)^2}.
  \]
\end{theorem}

We give the definition of \(V(\epsilon)\) in Eq.~\eqref{eq:veps} in
Appendix~\ref{app:sens}. Roughly, \(V(\epsilon)\) is related to the within-group
variance of the estimated risks \(V\),
\[
  V = \frac 1 n \sum_{j=1}^m n_j \cdot \operatorname{\textsc{Var}} \left( (\hat
  R_i)_{i \in \mathcal G_j} \right).
\]
In particular, \(V(\epsilon) \geq V - 4 \epsilon\). Moreover, \(V(\epsilon)\) is
positive whenever the base problem has a finite solution. In our principal
application, it takes a few seconds on standard hardware for the algorithm to
run for \(100\) values of \(\epsilon\) with \(m = 3\) race groups and \(n >
10^6\) observations with negligible approximation error.

As we noted previously, our approach to sensitivity analysis for risk-adjusted
regression is related to sensitivity analysis methods developed in the causal
inference literature, particularly to recent work on \(L_2\) sensitivity
bounds~\citep{zhang2022bounds, huang2022variance}. Our approach, however, is
distinct in important ways that make it uniquely  advantageous in our setting.
First, methods for sensitivity analysis in the causal inference literature that
can be adapted to our setting often involve a large number of
parameters---making them difficult to credibly calibrate---or require parametric
assumptions on the form of the confounding itself~\citep[e.g.,][]{%
  rosenbaum_1983a, simplerules, rosenbaum2002sensitivity, cinelli2020making%
}. In contrast, the single parameter in our sensitivity analysis---\(\epsilon\),
the average absolute deviation between the true and estimated risks---is
straightforward to reason about and explain, particularly to policymakers or
other non-technical stakeholders.

Further, most past sensitivity methods that do involve a single parameter would,
if suitably adapted to our setting, be parameterized in terms of an
\(L_\infty\)-bound \(\Gamma\) on the log odds ratio between true and estimated
risks~\citep[e.g.,][]{zhao2019sensitivity, dorn2022sharp}. However, in realistic
models of confounding, the change in the odds ratio typically cannot be bounded
in this way for all units: some---potentially very small---set of units will
have an estimated risk of approximately \(0\) but a true risk of approximately
\(1\) or \emph{vice versa}. As a result, one may need to set \(\Gamma\) to be
very large to strictly satisfy the assumptions of the approach, yielding
sensitivity bounds that are, in practice, quite conservative. More recent
methods involving \(L_2\) bounds avoid this problem~\citep{zhang2022bounds,
huang2022variance}, but would not yield tight bounds on the estimand of interest
in our setting.

\begin{algorithm}[t]
\raggedright
  \textbf{Input:} The estimated risk vector \(\hat {\mathbf R}\), the lower and
    upper bounds \(\boldsymbol \ell\) and \(\mathbf u\), the average absolute
    difference \(\epsilon\), and the step-size parameters \(\eta\) and
    \(\gamma\). \\
  \textbf{Output:} The minimum and maximum values of \(\hat\beta_j - \hat
    \beta_1\) for all groups \(j = 1, \ldots, m\).
  \begin{algorithmic}[1]
    \STATE Set \(\rho_j \gets \tfrac 1 {|\mathcal S_{j,1}|} \cdot \sum_{i \in
      \mathcal S_{j,1}} \hat R_i\) for \(j = 1, \ldots, m\)
    \STATE Generate a grid of \((\tau_1, \ldots, \tau_m)\) with step size
      \(\eta\)
    \REPEAT
      \STATE Calculate \(\mathbf R^*_{\min}\) minimizing the simplified problem
        with \(\rho_1, \ldots, \rho_m\) and \(\tau_1, \ldots, \tau_m\)
      \STATE Calculate \(\mathbf R^*_{\max}\) maximizing the simplified problem
        to within \(2 n m \gamma\) with \(\rho_1, \ldots, \rho_m\) and \(\tau_1,
        \ldots, \tau_m\)
      \STATE Calculate \(\hat \beta_j - \hat \beta_1\) using both \(\mathbf
        R^*_{\min}\) and \(\mathbf R^*_{\max}\)
    \UNTIL {The grid of \((\tau_1, \ldots, \tau_m)\) is exhausted}
    \RETURN The largest and smallest differences \(\beta_j - \beta_1\) observed
      for all \(j = 2, \ldots, m\)
  \end{algorithmic}
  \caption{(Sensitivity analysis)}
\label{algo:main}
\end{algorithm}

\section{An Application to Policing}
\label{sec:sqfny}

We apply our approach above to investigate the ``stop-and-frisk'' practices of
the New York City Police Department (NYPD).\footnote{%
    Reproduction materials for this analysis are available at
    \url{https://github.com/jgaeb/rar-repro}.
}
Police officers in the United States may stop and question pedestrians if they
have ``reasonable and articulable'' suspicion of criminal activity; officers may
additionally conduct a ``frisk'' (i.e., a brief pat-down of one's outer
garments) if they believe the stopped individual is carrying a weapon. Although
a policy of stopping and frisking individuals is not inherently illegal, in
\emph{Floyd v. City of New York}~(\citeyear{floyd}), a federal district court
ruled that the NYPD carried out such stops with racial animus, violating the
Equal Protection Clause of the Fourteenth Amendment.

The court in \emph{Floyd} was interested in assessing claims of disparate
treatment; here we re-analyze the data with a focus on disparate impact. We
specifically consider frisk decisions, as they have a clear goal of ensuring
officer safety by recovering weapons, and a well-measured outcome---whether a
weapon was in fact found. We study 2.2 million pedestrian stops that occurred
between 2008 and 2011. For each stop, we have detailed information on the date,
time, and location of the stop; the demographics of the stopped individual
(e.g., age, gender, and race); the suspected crime; the reasons prompting the
stop (e.g., ``furtive movements'' or ``suspicious bulge''); and additional
circumstances surrounding the stop (e.g., evasive responses to questioning,
witness reports, or evidence of criminal activity in the vicinity).\footnote{%
  This information is recorded in a standardized way on UF-250 forms that
  officers are required to complete after each stop. A copy of the form can be
  \href{https://www.prisonlegalnews.org/news/publications/blank-uf-250-form-stop-question-and-frisk-report-worksheet-nypd-2016/}{found online}.
}

To start, we note that 1.7\% of frisks turn up a weapon. White pedestrians are
frisked in 44\% of police stops, whereas Black and Hispanic pedestrians are
frisked in 57\% and 58\% of stops, respectively, a 13--14 percentage point gap.
These raw disparities are computed without adjusting for any potentially
explanatory variables, and so represent an extreme case of omitted-variable
bias. At the other extreme is the ``kitchen sink regression'', which adjusts for
all observed pre-frisk covariates in a standard linear probability model. In
this case, stopped Black and Hispanic pedestrians are 3--4 percentage points
more likely to be frisked, relative to white pedestrians with similar recorded
characteristics. These kitchen-sink disparities are suggestive of disparate
treatment (and similar evidence was indeed presented to the court in
\emph{Floyd} to support such an allegation), but they may understate the extent
to which the policy imposes an unjustified disparate impact on racial
minorities, due to included-variable bias.

\subsection{Estimating risk-adjusted disparities}%
\label{sec:sqf_rar}

The key ingredient in applying risk-adjusted regression is estimating the risk
\(R = \Pr(W = 1 \mid X)\), as in Eq.~\eqref{eq:risk}, where \(W\) indicates
whether a stopped individual has a weapon and \(X\) is the information available
to the officer when making their frisk decision. There are, however, two
challenges to estimating this quantity. First, the information available to
officers typically differs from that recorded in administrative datasets
available to analysts. Second, \(W\) remains unobserved for individuals who were
not frisked. These challenges can be mitigated, but not eliminated. Our approach
is thus to estimate these risks as best we can, and then gauge the robustness of
our results to estimation error. For ease of exposition, we adopt a simple
estimation strategy, regressing weapon possession (\(W\)) on the covariates
\(\tilde X\) in the recorded data, fit on the subset of individuals who were, in
reality, frisked.\footnote{%
    Our approach to estimating risk is not the only one, or even the best in any
    given instance. For example, one could, in theory, partially address the
    selection problem by requiring that some subset of individuals be randomly
    frisked, as is sometimes done at roadside checkpoints and airports---though
    this raises legal, ethical, and logistical difficulties. Alternatively, one
    could look for officers who frisk almost everyone they stop---similar to
    recent approaches for studying judicial
    decision-making~\citep[e.g.,][]{kleinberg2018human, dobbie2018effects}. Both
    of these alternatives come with their own limitations, meaning that
    sensitivity analysis is still critical to credible inference.
}
These estimates suffer from some degree of omitted-variable bias and selection
effects, an issue we return to in our sensitivity analysis below.

We start by dividing the original stops into two subsets: (1) the approximately
1 million stops that occurred in 2008 and 2009, which we use to train a risk
model; and (2) the remaining 1.2 million stops that occurred in 2010 and 2011,
to which we apply the fitted risk model to   estimate disparities in frisk
decisions. To estimate risk of weapon possession, we use gradient boosted
decision trees, a non-linear model popular in the machine learning community for
its predictive performance, restricting to stops in the first subset of data in
which a frisk was conducted. Predictive performance and model checks presented
in Figure~\ref{fig:modelchecks} in Appendix~\ref{app:figures} indicate that the
model yields predictions with reasonable performance and that predictions are
well-calibrated across groups. We use this fitted model to produce an estimate
of \emph{ex ante} risk \(\hat R_i\) for every pedestrian stopped in the second
subset of the data, including those who were not frisked. These risk estimates
are shown in Figures~\ref{fig:risk} and~\ref{fig:risk_by_outcome}

\begin{figure}[t]
  \begin{center}
    \includegraphics{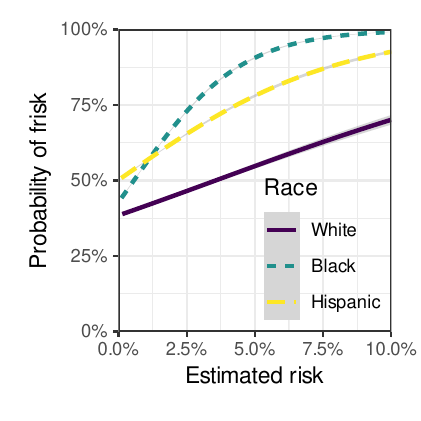}
  \end{center}
  \caption{\emph{%
    Frisk rates vs.\ risk, as estimated via logistic regression curves fit
    separately for each race group. Across risk levels, stopped Black and
    Hispanic pedestrians are frisked substantially more frequently than
    comparably risky white individuals, indicative of disparate impact.
  }}%
\label{fig:policy}
\end{figure}

\begin{figure*}[t]
  \begin{center}
      \includegraphics{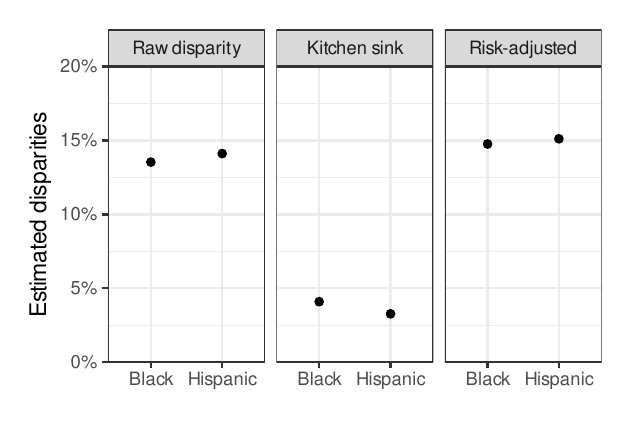}
  \end{center}
  \caption{\emph{%
    Racial gaps in frisk rates adjusting for different sets of covariates, where
    the \(y\)-axis shows the percentage point difference relative to stopped white
    individuals. The left panel shows the raw disparities in frisk rates. As a
    measure of discrimination, raw disparities suffer from omitted-variable
    bias: there may, in theory, be legitimate reasons why Black and Hispanic
    pedestrians are more likely to be frisked. The middle panel shows the
    estimated race effects in a kitchen-sink regression, adjusting for all
    pre-frisk covariates. These estimates suffer from included-variable bias
    because they adjust for features that are correlated with race but unrelated
    to risk. The right panel shows the results of our risk-adjusted regression,
    adjusting exclusively for estimated risk of weapon possession. In all cases,
    estimated standard errors are less than 0.2 percentage points, and so are
    not visible in the plot.
  }}%
\label{fig:comparison}
\end{figure*}

Figure~\ref{fig:policy} shows frisk rates as a function of estimated risk,
disaggregated by race. At every level of risk, stopped Black and Hispanic
pedestrians are frisked at a much higher rate than  stopped white individuals, a
gap that is suggestive of disparate impact in frisk decisions. To add
quantitative detail to these qualitative results, we now compute risk-adjusted
disparities, fitting the linear probability model in Eq.~\eqref{eq:rar} on the
second half of the NYPD data, computing estimates for the Black-white disparity
\(\hat\beta_{\text{Black}} - \hat\beta_{\text{White}}\) and Hispanic-white
disparity \(\hat\beta_{\text{Hispanic}} - \hat\beta_{\text{White}}\).
Figure~\ref{fig:comparison} shows the results, together with the raw disparities
and those estimated from a kitchen-sink model. We find that stopped Black and
Hispanic pedestrians were about 15 percentage points more likely to be frisked
than white pedestrians who were equally likely to be carrying a weapon. Further,
the risk-adjusted disparities are in fact \emph{greater} than the raw
disparities in frisk rates. To understand why, we note that stopped white
pedestrians were, on average, more likely to be carrying a weapon yet less
likely to be frisked than racial minorities; as a result, the risk-adjusted gap
in frisk rates is even larger than the raw, unadjusted gap. Finally, we see that
the kitchen-sink regression dramatically underestimates the extent of disparate
impact faced by minorities. In this case, the kitchen-sink model adjusts for a
variety of features---including whether the stopped individual made ``furtive
movements''---that are correlated with race but are poor predictors of weapon
possession, skewing estimates of disparate impact.\footnote{%
  We exclude hair color and eye color from the kitchen-sink model, since these
  are obvious proxies for race that are effectively unrelated to risk, and would
  thus be excluded in most traditional legal and statistical analyses of
  discrimination. As we would expect, including these variables as controls
  exacerbates the problem of included-variable bias, but our results show that
  such bias can occur even if obviously problematic variables are excluded.
}

\subsection{Sensitivity analysis}%
\label{ssec:grounding}

\begin{figure}[t]
  \begin{center}
    \includegraphics{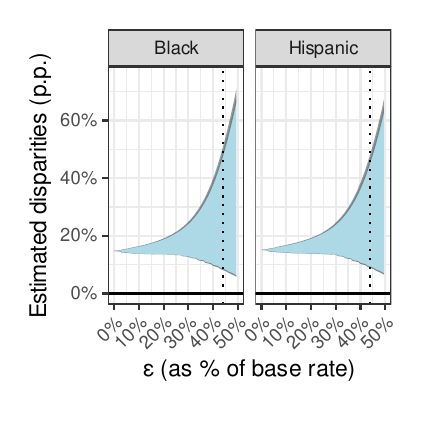}
  \end{center}
  \caption{\emph{%
    Sensitivity of the risk-adjusted disparities in frisk decisions to
    mismeasurement of risk. The blue bands bound our estimates of disparate
    impact as a function of the average absolute difference between the true and
    estimated risks \(\epsilon\), relative to the base rate (1.7\%). The dotted
    line at 44\% (\(\epsilon = 0.7\ \mathrm{p.p.}\)) corresponds to a simulated
    situation with severe confounding. The small grey bands along the top and
    bottom of the blue bands represent 95\% percentile bootstrapped confidence
    intervals~\citep[\(N = 1000\); ][]{zhao2019sensitivity}. The step size for
    the grid search over average risks of the unobserved strata was \(\eta =
    0.05\ \mathrm{p.p.}\) for all groups, and \(\gamma = 0.01\ \mathrm{p.p.}\)
    was the approximation parameter for the maximization routine; see
    Appendix~\ref{app:sens}.
  }}%
\label{fig:sensitivity}
\end{figure}

The disparities computed above aim to circumvent included-variable bias by
adjusting for each individual's estimated risk. But our estimates of risk may
themselves be skewed if officers observe factors that are predictive of risk but
are not recorded in our data. To account for this possibility,
Figure~\ref{fig:sensitivity} displays worst-case bounds on our estimate of
disparate impact as a function of the mismatch between true and estimated
risk---operationalized in terms of \(\epsilon\), as defined in
Eq.~\eqref{eq:epsilon}; see Appendix~\ref{app:sim} for further details on how
the bounds are computed. To ease interpretation, the horizontal axis in the plot
is expressed in terms of the \emph{relative} average absolute difference between
the true and estimated risks: \(\epsilon\) divided by the overall weapon
recovery rate among frisked individuals (1.7\%). Our analysis shows that large
risk-adjusted disparities remain even if we allow the true risk to differ
considerably from our risk estimates. In particular, we would find large
disparate impacts for both Black and Hispanic pedestrians even if the true risks
differed from our risk estimates by 50\% of the base rate.

It is impossible to know the precise extent to which our risk estimates differ
from the true risk. To understand the plausible magnitude of the discrepancy, we
conduct a simulation in which we remove a key set of risk-relevant covariates
from the data, estimate risk based on the reduced information, and then measure
differences between the original and new risk estimates. Specifically, we remove
variables listed in two sections of the UF-250 stop forms that describe the
``circumstances'' prompting the encounter. These sections each consist of 10
binary variables---including, for example, ``fits description'', ``actions
indicative of casing'', and ``changing direction at sight of officer''---that
are crucial for establishing the legal basis of the stop. We then compute the
average absolute difference between our original risk estimates based on the
full, uncensored data and the risk estimates based on the redacted
data.\footnote{%
    As above, we use gradient boosted decision trees to estimate the probability
    of frisk conditional on the remaining, uncensored covariates.
}
We find that this value is \(0.7\) percentage points---or 44\% of the base
rate---which we take as an estimate of \(\epsilon\) in a scenario with severe
unobserved confounding. As shown in Figure~\ref{fig:sensitivity}, this level of
error (indicated by the dashed vertical lines) would yield an estimate of
disparate impact that is, at a minimum, greater than \(7.5\) percentage points
for both Black and Hispanic individuals. This sensitivity analysis suggests our
results are robust to substantial unobserved confounding.

\section{Discussion}
\label{sec:discussion}

We have sought to develop a simple, intuitive framework for addressing the most
serious concerns of included- and omitted-variable bias in disparate impact
studies. On a detailed dataset of police stops, we found that these concerns are
more than hypothetical possibilities. In particular, regressions that adjust for
all available covariates---in line with common legal and statistical
convention---can substantially skew estimates of disparate impact.

Our risk-adjusted regression framework is subject to some important limitations.
First, our method requires access to an outcome to estimate risk. In some
instances, this information is not available to analysts. In other cases, it is
not even clear how to rigorously define the relevant outcome. For example, in
college admissions, decision makers often care about multiple factors in hard to
quantify ways~\citep{garg2021standardized, grossman2023disparate}. Second, to
credibly estimate risk, analysts need sufficiently rich covariate data. Our
sensitivity analysis helps mitigate omitted-variable bias, but it cannot replace
better data. Third, and relatedly, analysts may not have access to race or other
relevant protected characteristics, complicating the analysis. There has,
however, been recent progress on estimating disparities using auxiliary data
sources~\cite[e.g.,][]{kallus2022assessing, mccartan2023estimating}. Finally, we
have modeled decisions using linear probability models with constant slope
across groups. It is straightforward to estimate risk-adjusted disparities with
more flexible decision models. However, relaxing this assumption makes it harder
to gauge the sensitivity of the inferred disparities to inaccurate risk
estimates.

Throughout our analysis, we have estimated ``disparate impact'' by a regression
coefficient on protected-group identity in a model that adjusts for estimated
risk. This procedure is analogous to current practice in discrimination studies,
where we simply replace the usual set of control variables with a single
variable capturing risk. As seen by our formalization of disparate impact in
Eq.~\eqref{eq:nonpar}, we are effectively measuring a particular weighted
average of differences in decision rates across individuals of similar risk.
While intuitively reasonable, this definition raises subtle questions of law and
policy.

Consider, for example, Figure~\ref{fig:policy}, where we plot race-specific
frisk rates as a function of risk. Stopped Black and Hispanic pedestrians are
frisked more often than stopped white pedestrians at every level of risk. As a
result, one would find that racial minorities face disparate impact regardless
of how one averages across risk levels; the precise number might change, but the
qualitative conclusion would remain the same. However, comparing Black and
Hispanic pedestrians, the direction of the disparity depends on the risk level
one considers.\footnote{%
    Such a comparison between minority groups is unusual in disparate impact
    cases, but it illustrates the underlying theoretical issue.
}
Low-risk Hispanic individuals are frisked more often than low-risk Black
individuals, but high-risk Black individuals are frisked more often than their
high-risk Hispanic counterparts. Consequently, a conclusion of disparate impact
between Black and Hispanic pedestrians would depend heavily on the precise
definition applied. The analysis is further complicated if the risk
distributions differ substantially between groups. If, hypothetically, stopped
Hispanic pedestrians were mostly low-risk and stopped Black pedestrians mostly
high-risk, majorities of both groups could argue that they were treated more
harshly than members of the other group who were equally likely to be carrying a
weapon.\footnote{%
  Some scholars have similarly investigated interactions between race and other
  decision-making criteria. For example, \cite{espenshade2004admission} find
  that preferences for underrepresented minorities in college admissions is
  greatest for applicants with SAT scores in the 1200--1300 range, and the
  effect is attenuated at lower scores. That analysis, however, found no score
  ranges where minority applicants faced an absolute disadvantage relative to
  white applicants with equal scores. We do not know of any research that has
  found a change in the direction of the disparities like we see between Black
  and Hispanic pedestrians in Figure~\ref{fig:policy}.
}

The crossing of risk curves that we see in Figure~\ref{fig:policy} is a
potentially widespread phenomenon, and, to our knowledge, disparate impact law
has not yet resolved the underlying conceptual ambiguity it invokes. Many
discussions of disparate impact tacitly assume that policies either consistently
harm or help groups defined by protected traits. Such thinking can be seen in
the original \emph{Griggs} ruling, where the Supreme Court aimed to proscribe
policies that acted as ``built-in headwinds'' for racial minorities. But,
formally, disparate impact law concerns facially race-neutral policies, not
intentional discrimination, and there is no theoretical or empirical guarantee
that such policies will adversely impact all members of a particular group.

A related issue is the extent to which concern for unjustified disparities
compels decision makers to act optimally. This concern connects most closely to
the third element of legal tests of disparate impact outlined in
Section~\ref{ssec:background} above, namely, the existence of a less
discriminatory alternative policy. For example, Figure~\ref{fig:policy} suggests
that officers are only marginally responsive to risk, with the lowest-risk
individuals still frisked more than 40\% of the time. If, instead, officers
frisked only the people with high probability of carrying a weapon, they could
frisk far fewer individuals---and, in particular, far fewer minority
individuals---while recovering almost the same number of
weapons~\citep{goel_2016a}. A more efficient frisk strategy could thus reduce
the burdens of policing on racial minorities while still maintaining public
safety. Such efficiency is indeed one of the aims of statistical risk assessment
tools that are now used in the criminal justice system and beyond to guide
high-stakes decisions~\citep{%
  monahan2016risk, corbett2018measure, chouldechova2018case,
  shroff2017predictive, goel2018accuracy%
}.
If these tools are shown to reduce racial disparities, are policymakers
obliged---legally or ethically---to adopt them? The role of efficiency in
disparate impact claims has largely gone unanswered by the courts, adding yet
another subtlety to defining and measuring disparities. Researchers have only
recently taken up these questions~\citep{%
  corbett2018measure, raghavan2023should, bohren2022systemic,
  elzayn2023measuring, grossmanreconciling%
}.

By foregrounding the role of risk in understanding disparities, we have aimed to
clarify some of the thorny conceptual issues at the heart of disparate impact
analysis. While there are still important unresolved questions, we believe that
our statistical approach provides practitioners with a tractable way to assess
disparities in many domains while avoiding some important pitfalls of
traditional methods. Looking forward, we hope this work spurs further
theoretical and empirical research on discrimination at the intersection of
statistics, economics, law, and public policy.

\clearpage
\appendix

\section{Formalization of Griggs}%
\label{app:ex}

We begin by proving the results from Section~\ref{sec:formalization} for the
parametric formulation of the problem, and then use the parametric results to
prove the non-parametric claims.

\subsection{Linear Regression Formulation}

Recall that the OLS coefficient estimates are given in this case by
\[
  \left( \sum_{i=1}^n
  \begin{bmatrix}
    1
      & X_{0,i}
      & X_{1,i}
      & X_{0,i} \cdot X_{1,i}
      & C_i \\
    X_{0,i}
      & X_{0,i}^2
      & X_{0,i} \cdot X_{1,i}
      & X_{0,i}^2 \cdot X_{1,i}
      & X_{0,i} \cdot C_i \\
    X_{1,i}
      & X_{0,i} \cdot X_{1,i}
      & X_{1,i}^2
      & X_{0,i} \cdot X_{1,i}^2
      & X_{1,i} \cdot C_i \\
    X_{0,i} \cdot X_{1,i}
      & X_{0,i}^2 \cdot X_{1,i}
      & X_{0,i} \cdot X_{1,i}^2
      & X_{0,i}^2 \cdot X_{1,i}^2
      & X_{0,i} \cdot X_{1,i} \cdot C_i \\
    C_i
      & X_{0,i} \cdot C_i
      & X_{1,i} \cdot C_i
      & X_{0,i} \cdot X_{1,i} \cdot C_i
      & C_i^2
  \end{bmatrix} \right)^{-1}
  \left( \sum_{i=1}^n \begin{bmatrix}
    A_i \\
    X_{0,i} \cdot A_i \\
    X_{1,i} \cdot A_i \\
    X_{0,i} \cdot X_{1,i} \cdot A_i \\
    C_i \cdot A_i
  \end{bmatrix} \right)
\]
Note that because of the matrix inversion, we can divide both sums by \(n\).
Therefore, applying the continuous mapping theorem and the strong law of large
numbers, we see that this converges to
\[
  \begin{bmatrix}
    1
      & \mathbb E[E]
      & \mathbb E[H]
      & \mathbb E[E \cdot H]
      & \mathbb E[C] \\
    \mathbb E[E]
      & \mathbb E[E^2]
      & \mathbb E[E \cdot H]
      & \mathbb E[E^2 \cdot H]
      & \mathbb E[E \cdot C] \\
    \mathbb E[H]
      & \mathbb E[E \cdot H]
      & \mathbb E[H^2]
      & \mathbb E[E \cdot H^2]
      & \mathbb E[H \cdot C] \\
    \mathbb E[E \cdot H]
      & \mathbb E[E^2 \cdot H]
      & \mathbb E[E \cdot H^2]
      & \mathbb E[E^2 \cdot H^2]
      & \mathbb E[E \cdot H \cdot C] \\
    \mathbb E[C]
      & \mathbb E[E \cdot C]
      & \mathbb E[H \cdot C]
      & \mathbb E[E \cdot H \cdot C]
      & \mathbb E[C^2]
  \end{bmatrix}^{-1}
  \begin{bmatrix}
    \mathbb E[A] \\
    \mathbb E[E \cdot A] \\
    \mathbb E[H \cdot A] \\
    \mathbb E[E \cdot H \cdot A] \\
    \mathbb E[C \cdot A]
  \end{bmatrix}.
\]

Now, recalling the joint and marginal distributions of these variables, we have
that
\[
  \mathbb E[E] = \frac 1 2, \qquad \mathbb E[H] = \frac 1 2, \qquad \mathbb
  E[C] = \frac 1 2.
\]
Moreover,
\[
  \mathbb E[E^2] = \frac 1 3, \qquad \mathbb E[H^2] = \frac 1 2, \qquad
  \mathbb E[C^2] = \frac 1 2.
\]
By the definition of covariance,
\[
  \rho = \frac {\mathbb E[H \cdot C] - \mathbb E[H] \cdot \mathbb E[C]}
  {\sqrt{\operatorname{\textsc{Var}}(H) \cdot \operatorname{\textsc{Var}}(C)}} =
  \frac {\mathbb E[H \cdot C] - \frac 1 4} {\sqrt{\frac 1 4 \cdot \frac 1 4}} =
  4 \cdot \mathbb E[H \cdot C] - 1,
\]
whence \(\mathbb E[H \cdot C] = \tfrac {\rho + 1} 4\). By independence,
\begin{align*}
  \mathbb E[E \cdot H] &= \mathbb E[E] \cdot \mathbb E[H] = \frac 1 4,
    & \mathbb E[E \cdot C] &= \mathbb E[E] \cdot \mathbb E[C] = \frac 1 4, \\
  \mathbb E[E^2 \cdot H] &= \mathbb E[E^2] \cdot \mathbb E[H] = \frac 1 6,
    & \mathbb E[E \cdot H^2] &= \mathbb E[E \cdot H] = \frac 1 4, \\
  \mathbb E[E^2 \cdot H^2] &= \mathbb E[E^2 \cdot H] = \frac 1 6,
    & \mathbb E[E \cdot H \cdot C] &= \mathbb E[E] \cdot \mathbb E[H \cdot C] =
    \frac {\rho + 1} 8, \\
\end{align*}
Finally,
\begin{align*}
  \mathbb E[A] &= \mathbb E[\mathbb E[A \mid E, H]]
    & \mathbb E[E \cdot A] &= \mathbb E[\mathbb E[E \cdot A \mid E, H]] \\
               &= \mathbb E[E \cdot H] &
                           &= \mathbb E[E^2 \cdot H] \\
               &= \frac 1 4, &
                           &= \frac 1 6, \\ \\
  \mathbb E[H \cdot A] &= \mathbb E[\mathbb E[H \cdot A \mid E, H]]
    & \mathbb E[E \cdot H \cdot A] &= \mathbb E[\mathbb E[E \cdot H \cdot A \mid
    E, H]] \\
                        &= \mathbb E[E \cdot H^2] &
                                   &= \mathbb E[E^2 \cdot H^2] \\
                        &= \frac 1 4, &
                                   &= \frac 1 6, \\ \\
  \mathbb E[C \cdot A] &= \mathbb E[\mathbb E[C \cdot A \mid E, H, C]] \\
                       &= \mathbb E[E \cdot H \cdot C] \\
                       &= \frac {\rho + 1} 8.
\end{align*}

Thus, we have that
\[
  \begin{bmatrix}
    1
      & \mathbb E[E]
      & \mathbb E[H]
      & \mathbb E[E \cdot H]
      & \mathbb E[C] \\
    \mathbb E[E]
      & \mathbb E[E^2]
      & \mathbb E[E \cdot H]
      & \mathbb E[E^2 \cdot H]
      & \mathbb E[E \cdot C] \\
    \mathbb E[H]
      & \mathbb E[E \cdot H]
      & \mathbb E[H^2]
      & \mathbb E[E \cdot H^2]
      & \mathbb E[H \cdot C] \\
    \mathbb E[E \cdot H]
      & \mathbb E[E^2 \cdot H]
      & \mathbb E[E \cdot H^2]
      & \mathbb E[E^2 \cdot H^2]
      & \mathbb E[E \cdot H \cdot C] \\
    \mathbb E[C]
      & \mathbb E[E \cdot C]
      & \mathbb E[H \cdot C]
      & \mathbb E[E \cdot H \cdot C]
      & \mathbb E[C^2]
  \end{bmatrix}^{-1}
\]
equals
\[
  \begingroup
  \def\arraystretch{2.5}
  \begin{bmatrix}
    1          & \dfrac 1 2 & \dfrac 1 2           & \dfrac 1 4          & \dfrac 1 2 \\
    \dfrac 1 2 & \dfrac 1 3 & \dfrac 1 4           & \dfrac 1 6          & \dfrac 1 4 \\
    \dfrac 1 2 & \dfrac 1 4 & \dfrac 1 2           & \dfrac 1 4          & \dfrac {\rho + 1} 4 \\
    \dfrac 1 4 & \dfrac 1 6 & \dfrac 1 4           & \dfrac 1 6          & \dfrac {\rho + 1} 8 \\
    \dfrac 1 2 & \dfrac 1 4 & \dfrac {\rho + 1} 4  & \dfrac {\rho + 1} 8 & \dfrac 1 2
  \end{bmatrix}^{-1} =
  \endgroup
  \begingroup
  \def\arraystretch{2}
  \begin{bmatrix}
    & \frac {7 \rho + 9} {\rho + 1}   & -12 & - \frac {6 \rho + 8} {\rho + 1}     & 12  & - \frac 2 {\rho + 1} \\
    & -12                             &  24 & 12                                  & -24 & 0 \\
    & - \frac {6 \rho + 8} {\rho + 1} &  12 & \frac {12*\rho^2 - 16} {\rho^2 - 1} & -24 & \frac {4 \rho} {\rho^2 - 1} \\
    & 12                              & -24 & -24                                 & 48  & 0 \\
    & - \frac 2 {\rho + 1}            &   0 & \frac {4 \rho} {\rho^2 - 1}         & 0   & - \frac 4 {\rho^2 - 1}
  \end{bmatrix}
  \endgroup
\]
and so the OLS regression coefficients converge almost surely to
\[
  \begingroup
  \def\arraystretch{2}
  \begin{bmatrix}
    & \frac {7 \rho + 9} {\rho + 1}   & -12 & - \frac {6 \rho + 8} {\rho + 1}     & 12  & - \frac 2 {\rho + 1} \\
    & -12                             &  24 & 12                                  & -24 & 0 \\
    & - \frac {6 \rho + 8} {\rho + 1} &  12 & \frac {12*\rho^2 - 16} {\rho^2 - 1} & -24 & \frac {4 \rho} {\rho^2 - 1} \\
    & 12                              & -24 & -24                                 & 48  & 0 \\
    & - \frac 2 {\rho + 1}            &   0 & \frac {4 \rho} {\rho^2 - 1}         & 0   & - \frac 4 {\rho^2 - 1}
  \end{bmatrix}
  \endgroup
  \begingroup
  \def\arraystretch{2.5}
  \begin{bmatrix}
    \dfrac 1 4 \\
    \dfrac 1 6 \\
    \dfrac 1 4 \\
    \dfrac 1 6 \\
    \dfrac {\rho + 1} 8
  \end{bmatrix}
  = \begin{bmatrix}
    \;0\; \\
    0 \\
    0 \\
    1 \\
    0
  \end{bmatrix}.
  \endgroup
\]

On the other hand, if the analyst were to drop high school graduation from their
regression, then we would have that the regression coefficients converge almost
surely to
\[
  \begin{bmatrix}
    1
      & \mathbb E[E]
      & \mathbb E[C] \\
    \mathbb E[E]
      & \mathbb E[E^2]
      & \mathbb E[E \cdot C] \\
    \mathbb E[C]
      & \mathbb E[E \cdot C]
      & \mathbb E[C^2]
  \end{bmatrix}^{-1}
  \begin{bmatrix}
    \mathbb E[A] \\
    \mathbb E[E \cdot A] \\
    \mathbb E[C \cdot A]
  \end{bmatrix} =
  \begingroup
  \def\arraystretch{2.5}
  \begin{bmatrix}
    1         & \dfrac 1 2 & \dfrac 1 2 \\
    \dfrac 1 2 & \dfrac 1 3 & \dfrac 1 4 \\
    \dfrac 1 2 & \dfrac 1 4 & \dfrac 1 3
  \end{bmatrix}^{-1}
  \begin{bmatrix}
    \dfrac 1 4 \\
    \dfrac 1 6 \\
    \dfrac {\rho + 1} 8
  \end{bmatrix} =
  \begin{bmatrix}
    - \dfrac \rho 4 \\
    \dfrac 1 2 \\
    \dfrac \rho 2
  \end{bmatrix}.
  \endgroup
\]

\subsection{Nonparametric Formulation}

Recall the two different non-parametric estimands:
\[
  \mathbb E[\mathbb E[A \mid E, H, C = 1] - \mathbb E[A \mid E, H, C = 0]]
\]
and
\[
  \mathbb E[\mathbb E[A \mid E, C = 1] - \mathbb E[A \mid E, C = 0]].
\]

We note that in the first case, since \(\mathbb E[A \mid E, H, C] = E \cdot H\),
we have that
\[
  \mathbb E[\mathbb E[A \mid E, H, C = 1] - \mathbb E[A \mid E, H, C = 0]] =
  \mathbb E[E \cdot H - E \cdot H] = 0.
\]
In the second case, using the expectations calculated above, we have that
\begin{align*}
  \mathbb E[A \mid E, C = c] &= \mathbb E[A \mid E, H = 1, C = c] \cdot \Pr(H =
  1 \mid E, C = c)
    \\
    &\hspace{1cm} + \mathbb E[A \mid E, H = 0, C = c] \cdot \Pr(H = 0 \mid E, C
    = c) \\
    &= E \cdot \Pr(H = 1 \mid E, C = c) \\
    &= E \cdot \Pr(H = 1 \mid C = c) \\
    &= E \cdot \frac {\mathbb E[H \cdot C]} {\mathbb E[C]} \\
    &= E \cdot \frac {\frac {1 - (-1)^c \cdot \rho} 4} {\frac 1 2} \\
    &= E \cdot {\frac {1 - (-1)^c \cdot \rho} 2}.
\end{align*}
Here we have used the fact that \(A = 0\) when \(H = 0\) in the second equality,
the independence of \(H\) from \(E\) and \(C\) in the third, and various
expectations calculated above in the remaining equalities. Thus, it follows that
\[
  \mathbb E[\mathbb E[A \mid E, C = 1] - \mathbb E[A \mid E, C = 0]] =
  \mathbb E \left[E \cdot \frac {1 + \rho} 2 - E \cdot \frac {1 - \rho} 2
  \right] = \mathbb E[E \cdot \rho] = \frac \rho 2.
\]
This exactly equals \(\beta_C'\) in the second regression.

\section{Sensitivity Analysis}%
\label{app:sens}

In what follows, we formally justify the sensitivity analysis laid out in
Section~\ref{sec:sens}, deriving algorithms to efficiently solve the
optimization problem it gives rise to and rigorously verifying their correctness
and run-times. This discussion is structured in three parts. In the first part,
by introducing a new low-dimensional family of parameters, we show how to reduce
the base optimization problem of Eq.~\eqref{eq:basic-optim} into the simplified
problem of Eq.~\eqref{eq:simp-optim}. In the second and third parts, we show how
to efficiently find the minimum and maximum values of the simplified
optimization problem, respectively. 

\begin{remark}%
\label{rmk:notation}
  In the main text, to for expositional clarity, we parameterized our
  optimization problems in terms of \emph{average} risks. However, it is
  mathematically more natural to work with the \emph{total} risks. Therefore,
  throughout this appendix we let \(\rho_j\) and \(\tau_j\) denote the
  \emph{total} risk in each stratum, i.e., we rewrite the relevant constraints
  of the base problem and the simplified problem as follows:
  \[
    \sum_{i \in \mathcal S_{j,1}} R_i = \rho_j, \qquad \sum_{i \in \mathcal
    S_{j,0}} R_i = \tau_j.
  \]
\end{remark}

\subsection{Simplifying the objective}%
\label{ssec:simp}

To simplify the optimization problem in Eq.~\eqref{eq:basic-optim}, we can avail
ourselves of the fact that OLS has a closed-form solution. We augment the
notation in Section~\ref{sec:sens} as follows:
\[
  \mathcal G_j = \mathcal S_{j,0} \cup \mathcal S_{j,1} = \{\, i : C_i = j \,\},
\]
i.e., \(\mathcal G_j\) denotes the set of indices belonging to the \(j\)-th
group.

\begin{lemma}%
\label{lem:est}
  Let \(n_j = |\mathcal G_j| > 0\) denote the number of individuals in the
  \(j\)-th group; let \(\sigma_j\) denote the search rate in group \(j\), i.e.,
  \(\sigma_j = \tfrac 1 {n_j} \sum_{i \in \mathcal G_j} A_i\); and let \(r_j\)
  and \(t_j\) denote
  \[
    r_j = \sum_{i \in \mathcal S_{j,1}} R_i, \qquad t_j = \sum_{i \in \mathcal
    S_{j,0}} R_i,
  \]
  i.e., the total risk of the searched and unsearched individuals belonging to
  the \(j\)-th group, respectively.

  If \(R_i\) is constant within each of the \(m\) groups, then the OLS estimate
  need not exist due to collinearity. Otherwise, the OLS estimate of the
  coefficients \(\hat {\boldsymbol \beta}\) in Eq.~\eqref{eq:rar} satisfies
  \begin{equation}
  \label{eq:est}
    \hat {\boldsymbol \beta} = \begin{bmatrix}
      \sigma_1 \\
      \vdots \\
      \sigma_m \\
      0
    \end{bmatrix}
    + \frac
      {\frac 1 n \left[ \sum_{j=1}^m \sigma_j \cdot t_j - (1 - \sigma_j) \cdot
        r_j \right]}
      {\frac 1 n \left[ \sum_{i=1}^n R_i^2 - \sum_{j=1}^m \frac {(r_j +
        t_j)^2} {n_j} \right]}
      \begin{bmatrix}
        \frac {r_1 + t_1} {n_1} \\
        \vdots \\
        \frac {r_m + t_m} {n_m} \\
        -1
    \end{bmatrix}.
  \end{equation}
\end{lemma}

\begin{proof}
  Recall that our design matrix and outcome variables take the form
  \begin{align*}
    \mathbf{X} &= \begin{bmatrix}
      \mathbf 1 (C_1 = 1)
        & \> \hdots \>
        & \mathbf 1(C_1 = m)
        & R_1 \\
      \vdots
        & \ddots
        & \vdots 
        &\vdots \\
      \mathbf 1 (C_n = 1)
        & \hdots
        & \mathbf 1 (C_n = m)
        & R_n
    \end{bmatrix},
      &
    \mathbf{Y} &= \begin{bmatrix}
      A_1 \\
      \vdots \\
      A_n
    \end{bmatrix}.
  \end{align*}
  The OLS estimate of the coefficients \(\hat {\boldsymbol \beta}\) is then
  simply \((\mathbf{X}^\top \mathbf{X})^{-1} \mathbf{X} ^\top\mathbf{Y}\).
  
  Note that by definition, we have that
  \begin{equation*}
  \renewcommand*{\arraystretch}{1.2}
    \mathbf{X}^\top \mathbf{X}
    = \begin{bmatrix}
      n_1
        &
        &
        & \sum_{i \in \mathcal G_1} R_i\\
      {}
        & \ddots
        &
        & \vdots \\
      {}
        &
        & n_m
        & \sum_{i \in \mathcal G_m} R_i \\
      \sum_{i \in \mathcal G_1} R_i
        & \hdots
        & \sum_{i \in \mathcal G_m} R_i
        & \sum_{i=1}^n R_i^2 
    \end{bmatrix}
    = \begin{bmatrix}
     n_1
       &
       &
       & r_1 + t_1 \\
     {}
       & \ddots
       &
       & \vdots \\
     {}
       &
       & n_m
       & r_m + t_m \\
     r_1 + t_1
       & \hdots
       & r_m + t_m
       & \sum_{i=1}^n R_i^2
    \end{bmatrix}.
  \end{equation*}
  
  Similarly,
  \begin{equation*}
  \renewcommand*{\arraystretch}{1.2}
    \mathbf{X}^\top \mathbf{Y} = \begin{bmatrix}
      \sum_{i \in \mathcal G_1} A_i \\
      \vdots \\
      \sum_{i \in \mathcal G_m} A_i \\
      \sum_{i=1}^n R_i \cdot A_i
    \end{bmatrix}
    = \begin{bmatrix}
      n_1 \cdot \sigma_1 \\
      \vdots \\
      n_m \cdot \sigma_m \\
      \sum_{j=1}^m r_j
    \end{bmatrix}.
  \end{equation*}

  Note that \(\mathbf{X}^\top \mathbf{X}\) has the form
  \begin{equation*}
    \begin{bmatrix}
      D   & v^\top \\
      v   & \sum_{i=1}^n R_i^2,
    \end{bmatrix}
  \end{equation*}
  where \(D\) is diagonal. Then, we see that since \(D\) is invertible, it
  follows that \(\mathbf{X}^\top \mathbf{X}\) is invertible if the Schur
  complement of the \(D\) is invertible, i.e., non-zero; that is, if
  \begin{equation*}
    \sum_{i=1}^n R_i^2 - \sum_{j=1}^m \frac{(r_j + t_j)^2} {n_j} \neq 0.
  \end{equation*}
  Recalling the definitions of \(r_j\) and \(t_j\), we see that the left-hand
  expression equals
  \begin{equation*}
    \sum_{j=1}^m \sum_{i \in \mathcal G_j} \left( R_i - \frac {r_j + t_j} {n_j}
    \right)^2 = \sum_{j=1}^m n_j \cdot \operatorname{\textsc{Var}} \left(
    (R_i)_{i \in \mathcal G_j} \right).
  \end{equation*} 
  Now, the variance of a set is only zero if all the elements of the set are
  equal, i.e., if for all \(i \in \mathcal G_j\), \(R_i\) equals the average
  risk in that group, \(\tfrac {r_j + t_j} {n_j}\).
  
  Assume this is not the case, i.e., that \(\mathbf{X}^\top \mathbf{X}\) is
  invertible. Since \(\mathbf{X}^\top \mathbf{X}\) is an arrowhead matrix, we
  can invert it using the Sherman-Morrison Formula. In particular, we obtain
  that
  \begin{equation}
  \label{eq:inverse}
    (\mathbf{X}^\top \mathbf{X})^{-1} =
    \begin{bmatrix}
     \frac 1 {n_1}
       &
       &
       & \\
     {}
       & \ddots
       &
       & \\
     {}
       &
       & \frac 1 {n_m}
       & \\
     {}
       &
       &
       & 0
    \end{bmatrix}
    + \frac 1 {\sum R_i^2 - \sum_{j=1}^m \frac {(r_j + t_j)^2} {n_j}}
    \begin{bmatrix}
      \frac {r_1 + t_1} {n_1} \\
      \vdots \\
      \frac {r_m + t_m} {n_m} \\
      -1
    \end{bmatrix}
    \begin{bmatrix}
      \frac {r_1 + t_1} {n_1} \\
      \vdots \\
      \frac {r_m + t_m} {n_m} \\
      -1
    \end{bmatrix}^\top.
  \end{equation}
  Combining this with the expression derived for \(\mathbf{X}^\top\mathbf{Y}\)
  above and dividing the numerator and denominator by \(n\) yields that
  \begin{equation}
    \hat {\boldsymbol \beta} =
    \begin{bmatrix}
      \sigma_1 \\
      \vdots \\
      \sigma_m \\
      0
    \end{bmatrix}
    +
    \frac
    {\frac 1 n \left[ \sum_{j=1}^m \sigma_j \cdot t_j - (1 - \sigma_j) \cdot r_j
      \right]}
    {\frac 1 n \left[ \sum R_i^2 - \sum_{j=1}^m \frac {(r_j + t_j)^2} {n_j}
      \right]}
    \begin{bmatrix}
      \frac {r_1 + t_1} {n_1} \\
      \vdots \\
      \frac {r_m + t_m} {n_m} \\
      -1
    \end{bmatrix}.
  \end{equation}
\end{proof}

We note that Lemma~\ref{lem:est} provides the following useful condition on when
the base problem has a meaningful solution:

\begin{corollary}%
\label{cor:well-def}
  The maximum and minimum values of the base problem are both finite if and only
  if it is not the case that \(\ell_i \leq \rho_j \leq u_i\) for all \(i \in
  \mathcal G_j\) and \(j = 1, \ldots, m\), and in addition
  \[
    \frac 1 n \sum_{j=1}^m \sum_{i \in \mathcal G_j} \left| \hat R_i - \frac
    {\rho_j} {n_{j,1}} \right| \leq \epsilon.
  \]
\end{corollary}

\begin{proof}
  The proof follows immediately from noting that since the averages of the risks
  on the observed strata \(\mathcal S_{j,1}\) are fixed at \(\rho_j\), the risks
  for the whole \(j\)-th group can only all be equal if the risks of the
  observed \emph{and} the unobserved individuals are uniformly equal to
  \(\rho_j\).
\end{proof}

The following corollary completes the simplification of the optimization problem
in Eq.~\eqref{eq:basic-optim} in the case where the average true risk in the
unobserved strata is also assumed known (\textbf{``the parameterized problem''}
of Section~\ref{sec:sens}),\footnote{%
  See Remark~\ref{rmk:notation} above for the slight difference in notation
  between here and Section~\ref{sec:sens}.
}
i.e., of
\begin{equation}
\label{eq:fixed-optim}
  \arraycolsep=1.4pt
  \begin{array}{rrclc}
    \mathop\mathrm{Optimize}\limits_{\mathbf R \in \mathbb R^n}
      & \multicolumn{3}{c}{\hat\beta_{j^*} - \hat\beta_1} \\
    \mathrm{s.t.} & \frac 1 n \sum_{i=1}^n |R_i - \hat R_i|
                        &\leq& \epsilon, \\
                  & \sum_{i \in \mathcal S_{j,1}} R_i
                        &=&     \rho_j, & (j = 1, \ldots, m) \\[4pt]
                  & \sum_{i \in \mathcal S_{j,0}} R_i
                        &=&     \tau_j,    & (j = 1, \ldots, m) \\
                  & R_i &\leq&  u_i,    & (i = 1, \ldots, n) \\
                  & R_i &\geq&  \ell_i. & (i = 1, \ldots, n) \\
  \end{array}
\end{equation}

\begin{corollary}%
\label{cor:simp}
  The optimal solutions of the parameterized problem in
  Eq.~\eqref{eq:fixed-optim}, if they are both well-defined, are also the
  optimal solutions of the simplified problem in Eq.~\eqref{eq:simp-optim}.
\end{corollary}

\begin{proof}
  Let \(\hat {\boldsymbol \beta}\) be as in Eq.~\eqref{eq:rar}, and \(r_j\) and
  \(t_j\) as in Lemma~\ref{lem:est}.\footnote{%
    In particular, strictly speaking, \(\hat {\boldsymbol \beta}\), \(r_j\), and
    \(t_j\) should be written \(\hat {\boldsymbol \beta}(\mathbf R)\),
    \(r_j(\mathbf R)\), and \(t_j(\mathbf R)\), since they depend on \(\mathbf
    R\).
  }
  Then, it follows directly from Eq.~\eqref{eq:est} that we have that for all
  \(\mathbf{R}\),
  \begin{equation*}
    \hat \beta_j - \hat \beta_1 = \sigma_j - \sigma_1
    + \frac
    {\frac 1 n \left[ \sum_{j=1}^m \sigma_j \cdot t_j - (1 - \sigma_j) \cdot r_j
      \right]}
    {\frac 1 n \left[ \sum_{i=1}^n R_i^2 - \sum_{j=1}^m \frac {(r_j + t_j)^2}
    {n_j} \right]}
    \left( \frac {r_j + t_j} {n_j} - \frac {r_1 + t_1} {n_1} \right).
  \end{equation*}
  Since the constraints of the parameterized problem fix \(r_j = \rho_j\) and
  \(t_j = \tau_j\) for all feasible \(\mathbf R\), the objective function of the
  parameterized problem has the form
  \begin{equation}
  \label{eq:simp}
    a + \frac b {x - c},
  \end{equation}
  where
  \begin{align*}
    a &= \sigma_j - \sigma_1, \\
    b &= \frac 1 n \left[ \sum_{j=1}^n \sigma_j \cdot \tau_j - (1 - \sigma_j)
      \cdot \rho_j \right] \cdot \left( \frac {\rho_j + \tau_j} {n_j} - \frac
      {\rho_1 + \tau_1} {n_1} \right), \\
    c &= \frac 1 n \sum_{j=1}^m \frac {(\rho_j + \tau_j)^2} {n_j}, \\
    x &= \frac 1 n \sum_{i=1}^n R_i^2.
  \end{align*}
  If \(b = 0\), then the objective does not depend on \(\mathbf R\), and so
  there is nothing to prove. Otherwise, by continuity and the fact that the
  feasible region is convex, and hence connected, and compact, the range of the
  objective in Eq.~\eqref{eq:simp-optim} over the feasible region is a closed
  interval \(I\). As shown in the proof of Lemma~\ref{lem:est} and
  Corollary~\ref{cor:well-def}, \(x \geq c\), and \(x > c\) if the optima are
  both well-defined. Consequently, we can assume the interval is strictly
  positive, i.e., \(I \subseteq (c, \infty)\). The derivative of
  Eq.~\eqref{eq:simp} exists everywhere on \(I\) and is, moreover, non-zero and
  continuous, and so \(a + \tfrac b {x - c}\) is strictly monotone on \(I\).
  Therefore, the maximum and minimum must occur at the endpoints, i.e., at the
  maximum and minimum values of \(x\).
\end{proof}

We note that since \(b\) is not guaranteed to be positive, the maximum of the
simplified problem in Eq.~\eqref{eq:simp-optim} does not necessarily correspond
to the maximum of the base problem in Eq.~\eqref{eq:basic-optim} and \emph{vice
versa}.

\subsection{Minimization}%
\label{ssec:min}

In the case of minimization, the simplified problem in Eq.~\eqref{eq:simp-optim}
defines a convex optimization problem, which, as noted in Section~\ref{sec:sens}
above, can consequently be solved efficiently using standard interior-point
methods. However, these results can be improved in a problem-specific way to
yield a simpler, and, in practice, more computationally tractable method of
finding exact solutions. Moreover, they parallel the solution method for
maximization, which is not a convex problem, and hence to which standard
interior point methods cannot be applied.

\subsubsection{Optimizing over a single stratum}

We begin by solving a simpler version of the problem, where we restrict to a
single stratum. In the case of minimization---as in the case of maximization, as
we show below---solutions to the minimization problem must be in a certain
normal form. While searching over all risk vectors is prohibitively difficult,
searching over normal form risk vectors can be done straightforwardly in linear
time.

\begin{definition}[Minimization normal form]
  We say that a risk vector \(\mathbf R\) is in \emph{minimization normal form}
  if there exist thresholds \(t^{\mathrm{lwr}}_{a,j}\) and
  \(t^{\mathrm{upr}}_{a,j}\) such that for \(i \in \mathcal S_{j,a}\),
  \begin{equation*}
    R_i = \begin{cases}
      \min(u_i, t^{\mathrm{lwr}}_{a,j}) & \hat R_i \leq t^{\mathrm{lwr}}_{a,j}, \\
      \hat R_i & t^{\mathrm{lwr}}_{a,j} \leq \hat R_i \leq t^{\mathrm{upr}}_{a,j}, \\
      \max(\ell_i, t^{\mathrm{upr}}_{a,j}) & t^{\mathrm{upr}}_{a,j} \leq \hat R_i,
    \end{cases}
  \end{equation*}
  and, in addition, \(t^{\mathrm{upr}}_{a,j} - t^{\mathrm{lwr}}_{a,j}\) is equal
  to some \(\Delta\) for all \(a = 0, 1\) and \(j = 1, \ldots, m\).
\end{definition}

In the context of a single stratum, we refer to \(t^{\mathrm{upr}}\) and
\(t^{\mathrm{lwr}}\) for notational simplicity. We also assume that \(\mathcal
S_{j,a} = \{1, \ldots, n\}\) for the same reason.

Minimization normal form results from ``pushing'' \(\hat R_i\) up to
\(t^{\mathrm{lwr}}\) or down to \(t^{\mathrm{upr}}\) as far as possible. An
example of a risk vector in minimization normal form is shown in
Figure~\ref{fig:min-norm}. As the name suggests, risk vectors minimizing the
objective must be in minimization normal form.

\begin{lemma}%
\label{lem:minimize}
  Consider the optimization problem (\textbf{``the single stratum minimization
  problem''})
  \begin{equation}
  \label{eq:ss-optim}
    \arraycolsep=1.4pt
    \begin{array}{rrclc}
      \mathop{\mathrm{Minimize}}\limits_{\mathbf R \in \mathbb R^n}
        & \multicolumn{3}{c}{\displaystyle\frac 1 n \displaystyle\sum_{i=1}^n R_i^2} \\
      \mathrm{s.t.} & \frac 1 n \sum_{i=1}^n |R_i - \hat R_i|
                                       &\leq& \epsilon, \\
                    & \sum_{i=1}^n R_i &=&    \mu, \\
                    & R_i              &\leq&  u_i,    & (\forall i) \\
                    & R_i              &\geq&  \ell_i. & (\forall i) \\
      \end{array}
  \end{equation}
  Suppose that \(\mathbf R^*\) is a minimizer. Then, \(\mathbf R^*\) is unique
  and in minimization normal form. Moreover, \(\mathbf R^*\) exhausts the
  \(L_1\) budget, in the sense that either \(\tfrac 1 n \|\mathbf R^* - \hat
  {\mathbf R}\|_1 = \epsilon\) or \(t^{\mathrm{lwr}} = t^{\mathrm{upr}}\).
\end{lemma}

\begin{proof}
  The proof proceeds by examining the first-order KKT conditions to derive weak
  necessary conditions that solutions must satisfy. These conditions are then
  strengthened to minimization normal form by directly comparing the objective
  function at different points satisfying the weak conditions. Finally, we show
  that the minimum is the point satisfying the ``strengthened'' conditions which
  is most distant from \(\hat {\mathbf R}\), i.e., that exhausts the budget.
    
  The problem is simplified  by rewriting the first constraint as a collection
  of linear (and hence everywhere differentiable) constraints. In particular,
  \(\tfrac 1 n \|\mathbf R - \hat {\mathbf R}\|_1 \leq \epsilon\) is equivalent
  to the \(2^n\) constraints of the form \(\mathbf S^\top (\mathbf R - \hat
  {\mathbf R}) \leq \epsilon \cdot n\) for all \(\mathbf S \in \{-1, 1\}^n\). We
  assume throughout, without loss of generality, that \(\ell_i < u_i\).

  \paragraph*{Weak conditions}
  Let \(\mathbf e_i\) denote the \(i\)-th standard basis vector and \(\mathbf 1
  = \sum_{i=1}^n \mathbf e_i\). Note that:
  \begin{itemize}
    \item The gradient of the objective is \(\tfrac 2 n \cdot \mathbf R\);
    \item The gradient of \(\mathbf S^\top_{k} (\mathbf R - \hat {\mathbf R}) -
      \tfrac{\epsilon}{n}\), where \(\{\mathbf S_1, \ldots, \mathbf S_{2^n} \} =
      \{-1, 1\}^n\), is  \(\mathbf S_{k}\);
    \item The gradient of \(\sum_{i=1}^n R_i\) is \(\mathbf 1\);
    \item The gradient of \(R_i - u_i\) is \(\mathbf e_i\) \(\mathbf e_i\);
    \item The gradient of \(\ell_i - R_i\) is \(- \mathbf e_i\).
  \end{itemize}
  The first-order necessary KKT conditions therefore require that\footnote{%
    For notational simplicity, we have multiplied through by \(n\) and absorbed
    constants into the corresponding Lagrange multipliers.
  }
  \begin{equation}
  \label{eq:kkt-min}
      2 \cdot \mathbf R^* - \lambda \cdot \mathbf 1 + \sum_{i=1}^n \mu_{0,i}
      \cdot \mathbf e_i - \sum_{i=1}^n \mu_{1,i} \cdot \mathbf e_i +
      \sum_{k=1}^{2^n} \nu_k \cdot \mathbf S_k = 0
  \end{equation}
  for some arbitrary \(\lambda\), and for non-negative \(\mu_{0,i}\),
  \(\mu_{1,i}\), and \(\nu_k\) for \(k = 1, \ldots, 2^n\) satisfying
  complementary slackness, i.e., such that \(\mu_{0,i} \cdot (R_i^* - u_i) =
  0\), \(\mu_{1,i} \cdot (\ell_i - R_i^*) = 0\), and \(\nu_k \cdot (\mathbf
  S_k^\top (\mathbf R^* - \hat {\mathbf R}) - n \cdot \epsilon) = 0\) for all
  \(i = 1, \ldots, n\) and \(k = 1, \ldots, 2^n\).

  These consequences allow us to prove the following weak characterization of
  \(\mathbf R^*\). Let \(\Delta = \sum_{k=1}^{2^n} \nu_k\). Then, for all \(i\),
  \begin{equation}
  \label{eq:weak}
    R_i^*  \in \left\{ \hat R_i, u_i, \ell_i, \frac 1 2 \cdot [\lambda -
    \Delta], \frac 1 2 \cdot [\lambda + \Delta] \right\}.
  \end{equation}
  For, from \eqref{eq:kkt-min}, we get that
  \begin{equation}
  \label{eq:kkt-component}
      2 \cdot R_i^* - \lambda + \mu_{0,i} - \mu_{1,i} + \sum_{k=1}^{2^n} \nu_k
      \cdot S_{k,i} = 0,
  \end{equation}
  where \(S_{k,i}\) refers to the \(k\)-th component of \(\mathbf S_k\). If
  \(R_i^* = \hat R_i\), there is nothing to prove. Therefore, assume that
  \(R_i^* \neq \hat R_i\). We note that by complementary slackness, \(\nu_k >
  0\) only if \(S_{k,i} \cdot (R_i^* - \hat R_i) \geq 0\) for all \(i = 1,
  \ldots, n\). For, supposing without loss of generality that \(R_i^* > \hat
  R_i\), if \(S_{k,i} \cdot (R_i^* - \hat R_i) < 0\), then, since \(\mathbf S_k
  + 2 \mathbf e_i \in \{-1,1\}^n\), we would have that
  \begin{equation*}
    (\mathbf S_k + 2 \mathbf e_i)^\top (\mathbf R^* - \hat {\mathbf R}) =
    \epsilon \cdot n + 2 (R_i^* - \hat R_i) > \epsilon \cdot n,
  \end{equation*}
  which violates the constraints. Hence \(S_{k,i} = S_{k',i}\) for all \(k\) and
  \(k'\) such that \(\nu_k, \nu_{k'} > 0\). Therefore, our expression simplifies
  to
  \begin{equation*}
    2 \cdot R_i^* - \lambda + \mu_{0,i} - \mu_{1,i} + \Delta \cdot s = 0,
  \end{equation*}
  where \(s = \pm 1\). By complementary slackness, if \(\mu_{0,i} > 0\), then
  \(R_i^* = u_i\); if \(\mu_{1,i} > 0\), then \(R_i^* = \ell_i\). Therefore, we
  need only consider the case where \(\mu_{0,i} = \mu_{1,i} = 0\), i.e., where
  \begin{equation*}
    2 \cdot R_i^* - \lambda + s \cdot \Delta = 0,
  \end{equation*}
  whence \(R_i^* = \tfrac 1 2 [\lambda + s \cdot \Delta]\). Therefore, for all
  \(i\), \eqref{eq:weak} holds.

  \paragraph*{Strengthening conditions}
  Next, we strengthen \eqref{eq:weak} to minimization normal form. First, we
  show that if \(R_{i_0}^* > \hat R_{i_0}\) and \(R_{i_0}^* > R_{i_1}^*\), then
  \(R_{i_1}^* = u_{i_1}\). For, suppose not. Then, there exists some \(\delta >
  0\) such that (1) \(R_{i_0}^* - \delta > \hat R_{i_0}\), (2) \(R_{i_1}^* +
  \delta < u_{i_1}\), and (3) \(R_{i_0}^* - \delta > R_{i_1}^*\). Define
  \(\mathbf R' = \mathbf R^* + \delta \cdot (\mathbf e_{i_1} - \mathbf
  e_{i_0})\). Then, we note that
  \begin{align*}
    \|\mathbf R' - \hat {\mathbf R}\|_1 - \|\mathbf R^* - \hat {\mathbf R}\|_1
      &= \sum_{i=1}^n |R'_i - \hat R_i| - |R^*_i - \hat R_i| \\
      &= |R'_{i_0} - \hat R_{i_0}| - |R^*_{i_0} - \hat R_{i_0}| + |R'_{i_1} -
        \hat R_{i_1}| - |R^*_{i_1} - \hat R_{i_1}| \\
      &= (R^*_{i_0} - \delta - \hat R_{i_0}) - (R^*_{i_0} - \hat R_{i_0}) +
        |R^*_{i_1} + \delta - \hat R_{i_1}| - |R^*_{i_1} - \hat R_{i_1}| \\
      &\leq -\delta + |R^*_{i_1} - \hat R_{i_1}| + \delta - |R^*_{i_1} - \hat R_{i_1}| \\
      &= 0,
  \end{align*}
  so \(\mathbf R'\) satisfies the \(L_1\)-distance constraint. Here, we have
  used the fact that both \(R_{i_0}^*\) and \(R_{i_0}^* - \delta\) are greater
  than \(\hat R_{i_0}\) in the third equality, and the triangle inequality in
  the inequality. The remaining constraints also hold by construction, so
  \(\mathbf R'\) is feasible.

  Since \(\mathbf R'\) is feasible, we will arrive at a contradiction if we can
  show that it achieves a greater objective. However,
  \begin{align*}
    \sum_{i=1}^n (R_i^*)^2 - (R_i')^2
      &= (R_{i_0}^*)^2 - ((R_{i_0}^*)^2 + \delta^2 - 2 \cdot R_{i_0}^* \cdot
        \delta) + (R_{i_1}^*)^2 \\
      &\qquad- ((R_{i_1}^*)^2 + \delta^2 + 2 \cdot (R_{i_1}^*) \cdot \delta) +
        \sum_{i \neq i_0, i_1} (R_i^*)^2 - (R_i^*)^2 \\
      &= 2 \cdot \delta \cdot (R_{i_0}^* - R_{i_1}^* - \delta) \\
      &> 0,
  \end{align*}
  where the inequality follows from the fact that \(R_{i_0}^* - \delta >
  R_{i_1}^*\). Therefore \(\mathbf R^*\) is not a minimum, contrary to
  hypothesis. In exactly the same manner, we see that if \(R_{i_0}^* < \hat
  R_i\) and \(R_{i_1}^* < R_{i_0}^*\), then \(R_{i_1}^* = \ell_{i_1}\). 

  Combining the claim---i.e., that if \(R_{i_0}^* > \hat R_{i_0}\) and
  \(R_{i_0}^* > R_{i_1}^*\), then \(R_{i_1}^* = u_{i_1}\)---with \eqref{eq:weak}
  yields that for all \(i\), if \(R_i^* \leq \tfrac 1 2 \cdot [\lambda -
  \Delta]\), then \(R_i^* = \min(u_i, \tfrac 1 2 \cdot [\lambda - \Delta])\); if
  \(R_i^* \geq \tfrac 1 2 \cdot [\lambda + \Delta]\), then \(R_i^* =
  \max(\ell_i, \tfrac 1 2 \cdot [\lambda + \Delta])\); and otherwise \(R_i^* =
  \hat R_i\). Taking \(t^{\textrm{lwr}} = \tfrac 1 2 \cdot [\lambda - \Delta]\)
  and \(t^{\textrm{upr}} = \tfrac 1 2 \cdot [\lambda + \Delta]\) gives the
  result.

  \paragraph*{Uniqueness and budget exhaustion} If the budget is not exhausted,
  then, by complementary slackness, \(\nu_k = 0\) for all \(k = 1, \ldots,
  2^k\), whence \(\Delta = 0\). This immediately implies that \(t^{\mathrm{lwr}}
  = t^{\mathrm{upr}} = \tfrac 1 2 \lambda\).

  To see uniqueness, suppose that \(\mathbf R^{(0)}\) and \(\mathbf R^{(1)}\)
  are two distinct minima, both in normal form, with corresponding thresholds
  \(t^{\mathrm{lwr}}_k\) and \(t^{\mathrm{upr}}_k\) for \(k = 0, 1\). We note
  that since \(\sum_{i=1}^n R^{(0)}_i = \sum_{i=1}^n R^{(1)}_i = \mu\), we must
  have, without loss of generality, that \(t_0^{\textrm{lwr}} \leq
  t_1^{\textrm{lwr}}\) and \(t_0^{\textrm{upr}} \geq t_1^{\textrm{upr}}\), with
  at least one of the two inequalities strict. However, it follows that
  \(\|\mathbf R^{(0)} - \hat {\mathbf R}\|_1 < \|\mathbf R^{(1)} - \hat {\mathbf
  R}\|_1\), which is impossible by the preceding paragraph.
\end{proof}

Lemma~\ref{lem:minimize} suggests a natural algorithm for solving the
optimization problem in Eq.~\eqref{eq:simp-optim}: we can sweep over all
possible risk vectors in minimization normal form simply by sweeping over
\(t^{\mathrm{lwr}}\). In particular, as we increase \(t^{\mathrm{lwr}}\), the
sum constraint---i.e., that \(\sum_{i=1}^n R_i = \mu\)---forces
\(t^{\mathrm{upr}}\) to decrease so as to counterbalance it exactly.

More precisely, consider an index \(i\) to be ``active'' if one of the
thresholds is between \(\ell_i\) and \(u_i\). If \(k^{\mathrm{lwr}}\) represents
the number of indices that are active because of \(t^{\mathrm{lwr}}\)---and
\(k^{\mathrm{upr}}\) is also defined accordingly---then, locally, if we increase
\(t^{\mathrm{lwr}}\) at unit rate, the sum of the risks increases at the rate of
\(k^{\mathrm{lwr}}\). Therefore, the rates of increase of the two thresholds,
\(r^{\mathrm{lwr}}\) and \(r^{\mathrm{upr}}\), must satisfy\footnote{%
  If either \(k^{\mathrm{lwr}}\) and \(k^{\mathrm{upr}}\), then the
  corresponding threshold can be changed at any rate without affecting the sum
  constraint. However, this is because changing that threshold in this case does
  not actually change the risk vector, since no indices are active, and so we
  can ignore this case in the subsequent discussion.
}
\begin{equation}
\label{eq:balance}
  {r^{\mathrm{lwr}}} \cdot {k^{\mathrm{lwr}}} = {r^{\mathrm{upr}}} \cdot
  {k^{\mathrm{upr}}}.
\end{equation}
In particular, we know that we have reached the minimum---and hence solved the
optimization problem---once either the budget has been exhausted or the two
thresholds have become equal. For fixed \(r^{\mathrm{lwr}}\) and
\(r^{\mathrm{upr}}\), this would be straightforward to calculate; however, these
rates can, in principle, change whenever \(k^{\mathrm{upr}}\) and
\(k^{\mathrm{lwr}}\) change. We call these change points ``distinguished
point''---that is, points at which a threshold \(t\) equals \(\ell_i\), \(\hat
R_i\), or \(u_i\) for some \(i\). Therefore, the algorithm consists of repeating
the following steps:
\begin{enumerate}
  \item Determining the ``next'' distinguished points that either the lower or
    upper threshold will reach.
  \item Checking whether the budget will be exhausted before that point is
    reached.
  \item If not, advancing to that point and recalculating \(k^{\mathrm{lwr}}\)
    or \(k^{\mathrm{upr}}\) as appropriate.
\end{enumerate}
In addition, since \(\sum_{i=1}^n R_i\) may not equal \(\mu\), a
``preprocessing'' step, where \(t^{\mathrm{lwr}}\) is increased or
\(t^{\mathrm{upr}}\) is decreased may be needed to find the ``initial'' risk
vector in minimization normal form satisfying the sum constraint.

We make two observations about the informal algorithm sketch given above needed
for its extension to the case of multiple strata. First, we note that every risk
vector \(\mathbf R^*\) in minimization normal form is the minimizer for the
corresponding single stratum minimization problem with \(\epsilon = \tfrac 1 n
\|\mathbf R^* - \hat {\mathbf R}\|_1\) and \(\mu = \sum_{i=1}^n R_i^*\).
Consequently, although intended to find the minimizer for a specific
\(\epsilon\), the algorithm sketched above actually sweeps over the minimizers
for \emph{all possible} \(\epsilon\). As a result, it is not any harder to solve
a single instance of the single stratum minimization problem than it is to solve
it generally for all possible \(\epsilon\), given some fixed \(\mu\). Secondly,
while the natural way to parameterize the risk vectors in the discussion above
is in terms of \(t^{\mathrm{lwr}}\), when optimizing over multiple strata, it is
actually more natural to reparameterize the risk vectors in terms of the gap
\(\Delta = t^{\mathrm{upr}} - t^{\mathrm{lwr}}\). If in addition to
Eq.~\eqref{eq:balance} we impose the condition that \(r^{\mathrm{lwr}} +
r^{\mathrm{upr}} = 1\), so that \(\Delta\) decreases at a constant rate, then we
have that\footnote{%
  We note that by Lemma~\ref{lem:minimize}, for every \(\delta\) there is a
  unique \(\mathbf R_\delta\), but that the reverse is not necessarily true,
  since there may not be any \(i\) such that \(R_i = t^{\mathrm{lwr}}\) or
  \(t^{\mathrm{upr}}\), in which case different thresholds can be chosen without
  altering the underlying risk vector.
}
\begin{equation}
\label{eq:rates}
  r^{\mathrm{lwr}} = \frac {k^{\mathrm{upr}}} {k^{\mathrm{lwr}} +
  k^{\mathrm{upr}}}, \qquad r^{\mathrm{upr}} = \frac {k^{\mathrm{lwr}}}
  {k^{\mathrm{lwr}} + k^{\mathrm{upr}}}.
\end{equation}

To solve the single stratum minimization problem across the whole range of
possible \(\epsilon\), let \(\mathbf R_\Delta\) be the unique solution we can
associate with a given gap \(\Delta\) between \(t^{\mathrm{lwr}}\) and
\(t^{\mathrm{upr}}\).\footnote{%
  For every \(\Delta\) there is a unique \(\mathbf R_\Delta\). In particular,
  any other distinct risk vector in minimization normal form with the same gap
  must be strictly greater or lesser, as argued at the end of
  Lemma~\ref{lem:minimize}, meaning that it cannot satisfy the sum constraint.
  However, the reverse is not necessarily true---that is, \(\mathbf R_{\Delta}\)
  may equal \(\mathbf R_{\Delta'}\) for \(\Delta \neq \Delta'\)---since
  achieving a given gap may require that \(k^{\mathrm{lwr}} = k^{\mathrm{upr}} =
  0\), in which case different thresholds can be chosen without altering the
  underlying risk vector.
}
Then, we can study two functions:
\begin{equation*}
  \epsilon(\Delta) = \|\mathbf R_\Delta - \hat {\mathbf R}\|_1, \qquad
  \Sigma(\Delta) = \sum_{i=1}^n R_{\Delta, i}^2 - \hat R_i^2.
\end{equation*}
By the preceding discussion we readily derive that \(\epsilon(\Delta)\) is a
piecewise linear function of \(\Delta\), whose slope at a given value of
\(\Delta\) is equal to\footnote{%
  Or zero if both \(k^{\mathrm{lwr}}\) and \(k^{\mathrm{upr}}\) are zero.
}
\[
  K = 2 \cdot \frac {k^{\mathrm{lwr}} \cdot k^{\mathrm{upr}}} {k^{\mathrm{lwr}} + k^{\mathrm{upr}}},
\]
and \(\Sigma(\Delta)\) is a piecewise quadratic function of \(\Delta\)
satisfying
\begin{align*}
  \Sigma(\Delta + t)
    &= \Sigma(\Delta) + k^{\mathrm{lwr}} \cdot [(t^{\mathrm{lwr}} +
      r^{\mathrm{lwr}} \cdot t)^2 - (t^{\mathrm{lwr}})^2] + k^{\mathrm{upr}} \cdot
      [(t^{\mathrm{upr}} - r^{\mathrm{upr}} \cdot t)^2 - (t^{\mathrm{lwr}})^2] \\
    &= \Sigma(\Delta) + k^{\mathrm{lwr}} \cdot r^{\mathrm{lwr}} \cdot t \cdot
      (r^{\mathrm{lwr}} \cdot t + 2 t^{\mathrm{lwr}}) + k^{\mathrm{upr}} \cdot
      r^{\mathrm{upr}} \cdot t \cdot (r^{\mathrm{upr}} \cdot t - 2
      t^{\mathrm{upr}}) \\
    &= \Sigma(\Delta) - 2 \cdot \frac {k^{\mathrm{lwr}} \cdot k^{\mathrm{upr}}}
      {k^{\mathrm{lwr}} + k^{\mathrm{upr}}} \cdot (t^{\mathrm{upr}} -
      t^{\mathrm{lwr}}) \cdot t + \frac {k^{\mathrm{lwr}} (k^{\mathrm{upr}})^2 +
      k^{\mathrm{upr}} (k^{\mathrm{lwr}})^2} {(k^{\mathrm{lwr}} +
      k^{\mathrm{upr}})^2} \cdot t^2 \\
    &= \Sigma(\Delta) - K \Delta t + \frac K 2 t^2,
\end{align*}
as long as \(t\) is sufficiently small that the number of active indices does
not change. It follows that \(\epsilon(\Delta)\) and \(\Sigma(\Delta)\) are
fully determined by the following collections:
\[
  \boldsymbol \Delta = (\Delta_1, \ldots, \Delta_N),\qquad 1 = \Delta_1 < \cdots
  < \Delta_N = 0,
\]
where \(\Delta_k\) represents the \(k\)-th value of \(\Delta\) at which
\(k^{\mathrm{lwr}}\) and \(k^{\mathrm{upr}}\) change; and
\[
  \mathbf K = (K_1, \ldots, K_N),\qquad K_1 = 0, \quad K_N = 0,
\]
where \(K_k\) denotes the value of \(K\) beginning at \(\Delta = \Delta_k\).
From \(\boldsymbol \Delta\) and \(\mathbf K\), the \(\Delta^*\) such that
\(\|\mathbf R_{\Delta^*} - \hat {\mathbf R}\|_1 = \epsilon\) can be calculated
in linear time. For completeness, this algorithm is given in
Algorithm~\ref{algo:K-delta-solve}.

Thus, a complete solution for the optimization problem for a single stratum
across \emph{all possible \(\epsilon\)} requires only calculating \(\boldsymbol
\Delta\) and \(\mathbf K\). As described above, this can be calculated by moving
the thresholds toward each other at the prescribed rates \(r^{\mathrm{lwr}}\)
and \(r^{\mathrm{upr}}\), updating the rates each time an index is activated or
deactivated, until the gap between the thresholds is zero. Similar to evaluating
\(\epsilon(\Delta)\) and \(\Sigma(\Delta)\), this can be completed in linear
time, as described in Algorithm~\ref{algo:min-one}. (As above, we note that if
\(\tfrac 1 n \sum_{i=1}^n \hat R_i \neq \mu\), it may be necessary to do a
preprocessing step, also in linear time, as shown in
Algorithm~\ref{algo:min-adjust}.)

\subsubsection{Optimizing over all strata}

With a complete solution to the problem of minimizing the sum of squares for a
single stratum, the problem of minimizing the sum of squares across all strata
is straightforward. We begin by characterizing solutions to the minimization
problem in the general case.

\begin{lemma}%
\label{lem:minimize-all}
  Consider the simplified minimization problem in Eq.~\eqref{eq:simp-optim}.
  Suppose that \(\mathbf R^*\) is a solution. Then, the restriction of \(\mathbf
  R^*\) to any stratum is in minimization normal form. Moreover, \(\mathbf R^*\)
  exhausts the \(L_1\) budget, in that either \(\tfrac 1 n \|\mathbf R - \hat
  {\mathbf R}\|_1 = \epsilon\) or \(t^{\mathrm{lwr}} = t^{\mathrm{upr}} =
  \mu_{a,j}\), where \(\mu_{a,j}\) equals either \(\rho_j\) or \(\tau_j\)
  depending on whether \(a\) equals \(1\) or \(0\).
\end{lemma}

\begin{proof}
  The proof is virtually identical to the proof of Lemma~\ref{lem:minimize}. The
  only difference is that the first-order necessary KKT conditions take the form 
  \begin{equation}
  \label{eq:kkt-min-all}
    2 \cdot \mathbf R^* - \left( \sum_{j=1}^m \lambda_{0,j} \cdot \mathbf
    1_{0,j} + \lambda_{1,j} \cdot \mathbf 1_{1,j} \right) + \sum_{i=1}^n
    \mu_{0,i} \cdot \mathbf e_i - \sum_{i=1}^n \mu_{1,i} \cdot \mathbf e_i +
    \sum_{k=1}^{2^n} \nu_k \cdot \mathbf S_k = 0,
  \end{equation}
  where \(\mathbf 1_{a,j} = \sum_{C_i = c_j, A_i = a} \mathbf e_i\). Restricting
  to a single \(i\) gives the following minor variant of
  Eq.~\eqref{eq:kkt-component}:
  \[
    2 \cdot R_i^* - \lambda_{a,j} + \mu_{0,i} - \mu_{1,i} + \sum_{k=1}^n \nu_k
    \cdot S_{k,i} = 0.
  \]
  The proof then proceeds identically; the only difference is to note that while
  \(\lambda_{a,j}\) varies by stratum, \(\Delta = \sum_{k=1}^{2^n} \nu_k\) does
  not.
\end{proof}

It follows that solving the minimization problem across all strata can be
carried out in almost the same way as across a single stratum:
\begin{enumerate}
  \item Construct piecewise linear functions \(\epsilon_{a,j}(\Delta)\) and
    piecewise quadratic functions \(\Sigma_{a,j}(\Delta)\) for each stratum;
  \item Note that because the sums \(\sum_{a = 0}^1 \sum_{j=1}^m
    \epsilon_{a,j}(\Delta)\) and \(\sum_{a=0}^1 \sum_{j=1}^m
    \Sigma_{a,j}(\Delta)\) are also piecewise linear and quadratic,
    respectively, they can also be evaluated using
    Algorithm~\ref{algo:K-delta-solve};
  \item Find \(\Delta^*\) such that \(\sum_{a = 0}^1 \sum_{j=1}^m
    \epsilon_{a,j}(\Delta^*) = \epsilon\), and evaluate \(\sum_{a=0}^1
    \sum_{j=1}^m \Sigma_{a,j}(\Delta^*)\).
\end{enumerate}
Consequently, beyond the machinery established in
Algorithms~\ref{algo:K-delta-solve},~\ref{algo:min-adjust},
and~\ref{algo:min-one}, we only need a way of calculating the sums of these
piecewise functions. This can be carried out straightforwardly using a variation
on the standard merge-sort algorithm, which we give in
Algorithm~\ref{algo:min-all} for completeness.

Putting this all together, we obtain the following lemma summarizing the results
of this section.

\begin{lemma}
\label{lem:min-runtime}
  If \(\boldsymbol \ell\), \(\hat {\mathbf R}\), and \(\boldsymbol u\) have been
  sorted, then there exists an \(O(\log(m) \cdot n)\) algorithm solving the
  simplified minimization problem (Eq.~\eqref{eq:simp-optim}).
\end{lemma}

\begin{proof}
  The proof is straightforward. We begin by noting that
  Algorithm~\ref{algo:K-delta-solve} requires linear time: the index \(i\)
  increases by one in the while loop on lines 6 through 10 on each iteration,
  and must be less than the length of the input. Similarly, in
  Algorithm~\ref{algo:min-adjust}, the while loop on lines 11 through 22
  increments \(i^{\mathrm{lwr}}\) by one on each loop, and limits it to three
  times the length of the input. The same analysis applies to
  Algorithm~\ref{algo:min-one}, in which either \(i^{\mathrm{upr}}\) or
  \(i^{\mathrm{lwr}}\) increases by one on each iteration. We note that the run
  time of Algorithm~\ref{algo:min-all} is proportional to the sum of the lengths
  of its inputs. Using a standard divide-and-conquer approach, we can accomplish
  the combination of the \(2m\) strata in \(m\) total calls, where each element
  of the input appears in \(1 + \lceil \log_2(m) \rceil\) of the function calls,
  for a total time complexity of \(O(\log(m) \cdot n)\). Putting this all
  together, the total complexity is \(O(\log(m) \cdot n)\).
\end{proof}

\subsection{Maximization}%
\label{ssec:max}

Even in the simplified problem in Eq.~\eqref{eq:simp-optim}, maximization is not
a convex problem. In general, maximizing a quadratic objective over a linearly
constrained convex set is NP-hard~\citep{sahni1974computationally}. However, the
restricted forms of the objective and constraints allow us efficiently to solve
the optimization problem exactly in the case of a single stratum and
approximately in the case of multiple strata.

\subsubsection{Optimizing over a single stratum}

As in the case of minimization, risk vectors that maximize the single-stratum
maximization problem have a special normal form.

\begin{definition}[Maximization normal form]
  We say that a risk vector \(\mathbf R\) is in \emph{maximization normal form}
  if there exist indices (``pivots'') \(i^{\mathrm{lwr}}\) and
  \(i^{\mathrm{upr}}\) such that
  \[
    R_i = \begin{cases}
      \ell_i & i < i^{\mathrm{lwr}}, \\
      \hat R_i & i^{\mathrm{lwr}} < i < i^{\mathrm{upr}}, \\
      u_i & i^{\mathrm{upr}} < i.
    \end{cases}
  \]
  We additionally require that the risk values at the pivots themselves satisfy
  \(R_{i^{\mathrm{lwr}}} \in [\ell_{i^{\mathrm{lwr}}}, \hat
  R_{i^{\mathrm{lwr}}}]\) and \(R_{i^{\mathrm{upr}}} \in [\hat
  R_{i^{\mathrm{upr}}}, u_{i^{\mathrm{upr}}}]\).\footnote{%
    Since it involves the ordering of the indices as well as the risk values,
    the appropriate generalization of maximization normal form is notationally
    awkward to express when multiple strata are involved. Fortunately, in
    contrast to the case of minimization, we will not have occasion to use such
    a generalization, and so do not give its definition.
  }\textsuperscript{,}\footnote{%
    We note that we allow \(i^{\mathrm{lwr}}\) and \(i^{\mathrm{upr}}\) to take
    the values \(0\) or \(n+1\) in addition to \(\{1, \ldots, n\}\) to cover
    cases where, e.g., \(R_i = \ell_i\) for all \(i = 1, \ldots, n\).
  }
\end{definition}

Maximization normal form is similar to minimization normal form, except that
instead of begin pushed toward the thresholds, values at indices \emph{below}
the lower pivot are pushed down, and values at indices \emph{above} the upper
pivot are pushed up. An illustration of maximum normal form is given in
Figure~\ref{fig:max-norm}.

\begin{figure}[t]
  \begin{center}
    \begin{tikzpicture}
      \draw (-1,0) -- (11,0);
      \node[anchor = east] at (-1, 0) {\(R = 0\)};
      \draw (-1, 5) -- (11, 5);
      \node[anchor = east] at (-1, 5) {\(R = 1\)};
  
      \node[anchor = north] at (3, 0) {\(i^{\mathrm{lwr}}\)};
      \node[anchor = north] at (7, 0) {\(i^{\mathrm{upr}}\)};
      
      \filldraw[black]  (0,1/2) circle (2pt);
      \filldraw[black]  (1,2/3) circle (2pt);
      \filldraw[black]  (2, 1)  circle (2pt);
      \filldraw[black]  (3,6/5) circle (2pt);
      \filldraw[black]  (4,3/2) circle (2pt);
      \filldraw[black]  (5,7/3) circle (2pt);
      \filldraw[black]  (6,10/3) circle (2pt);
      \filldraw[black]  (7,7/2)  circle (2pt);
      \filldraw[black]  (8,9/2)  circle (2pt);
      \filldraw[black]  (9,14/3) circle (2pt);
      \filldraw[black]  (10,19/4) circle (2pt);
  
      \draw (0,0) -- (0,1); \draw (-0.1,1) -- (0.1,1);
      \draw (1,0) -- (1,2); \draw (0.9,2) -- (1.1,2);
      \draw (2,0) -- (2,2.2); \draw (1.9, 2.2) -- (2.1, 2.2);
      \draw (3,0.2) -- (3, 2.4); \draw (2.9, 0.2) -- (3.1, 0.2); \draw (2.9, 2.4) -- (3.1, 2.4);
      \draw (4,0.2) -- (4,3); \draw (3.9, 0.2) -- (4.1, 0.2); \draw (3.9, 3) -- (4.1, 3);
      \draw (5,1) -- (5, 3.3); \draw (4.9, 1) -- (5.1, 1); \draw (4.9, 3.3) -- (5.1, 3.3);
      \draw (6, 7/3) -- (6, 13/3); \draw (5.9, 7/3) -- (6.1, 7/3); \draw (5.9, 13/3) -- (6.1, 13/3);
      \draw (7, 5/2) -- (7, 19/4); \draw (6.9, 5/2) -- (7.1, 5/2); \draw (6.9, 19/4) -- (7.1, 19/4);
      \draw (8, 3) -- (8, 5); \draw (7.9, 3) -- (8.1, 3); \draw (7.9, 5) -- (8.1, 5);
      \draw (9, 4) -- (9, 5); \draw (8.9, 4) -- (9.1, 4); \draw (8.9, 5) -- (9.1, 5);
      \draw (10, 4.5) -- (10, 5); \draw (9.9, 4.5) -- (10.1, 4.5); \draw (9.9, 5) -- (10.1, 5);
  
      \filldraw[red] (0.2, 0) circle (2pt);
      \filldraw[red] (1.2, 0) circle (2pt);
      \filldraw[red] (2.2, 0) circle (2pt);
      \filldraw[red] (3.2, 1/2) circle (2pt);
      \filldraw[red] (4.2, 3/2) circle (2pt);
      \filldraw[red] (5.2, 7/3) circle (2pt);
      \filldraw[red] (6.2, 10/3) circle (2pt);
      \filldraw[red] (7.2, 4.5) circle (2pt);
      \filldraw[red] (8.2, 5) circle (2pt);
      \filldraw[red] (9.2, 5) circle (2pt);
      \filldraw[red] (10.2, 5) circle (2pt);
  
      \draw[->, color = red] (0.2, 1/2) -- (0.2, 0.1);
      \draw[->, color = red] (1.2, 2/3) -- (1.2, 0.1);
      \draw[->, color = red] (2.2, 1) -- (2.2, 0.1);
      \draw[->, color = red] (3.2, 6/5) -- (3.2, 0.6);
      \draw[->, color = red] (7.2, 7/2) -- (7.2, 4.4);
      \draw[->, color = red] (8.2, 9/2) -- (8.2, 4.9);
      \draw[->, color = red] (9.2, 14/3) -- (9.2, 4.9); 
      \draw[->, color = red] (10.2, 19/4) -- (10.2, 4.9);
    \end{tikzpicture}
  \end{center}
  \caption{\emph{%
    Maximization normal form. Black dots represent estimated risks \(\hat
    R_i\), the error bars represent the allowable range \([\ell_i, u_i]\), and
    red dots indicate the corresponding value of the normal form vector
    \(R_i\). The pivots are indicated on the \(x\)-axis.
  }}%
\label{fig:max-norm}
\end{figure}

An important difference from the case of minimization is that studying the KKT
conditions does not yield the normal form directly; instead, it yields a weaker
characterization.

\begin{definition}[Weak normal form]
  We say that a risk vector is in \emph{weak normal form} if there exist
  thresholds \(t^{\mathrm{lwr}}\) and \(t^{\mathrm{upr}}\) such that
  \[
    R_i = \begin{cases}
      \ell_i & R_i < t^{\mathrm{lwr}}, \\
      t^{\mathrm{lwr}} & R_i = t^{\mathrm{lwr}}, \\
      \hat R_i & t^{\mathrm{lwr}} < R_i < t^{\mathrm{upr}}, \\
      t^{\mathrm{upr}} & R_i = t^{\mathrm{upr}}, \\
      u_i & t^{\mathrm{upr}} < R_i,
    \end{cases}
  \]
  and there exists at most one lower index \(i^{\mathrm{lwr}}\) such that
  \(R_{i^{\mathrm{lwr}}} = t^{\mathrm{lwr}}\); and similarly, there exists at
  most one upper index \(i^{\mathrm{upr}}\) such that \(R_{i^{\mathrm{upr}}} =
  t^{\mathrm{upr}}\)
\end{definition}

We note that, in the case of \emph{weak} maximization normal form, the indices
\(i^{\mathrm{lwr}}\) and \(i^{\mathrm{upr}}\) are analogous to the pivots in
maximization normal form; however, they do not share the key property, which is
that all of the indices \(i < i^{\mathrm{lwr}}\) index risks that have been
pushed down all the way to \(\ell_i\) and all of the indices \(i >
i^{\mathrm{upr}}\) index risks that have been pushed all the way up to \(u_i\);
see Figure~\ref{fig:weak-max-norm}.

\begin{figure}[t]
  \begin{center}
    \begin{tikzpicture}
      \draw (-1,0) -- (11,0);
      \node[anchor = east] at (-1, 0) {\(R = 0\)};
      \draw (-1, 5) -- (11, 5);
      \node[anchor = east] at (-1, 5) {\(R = 1\)};
  
      \node[anchor = east] at (-1, 2) {\(t^{\mathrm{lwr}}\)};
      \node[anchor = east] at (-1, 4) {\(t^{\mathrm{upr}}\)};
      \node[anchor = north] at (3, 0) {\(i^{\mathrm{lwr}}\)};
      \node[anchor = north] at (8, 0) {\(i^{\mathrm{upr}}\)};
      
      \filldraw[black]  (0,1/2) circle (2pt);
      \filldraw[gray]   (1,2/3) circle (2pt);
      \filldraw[gray]   (2, 1)  circle (2pt);
      \filldraw[gray]   (3,6/5) circle (2pt);
      \filldraw[gray]   (4,3/2) circle (2pt);
      \filldraw[blue]   (5,7/3) circle (2pt);
      \filldraw[blue]   (6,10/3) circle (2pt);
      \filldraw[blue]   (7,7/2)  circle (2pt);
      \filldraw[gray]   (8,9/2)  circle (2pt);
      \filldraw[gray]   (9,14/3) circle (2pt);
      \filldraw[black]  (10,19/4) circle (2pt);
  
      \draw (0,0) -- (0,1); \draw (-0.1,1) -- (0.1,1);
      \draw[gray] (1,0) -- (1,2); \draw[gray] (0.9,2) -- (1.1,2);
      \draw[gray] (2,0) -- (2,2.2); \draw[gray] (1.9, 2.2) -- (2.1, 2.2);
      \draw[gray] (3,0.2) -- (3, 2.4); \draw[gray] (2.9, 0.2) -- (3.1, 0.2); \draw[gray] (2.9, 2.4) -- (3.1, 2.4);
      \draw[gray] (4,0.2) -- (4,3); \draw[gray] (3.9, 0.2) -- (4.1, 0.2); \draw[gray] (3.9, 3) -- (4.1, 3);
      \draw[blue] (5,1) -- (5, 3.3); \draw[blue] (4.9, 1) -- (5.1, 1); \draw[blue] (4.9, 3.3) -- (5.1, 3.3);
      \draw[blue] (6, 7/3) -- (6, 13/3); \draw[blue] (5.9, 7/3) -- (6.1, 7/3); \draw[blue] (5.9, 13/3) -- (6.1, 13/3);
      \draw[blue] (7, 5/2) -- (7, 19/4); \draw[blue] (6.9, 5/2) -- (7.1, 5/2); \draw[blue] (6.9, 19/4) -- (7.1, 19/4);
      \draw[gray] (8, 3) -- (8, 5); \draw[gray] (7.9, 3) -- (8.1, 3); \draw[gray] (7.9, 5) -- (8.1, 5);
      \draw[gray] (9, 4) -- (9, 5); \draw[gray] (8.9, 4) -- (9.1, 4); \draw[gray] (8.9, 5) -- (9.1, 5);
      \draw (10, 4.5) -- (10, 5); \draw (9.9, 4.5) -- (10.1, 4.5); \draw (9.9, 5) -- (10.1, 5);
  
      \draw[dashed] (-1, 2) -- (11, 2);
      \draw[dashed] (-1, 4) -- (11, 4);
  
      \filldraw[red] (0.2, 0) circle (2pt);
      \filldraw[red] (1.2, 0) circle (2pt);
      \filldraw[red] (2.2, 0) circle (2pt);
      \filldraw[red] (3.2, 2) circle (2pt);
      \filldraw[red] (4.2, 0.2) circle (2pt);
      \filldraw[red] (5.2, 1) circle (2pt);
      \filldraw[red] (6.2, 10/3) circle (2pt);
      \filldraw[red] (7.2, 19/4) circle (2pt);
      \filldraw[red] (8.2, 4) circle (2pt);
      \filldraw[red] (9.2, 5) circle (2pt);
      \filldraw[red] (10.2, 5) circle (2pt);
  
      \draw[->, color = red] (0.2, 1/2) -- (0.2, 0.1);
      \draw[->, color = red] (1.2, 2/3) -- (1.2, 0.1);
      \draw[->, color = red] (2.2, 1) -- (2.2, 0.1);
      \draw[->, color = red] (3.2, 6/5) -- (3.2, 1.9);
      \draw[->, color = red] (4.2, 3/2) -- (4.2, 0.3);
      \draw[->, color = red] (5.2, 7/3) -- (5.2, 1.1);
      \draw[->, color = red] (7.2, 7/2) -- (7.2, 4.65);
      \draw[->, color = red] (8.2, 9/2) -- (8.2, 4.1);
      \draw[->, color = red] (9.2, 14/3) -- (9.2, 4.9); 
      \draw[->, color = red] (10.2, 19/4) -- (10.2, 4.9);
    \end{tikzpicture}
  \end{center}
  \caption{\emph{%
    Weak maximization normal form. Black, blue, and grey dots represent
    estimated risks \(\hat R_i\), the error bars represent the allowable range
    \([\ell_i, u_i]\), and red dots indicate the corresponding value of the
    normal form vector \(R_i\). The pivots are indicated on the \(x\)-axis.
    The risk \(R_i\) at indices where the estimated risk is between the
    thresholds and the bounds straddle the thresholds (shown in blue) can take
    on a number of values (e.g., \(R_i = \hat R_i\) or \(R_i = u_i\)).
    Likewise, \(i^{\mathrm{lwr}}\) and \(i^{\mathrm{upr}}\) can occur at any
    index \(i\) such that \(R_i = t^{\mathrm{lwr}}\) or \(R_i =
    t^{\mathrm{upr}}\) is feasible (shown in grey). As a result, the set of
    risk vectors in \emph{weak} maximization normal form is a very complex
    search space.
  }}%
\label{fig:weak-max-norm}
\end{figure}

\begin{lemma}%
\label{lem:maximize}
  Consider the single-stratum maximization problem---i.e., maximization in
  Eq.~\eqref{eq:ss-optim}. Suppose that \(\mathbf R^*\) is a solution. Then,
  \(\mathbf R^*\) is in weak normal form. Moreover, either \(\tfrac 1 n
  \|\mathbf R^* - \hat {\mathbf R}\|_1 = \epsilon\) or \(R_i^*\) equals
  \(\ell_i\) or \(u_i\) for all but (at most) one \(i\).
\end{lemma}

\begin{proof}
  The proof is quite similar to the proof of Lemma~\ref{lem:minimize}. We note
  that the first-order KKT conditions are almost the same, namely that 
  \begin{equation}
  \label{eq:kkt-max}
    2 \cdot \mathbf R^* - \lambda \cdot \mathbf 1 - \sum_{i=1}^n \mu_{0,i} \cdot
    \mathbf e_i + \sum_{i=1}^n \mu_{1,i} \cdot \mathbf e_i - \sum_{k=1}^{2^n}
    \nu_k \cdot \mathbf S_k = 0
  \end{equation}
  for some \(\lambda\) and \(\mu_{0,i}, \mu_{1,i}, \nu_k \geq 0\) satisfying
  complementary slackness. Virtually exactly as before, it follows from this
  that Eq.~\eqref{eq:weak} holds, i.e., that
  \begin{equation}
  \label{eq:weak2}
    R_i^*  \in \left\{ \hat R_i, u_i, \ell_i, \frac 1 2 \cdot [\lambda -
    \Delta], \frac 1 2 \cdot [\lambda + \Delta] \right\}.
  \end{equation}
  Again, as in the case of minimization, we can strengthen Eq.~\eqref{eq:weak2}.
  In particular, exactly the same argument used to strengthen the weak
  conditions in Lemma~\ref{lem:minimize} yields that if \(R_{i_0}^* > \hat
  R_{i_0}\) and \(R_{i_0}^* < R_{i_1}^*\), then \(R_{i_1}^* = u_{i_1}\); and,
  conversely, if \(R_{i_0}^* < \hat R_{i_0}\) and \(R_{i_0}^* > R_{i_1}^*\),
  then \(R_{i_1}^* = \ell_{i_1}\). The proof is then otherwise the same as in
  Lemma~\ref{lem:minimize}.

  To see that there can be at most one \(i\) such that such that \(\ell_i <
  R_i^* < \hat R_i\), suppose that \(\ell_{i_0} < R_{i_0}^* < \hat R_{i_0}\) and
  \(\ell_{i_1} < R_{i_1}^* < \hat R_{i_1}\) for \(i_0 \neq i_1\). By
  Eq.~\eqref{eq:weak2}, \(R_{i_0}^* = R_{i_1}^*\). Choose \(\delta > 0\)
  sufficiently small that \(R_{i_0}^* - \delta > \ell_{i_0}\) and \(R_{i_1}^* +
  \delta < \hat R_{i_1}\). Then, as in the case of minimization, it follows
  immediately that \(\mathbf R^* + \delta (\mathbf e_{i_1} - \mathbf e_{i_0})\)
  is feasible, but increases the objective by \(2 \delta^2\), contrary to the
  hypothesis that \(\mathbf R^*\) was a maximizer. In the same way, it follows
  that there exists at most one \(i\) such that \(u_i > R_i^* > \hat R_i\).
  
  Combining this with our previous necessary condition on \(\mathbf R^*\) and
  letting \(t^{\mathrm{lwr}} = \tfrac 1 2 \cdot [\lambda - \Delta]\) and
  \(t^{\mathrm{upr}} = \tfrac 1 2 \cdot [\lambda + \Delta]\) yields that for all
  \(i\) such that \(R_i^* < t^{\mathrm{lwr}}\), \(R_i^* = \ell_i\); that for all
  \(i\) such that \(R_i^* > t^{\mathrm{upr}}\), \(R_i^* = u_i\); that there is
  at most one index such that \(\hat R_i > R_i^* > t^{\mathrm{lwr}}\); and,
  finally, that there is at most one \(i'\) such that \(\hat R_{i'} < R_{i'}^* <
  t^{\mathrm{upr}}\).
\end{proof}

Lemma~\ref{lem:maximize} is, by itself, insufficient to form the basis of an
effective solution algorithm since it does not determine which indices are
\(i^{\mathrm{lwr}}\) and \(i^{\mathrm{upr}}\). Even if these indices were known,
they do not pin down the value of \(R_i\) for \(i \neq i^{\mathrm{lwr}},
i^{\mathrm{upr}}\), which could, almost without restriction, be \(\ell_i\),
\(\hat R_i\), or \(u_i\). As a result, the search space of vectors in weak
maximization normal form is extremely large.

Therefore, to construct an effective maximization algorithm, we must transform
vectors merely in weak maximization normal form into vectors that are in
full-fledged maximization normal form.

To do so, we will have need of the following elementary fact.

\begin{lemma}%
\label{lem:abcd}
  Suppose \(a \geq b\) and \(c \geq d\). Then
  \[
      |a - c| + |b - d| \leq |a - d| + |b - c|.
  \]
\end{lemma}

\begin{proof}
  The proof is not complicated, but is simplest to understand if expressed in
  geometric terms. Consider the points \(\mathbf p_0 = (a,b)\) and \(\mathbf p_1
  = (c,d)\). Then both \(\mathbf p_0\) and \(\mathbf p_1\) lie in the lower
  half-plane 
  \[
    H = \{ \, (x, y) : x \geq y \}.
  \]
  Let \(T : (x, y) \mapsto (y, x)\) be the linear transformation given by
  reflection across the line \(y = x\). Then, the claim of the lemma can be
  reframed as follows: for any \(\mathbf p_0, \mathbf p_1 \in H\),
  \begin{equation}
  \label{eq:abcd}
    \|\mathbf p_0 - \mathbf p_1\|_1 \leq \|\mathbf p_0 - T (\mathbf p_1)\|_1.
  \end{equation}
  This setup is shown in Figure~\ref{fig:abcd-1}. We divide into four cases
  according to whether \(a \geq c\) and \(b \geq d\), or, equivalently,
  depending on in which of the four regions shown in Figure~\ref{fig:abcd-2} the
  point \(\mathbf p_1\) lies.

  \paragraph*{Case (A): \(a \geq c\) and \(b \geq d\)}
  In this case, we have that
  \begin{align*}
    |a - c| + |b - d|
      &= a - c + b - d \\
      &= a - d + b - c \\
      &\leq |(a - d) + (b - c)| \\
      &\leq |a - d| + |b - c|.
  \end{align*}
  Therefore, Eq.~\eqref{eq:abcd} holds.

  \paragraph*{Case (B): \(a \leq c\) and  \(b \geq d\)}
  In this case,
  \begin{align*}
    |a - c| + |b - d|
      &= c - a + b - d \\
      &\leq c - a + b - d + 2 (a - b) \\
      &= c - b + a - d \\
      &\leq |(c - b) + (a - d)| \\
      &\leq |c - b| + |a - d|,
  \end{align*}
  so Eq.~\eqref{eq:abcd} holds in this case as well.

  Before completing the proof, we note that \(T\) has the useful property that
  \begin{equation}
  \label{eq:reflect}
    \|T (\mathbf p)\|_1 = \|\mathbf p\|_1.
  \end{equation}

  \paragraph*{Case (C): \(a \leq c\) and  \(b \leq d\)}
  We note that this case is the same as Case~(A) with the roles of \(\mathbf
  p_0\) and \(\mathbf p_1\) reversed. Consequently, by Eq.~\eqref{eq:abcd}, we
  have that
  \[
    \|\mathbf p_0 - \mathbf p_1\|_1 \leq \|T(\mathbf p_0) - \mathbf p_1\|_1.
  \]
  Applying Eq.~\eqref{eq:reflect} and using the fact that \(T^2\) is the
  identity, we have that
  \[
    \|T(\mathbf p_0) - \mathbf p_1\|_1 = \|T (T(\mathbf p_0) - \mathbf p_1) \|_1
    = \|\mathbf p_0 - T(\mathbf p_1)\|_1,
  \]
  which yields Eq.~\eqref{eq:abcd}.

  \paragraph*{Case (D): \(a \geq c\) and \(b \leq d\)}
  We argue as in the previous case, noting that this case is equivalent to
  Case~(B) with the roles of \(\mathbf p_0\) and \(\mathbf p_1\) reversed,
  whence
  \[
    \|\mathbf p_0 - \mathbf p_1\|_1 \leq \|T(\mathbf p_0) - \mathbf p_1\|_1 =
    \|\mathbf p_0 - T(\mathbf p_1)\|_1.
  \]
\end{proof}

\begin{figure}
  \begin{center}
    \begin{subfigure}{0.49\textwidth}
      \begin{center}
        \begin{tikzpicture}[yscale = -1, xscale = -1]
          \fill[fill = black!20] (-3, -3) -- (3, 3) -- (-3, 3);
          \draw (-3, -3) -- (3, 3);
          
          \draw[<-, dotted] (0, -3) -- (0, 3);
          \draw[<-, dotted] (-3, 0) -- (3, 0);

          \fill (-2, 2) circle (2pt); \node[anchor = west] at (-2, 2) {\(\mathbf p_0\)};
          \fill (1/2, 2) circle (2pt);  \node[anchor = south west] at (1/2, 2) {\(\mathbf p_1\)};
          \fill (2, 1/2) circle (2pt);
          \draw[dashed, ->] (1/2 + 0.1, 2 - 0.1) -- node[midway, pos = 2/3, above right] {\(T\)} (2 - 0.1, 1/2 + 0.1);

          \draw[->] (-2, 2) --  (1/2 - 0.1, 2);
          \draw[->] (-2, 2) -- (-2, 1/2) -- (2 - 0.1, 1/2);
        \end{tikzpicture}
      \end{center}
      \caption{Setup}
    \label{fig:abcd-1}
    \end{subfigure}
    \begin{subfigure}{0.49\textwidth}
      \begin{center}
        \begin{tikzpicture}[yscale = -1, xscale = -1]
          \fill[fill = black!20] (-3, -3) -- (3, 3) -- (-3, 3);
          \draw (-3, -3) -- (3, 3);
          
          \draw[<-, dotted] (0, -3) -- (0, 3);
          \draw[<-, dotted] (-3, 0) -- (3, 0);
  
          \fill (-1, 1) circle (2pt);
          \node[anchor = south east] at (-1, 1) {\(\mathbf p_0\)};
  
          \draw (-1, -1) -- (-1, 3);
          \draw(-3, 1) -- (1, 1);

          \node at (-1/3, 5/3) {(A)};
          \node at (-5/3, 5/3) {(B)};
          \node at (-5/3, 1/3) {(C)};
          \node at (-1/3, 1/3) {(D)};
        \end{tikzpicture}
      \end{center}
      \caption{Division into cases}
    \label{fig:abcd-2}
    \end{subfigure}
  \end{center}
  \caption{\emph{%
    An illustration of the proof of Lemma~\ref{lem:abcd}. The half-plane \(H\)
    is shown in gray, with the dotted lines representing the coordinate axes.
    \emph{Left:} The claim of the lemma is equivalent to the claim that the path
    joining \(\mathbf p_0\) to \(\mathbf p_1\) is no longer than the path
    joining \(\mathbf p_0\) to \(T(\mathbf p_1)\). \emph{Right:} The division
    into four cases, depending on where \(\mathbf p_0\) lies relative to
    \(\mathbf p_1\).
  }}%
\label{fig:abcd}
\end{figure}

A further difference from minimization is that the additional assumption of
sortability, Definition~\ref{defn:sortability}, is required to connect
maximizers to maximization normal form. If sortability fails, there may be
intervals \([\ell_i, u_i]\) that can be non-trivially nested for distinct \(i\),
which can cause \(i_0\) and \(i_1\) to ``jump between'' different indices a
potentially exponential number of times in the search for a maximizer. For
notational convenience, we assume throughout that sortable risk vectors are, in
fact, already sorted.

The first step to deriving an effective maximization algorithm is to remove
``trivial'' deviations from maximization normal form.

\begin{definition}[Trivial deviations from normality]
  We say that a risk vector in weak normal form has a \emph{trivial deviation
  from normality} relative to its \emph{bounds} if there are \(i_0 < i_1\) such
  that \(\ell_{i_0} = \ell_{i_1}\), \(\ell_{i_0} < R_{i_0}\), and \(R_{i_1} =
  \ell_{i_1}\); or there are \(i_1' > i_0'\) such that \(u_{i_0'} = u_{i_1'}\),
  \(R_{i_1'} < u_{i_1'}\), and \(R_{i_0'} = u_{i_0'}\).

  We say that a risk vector in weak normal form has a \emph{trivial deviation
  from normality} relative to its \emph{estimates} if there are \(i_0 < i_1\)
  such that \(\hat R_{i_0} = \hat R_{i_1}\) but \(R_{i_0} > R_{i_1}\).
\end{definition}

An illustration of trivial deviations from normality is shown in
Figure~\ref{fig:weak-norm}. As their name suggests, trivial deviations from
normality are straightforward to eliminate.

\begin{figure}[t]
  \begin{center}
    \begin{tikzpicture}
      \draw (-1, 0) -- (2, 0); \draw (3, 0) -- (6, 0); \draw (7, 0) -- (10, 0);
      \node at (2.5, 0) {\dots}; \node at (6.5, 0) {\dots};
      \draw (-1, 5) -- (2, 5); \draw (3, 5) -- (6, 5); \draw (7, 5) -- (10, 5);
      \node at (2.5, 5) {\dots}; \node at (6.5, 5) {\dots};

      \filldraw (0, 1.4) circle (2pt); \draw (0, 0.5) -- (0, 2.3); \draw (-0.1, 0.5) -- (0.1, 0.5); \draw (-0.1, 2.3) -- (0.1, 2.3);
      \filldraw (1, 1.6) circle (2pt); \draw (1, 0.5) -- (1, 2.7); \draw (0.9, 0.5) -- (1.1, 0.5); \draw (0.9, 2.7) -- (1.1, 2.7);
      \filldraw (4, 3) circle (2pt); \draw (4, 2) -- (4, 4.5); \draw (3.9, 2) -- (4.1, 2); \draw (3.9, 4.5) -- (4.1, 4.5);
      \filldraw (5, 3.5) circle (2pt); \draw (5, 2.5) -- (5, 4.5); \draw (4.9, 2.5) -- (5.1, 2.5); \draw (4.9, 4.5) -- (5.1, 4.5);
      \filldraw (8, 2.5) circle (2pt); \draw (8, 1) -- (8, 3.5); \draw (7.9, 1) -- (8.1, 1); \draw (7.9, 3.5) -- (8.1, 3.5);
      \filldraw (9, 2.5) circle (2pt); \draw (9, 1.5) -- (9, 4); \draw (8.9, 1.5) -- (9.1, 1.5); \draw (8.9, 4) -- (9.1, 4);

      \filldraw[red, inner sep = 2pt] (0.2, 1) circle (2pt); \draw[->, red] (0.2, 1.4) -- (0.2, 1.1);
      \filldraw[red, inner sep = 2pt] (1.2, 0.5) circle (2pt); \draw[->, red] (1.2, 1.6) -- (1.2, 0.6);
      \filldraw[red, inner sep = 2pt] (4.2, 4.5) circle (2pt); \draw[->, red] (4.2, 3) -- (4.2, 4.4);
      \filldraw[red, inner sep = 2pt] (5.2, 4) circle (2pt); \draw[->, red] (5.2, 3.5) -- (5.2, 3.9);
      \filldraw[red, inner sep = 2pt] (8.2, 3) circle (2pt); \draw[->, red] (8.2, 2.5) -- (8.2, 2.9);
      \filldraw[red, inner sep = 2pt] (9.2, 2) circle (2pt); \draw[->, red] (9.2, 2.5) -- (9.2, 2.1);

      \node[anchor = north] at (0, 0) {\(i_0\)};
      \node [anchor = north] at (1, 0) {\(i_1\)};
      \node[anchor = north] at (4, 0) {\(i_0'\)};
      \node[anchor = north] at (5, 0) {\(i_1'\)};
      \node[anchor = north] at (8, 0) {\(i_0^\dagger\)};
      \node[anchor = north] at (9, 0) {\(i_1^\dagger\)};
    \end{tikzpicture}
  \end{center}
  \caption{\emph{%
    An illustration of \emph{trivial deviations} from normality. As in
    Figures~\ref{fig:min-norm} and~\ref{fig:max-norm}, the black dots represent
    estimated risks \(\hat R_i\), the error bars represent the allowable range
    \([\ell_i, u_i]\), and the red dots represent the corresponding values of
    \(R_i\), with indices increasing from left to right on the \(x\)-axis. There
    is a trivial deviation from normality relative to the lower bound on the
    left because \(\ell_{i_0} = \ell_{i_1}\) but \(R_{i_1} = \ell_{i_1}\), and a
    similar illustration of a trivial deviation from normality relative to the
    upper bound in the center. On the right there is a trivial deviation from
    the estimates.
}}%
\label{fig:weak-norm}
\end{figure}

\begin{lemma}%
\label{lem:trivial-deviations}
  Suppose the constraints are sortable. If a feasible risk vector \(\mathbf R\)
  is in weak normal form with trivial deviations from normality, then there
  exists a feasible \(\mathbf R'\) in weak normal form with no trivial
  deviations from normality such that
  \[
    \frac 1 n \sum_{i=1}^n R_i^2 = \frac 1 n \sum_{i=1}^n (R_i')^2.
  \]
\end{lemma}

\begin{proof}
  We first define an algorithm for removing trivial deviations from normality,
  and then show that it must eventually terminate when all trivial deviations
  have been removed.
  
  \paragraph*{Algorithm}
  If there were any indices \(i_0\) and \(i_1\) representing a trivial deviation
  from normality, then we could eliminate them by swapping \(R_{i_0}\) and
  \(R_{i_1}\), i.e., by considering 
  \[
    R_i' = R_i \quad (i \neq i_0, i_1), \qquad R_{i_0}' = R_{i_1}, \qquad
    R_{i_1}' = R_{i_0}.
  \]
  In particular, it follows immediately that \(\mathbf R'\) is still in weak
  normal form and that \(\tfrac 1 n \sum_{i=1}^n R_i^2 = \tfrac 1 n \sum_{i=1}^n
  (R_i')^2\). Because \(\mathbf{R}\) is sorted, the only feasibility constraint
  on \(\mathbf{R'}\) that is not satisfied trivially is that \(\|\mathbf R' -
  \hat {\mathbf R}\|_1 \leq \epsilon\). However, this constraint follows by
  applying Lemma~\ref{lem:abcd} with \(a = R_{i_0}, b = R_{i_1}, c =
  \hat{R}_{i_1}\), and \(d = \hat{R}_{i_0}\), which yields
  \[
    \|\mathbf R' - \hat {\mathbf R}\|_1 \leq \|\mathbf R - \hat {\mathbf R}\|_1
    \leq \epsilon.
  \]

  \paragraph*{Termination}
  We call a pair of indices \((i_0, i_1)\) \emph{out of order} if \(i_0 < i_1\),
  but \(R_{i_0} > R_{i_1}\). We note that any trivial deviation from normality
  requires that the corresponding indices \(i_0\) and \(i_1\) are out of order.
  Then, we observe the following facts:
  \begin{enumerate}[label=\emph{Fact~\arabic*.}, leftmargin=1in]
    \item There are only finitely many pairs of out-of-order indices.
    \item As long as there is a trivial deviation of any kind, it is possible to
      swap indices so as to remove that trivial deviation.
    \item Any swap undertaken as part of the algorithm strictly reduces the
      number of out-of-order pairs of indices.
  \end{enumerate}
  We note that it follows immediately from these three facts that a greedy
  algorithm removing trivial deviations in any order whatsoever must eventually
  terminate.

  Fact~1 is immediate, and Fact~2 follows from the discussion of the algorithm
  above. To see that Fact~3 holds, it suffices to show that \emph{any} swap of
  out-of-order pairs of indices strictly reduces the number of out-of-order
  pairs of indices.

  To this end, suppose \(i_0 < i_1\) are out of order. As before, we use
  \(\mathbf R'\) to denote the risk vector \emph{after} swapping indices \(i_0\)
  and \(i_1\). First, suppose there is an index \(i^*\) such that \(i_0 < i^* <
  i_1\). Then, if \((i_0, i^*)\) is out of order \emph{after} swapping \(i_0\)
  and \(i_1\), it must have been out of order before swapping, since
  \[
    R_{i_0} > R_{i_1} = R_{i_0}' > R_{i^*}' = R_{i_*}.
  \]
  Likewise, if \((i^*, i_1)\) is out of order \emph{after} swapping \(i_0\) and
  \(i_1\), it too must have been out of order before swapping, since
  \[
    R_{i^*} = R_{i^*}' > R_{i_1}' = R_{i_0} > R_{i_1}.
  \]
  Therefore, swapping cannot increase the number of out-of-order pairs involving
  \(i_0 < i^* < i_1\).

  On the other hand, since \(R_{i_0}' < R_{i_1}'\), if \(i^* < i_0\), then there
  are three possibilities:
  \begin{enumerate}[label=\emph{Case~\arabic*.}, leftmargin=1in]
    \item After swapping, neither \((i^*, i_0)\) nor \((i^*, i_1)\) is out of order.
    \item After swapping, both \((i^*, i_0)\) and \((i^*, i_1)\) are out of order.
    \item After swapping, \((i^*, i_0)\) is out of order, but \((i^*, i_1)\) is not.
  \end{enumerate}
  In the first case, the number of out-of-order pairs cannot have increased, so
  we only need to consider the second and third cases. In Case~2, we must have
  that
  \[
    R_{i^*} = R_{i^*}' > R_{i_1}' = R_{i_0} > R_{i_1},
  \]
  so both \((i^*, i_0)\) and \((i^*, i_1)\) were out of order before swapping.
  In Case~3, we must have that
  \[
    R_{i_0} = R_{i_1}' \geq R_{i^*}' = R_{i_*} > R_{i_0}' = R_{i_1},
  \]
  so that before swapping \((i^*, i_1)\) was out of order, but \((i^*, i_0)\)
  was not. Therefore, it follows that swapping again cannot increase the number
  of out-of-order pairs involving \(i^* < i_0\). A similar argument shows that
  the number of out-of-order pairs involving \(i_1 < i^{*}\) does not increase
  after swapping. Therefore swapping strictly reduces the number of out-of-order
  pairs, since it at least eliminates the pair \((i_0, i_1)\), which is Fact~3.
\end{proof}

With these preliminaries in hand, we are prepared to solve the maximization
problem.

\begin{lemma}%
\label{lem:maximize-simp}
  Suppose the constraints are sortable. Then the single stratum maximization
  problem in Eq.~\eqref{eq:ss-optim} is maximized by a risk vector in
  maximization normal form and that either exhausts the budget or has
  \(i^{\mathrm{lwr}} = i^{\mathrm{upr}}\).
\end{lemma}

\begin{proof}
  We adopt a strategy similar to that used in the ``strengthening'' portion of
  the proof of Lemma~\ref{lem:minimize}. Namely, we show that any point
  satisfying the necessary conditions of Lemma~\ref{lem:maximize} but without
  pivots can be ``improved'' to a feasible point that achieves a greater
  objective.

  Let \(\mathbf R^*\) be a maximizer. By Lemma~\ref{lem:maximize}, \(\mathbf
  R^*\) is in weak maximization normal form; by
  Lemma~\ref{lem:trivial-deviations}, we may assume that \(\mathbf R^*\) has no
  trivial deviations from normality.
  
  To prove the theorem, we first observe that it suffices to prove the following
  two claims:
  \begin{enumerate}
    \item There do not exist \(i_0 < i_1\) such that \(R_{i_0}^* > \ell_{i_0}\)
      and \(R_{i_1}^* < \hat R_{i_1}\),
    \item There do not exist \(i_0 < i_1\) such that \(R_{i_1}^* < u_{i_1}\) and
      \(R_{i_0}^* > \hat R_{i_0}\). 
  \end{enumerate}
  First, we see how the theorem follows from the two claims. Define
  \(i^{\mathrm{lwr}}\) and \(i^{\mathrm{upr}}\) as follows:
  \[
    i^{\mathrm{lwr}} = \min \{\, i : R_i^* > \ell_i \,\}, \qquad
    i^{\mathrm{upr}} = \max \{\, i : R_i^* < u_i \,\},
  \]
  then, for all \(i < i^{\mathrm{lwr}}\), \(R_i^* = \ell_i\), and for all \(i >
  i^{\mathrm{upr}}\), \(R_i^* = u_i\). Moreover, for all \(i\) satisfying
  \(i^{\mathrm{lwr}} < i < i^{\mathrm{upr}}\), by Claim~1, \(R_i^* \geq \hat
  R_i\); and, by Claim~2, \(R_i^* \leq \hat R_i\). Consequently, \(R_i^* = \hat
  R_i\) for these indices. Moreover, if \(i^{\mathrm{lwr}} \neq
  i^{\mathrm{upr}}\), by Claim~2, \(R_{i^{\mathrm{lwr}}}^* \leq \hat
  R_{i^{\mathrm{lwr}}}\), i.e., \(R_{i^{\mathrm{lwr}}}^* \in
  [\ell_{i^{\mathrm{lwr}}}, \hat R_{i^{\mathrm{lwr}}}]\). Similarly, by Claim~1,
  we conclude that \(R_{i^{\mathrm{upr}}}^* \in [\hat R_{i^{\mathrm{upr}}},
  u_{i^{\mathrm{upr}}}]\).

  Therefore it suffices to prove the two claims. Since the proofs are virtually
  identical, we focus on Claim~1.

  The key observation is the following. Suppose there exist indices
  \(R_{i_0}^*\) and \(R_{i_1}^*\) with \(R_{i_1}^* \geq R_{i_0}^*\), and suppose
  that there is some \(\delta\) such that after decreasing \(R_{i_0}^*\) and
  increasing \(R_{i_1}^*\) by \(\delta\) the result is still feasible. Then
  \(\mathbf R^*\) cannot be maximal, since, letting \(\mathbf R' = \mathbf R^* +
  \delta (\mathbf e_{i_1} - \mathbf e_{i_0})\), we have that
  \begin{align}
  \label{eq:adjust}
    \begin{split}
      \left[ \sum_{i=1}^n (R_i')^2 \right] - \left[ \sum_{i=1}^n (R_i^*)^2
      \right]
        &= (R_{i_0}^* - \delta)^2 + (R_{i_1}^* + \delta)^2 - (R_{i_0}^*)^2 -
        (R_{i_1}^*)^2 \\
        &= 2 \delta (\delta + [R_{i_1}^* - R_{i_0}^*]) \\
        &\geq 2 \delta^2 \\
        &> 0.
    \end{split}
  \end{align}

  To actually prove the claim, we proceed by contradiction. In particular,
  suppose that \(i_0 < i_1\) represented an exception to Claim~1, so that
  \(R_{i_0}^* > \ell_{i_0}\) and \(R_{i_1}^* < \hat R_{i_1}\).

  If \(R_{i_0}^* \leq R_{i_1}^*\), then Eq.~\eqref{eq:adjust} immediately
  applies, since increasing \(R_{i_1}^*\) \emph{decreases} the \(L_1\) distance
  between \(\hat {\mathbf R}\) and \(\mathbf R^*\) and decreasing \(R_{i_0}^*\),
  at worst, \emph{increases} it at the same rate. More formally, for \(0 <
  \delta\) sufficiently small,
  \begin{align*}
    |(R_{i_0}^* - \delta) - \hat R_{i_0}| + |(R_{i_1}^* + \delta) - \hat
    R_{i_1}|
      &= |(R_{i_0}^* - \delta) - \hat R_{i_0}| + \hat R_{i_1} - (R_{i_1}^* +
      \delta) \\
      &\leq |R_{i_0}^* - \hat R_{i_0}| + \delta + \hat R_{i_1} - R_{i_1}^* -
      \delta \\
      &= |R_{i_0}^* - \hat R_{i_0}| + |R_{i_1}^* - \hat R_{i_1}|,
  \end{align*}
  and so \(\mathbf R'\) is feasible. Because \(R_{i_0}^* \leq R_{i_1}^*\), by
  Eq.~\eqref{eq:adjust}, \(\mathbf R'\) achieves a greater average sum of
  squares that \(\mathbf R^*\), contrary to the assumed maximality of \(\mathbf
  R^*\).
  
  On the other hand, if \(R_{i_0}^* > R_{i_1}^*\), then, by
  Lemma~\ref{lem:abcd}, the vector \(\mathbf R^\dagger\) given by switching the
  indices \(i_0\) and \(i_1\), i.e.,
  \[
    R^\dagger_i = R^*_i \quad (i \neq i_0, i_1), \qquad R^\dagger_{i_0} =
    R^*_{i_1}, \qquad R^\dagger_{i_1} = R^*_{i_0},
  \]
  satisfies \(\|\mathbf R^\dagger - \hat {\mathbf R}\|_1 \leq \epsilon\).
  Moreover, applying our sortability hypothesis, we have that
  \[
    u_{i_0} \geq R^*_{i_0} > R^*_{i_1} \geq \ell_{i_1} \geq \ell_{i_0},
  \]
  and
  \[
    u_{i_1} \geq u_{i_0} \geq R^*_{i_0} > R^*_{i_1} \geq \ell_{i_1},
  \]
  we have that for all \(i = 1, \ldots, n\), \(\ell_i \leq R^\dagger_i \leq
  u_i\). Therefore \(\mathbf R^\dagger\) is feasible. However, since \(\mathbf
  R^*\) has no trivial deviations from normality, it follows that
  \[
    R^\dagger_{i_0} = R_{i_1}^* \geq \ell_{i_1} > \ell_{i_0}, \qquad
    R^{\dagger}_{i_1} = R_{i_0}^* \leq u_{i_0} < u_{i_1}.
  \]
  Since \(R^\dagger_{i_0} < R^\dagger_{i_1}\), we would like to apply
  Eq.~\eqref{eq:adjust} to derive a contradiction. To that end, we require that
  \(\mathbf R^\dagger + \delta (\mathbf e_{i_1} - \mathbf e_{i_0})\) is feasible
  for sufficiently small \(\delta > 0\). If it were the case that
  \begin{equation}
  \label{eq:nice-case}
    \| \mathbf R^\dagger + \delta (\mathbf e_{i_1} - \mathbf e_{i_0})\|_1 \leq
    \|\mathbf R^\dagger\|_1,
  \end{equation}
  then the result follows immediately. However, Eq.~\eqref{eq:nice-case} holds
  except when \(R_{i_0}^\dagger \leq \hat R_{i_0}\) and \(R_{i_1}^\dagger \geq
  \hat R_{i_1}\); or, equivalently, \(R_{i_1}^* \leq \hat R_{i_0}\) and
  \(R_{i_0}^* \geq \hat R_{i_1}\)---see Figure~\ref{fig:bad-case}. However, in
  this case, we have by sortability that
  \begin{align*}
    |R_{i_0}^* - \hat R_{i_0}| + |R_{i_1}^* - \hat R_{i_1}|
      &= R_{i_0}^* - \hat R_{i_0} + \hat R_{i_1} - R_{i_1}^* \\
      &> (R_{i_0}^* - \hat R_{i_0} + \hat R_{i_1} - R_{i_1}^*) + 2 (\hat R_{i_0}
        - \hat R_{i_1}) \\
      &= R_{i_0}^* - \hat R_{i_1} + \hat R_{i_0} - R_{i_1}^* \\
      &= |R_{i_0}^\dagger - \hat R_{i_0}| + |R_{i_1}^\dagger - \hat R_{i_1}|,
  \end{align*}
  where we have used the fact that, because there are no trivial deviations from
  normality relative to the estimates, \(\hat R_{i_1} > \hat R_{i_0}\).
  Therefore \(\|\mathbf R^\dagger\|_1 < \epsilon\) and so we can choose
  \(\delta\) sufficiently small that \(\|\mathbf R^{\dagger} + \delta (\mathbf
  e_{i_1} - \mathbf e_{i_0})\|_1 \leq \epsilon\), completing the proof.

  Therefore, no such \(i_0 < i_1\) exist, and so \(\mathbf R^*\) is in
  maximization normal form.
\end{proof}

\begin{figure}
  \begin{center}
    \begin{tikzpicture}[xscale=0.9]
      \draw (-1, 0) -- (2, 0); \draw (3, 0) -- (6, 0); \draw (7, 0) -- (10, 0); \draw (11, 0) -- (14, 0);
      \draw (-1, 5) -- (2, 5); \draw (3, 5) -- (6, 5); \draw (7, 5) -- (10, 5); \draw (11, 5) -- (14, 5);

      \filldraw (0, 1.4) circle (2pt); \draw (0, 0.5) -- (0, 3.8); \draw (-0.1, 0.5) -- (0.1, 0.5); \draw (-0.1, 3.8) -- (0.1, 3.8); \node at (0, 0.2) {\(i_0\)};
      \filldraw (1, 2.6) circle (2pt); \draw (1, 0.8) -- (1, 4.1); \draw (0.9, 0.8) -- (1.1, 0.8); \draw (0.9, 4.1) -- (1.1, 4.1); \node at (1, 0.2) {\(i_1\)};
      \filldraw (4, 1.4) circle (2pt); \draw (4, 0.5) -- (4, 3.8); \draw (3.9, 0.5) -- (4.1, 0.5); \draw (3.9, 3.8) -- (4.1, 3.8); \node at (4, 0.2) {\(i_0\)};
      \filldraw (5, 2.6) circle (2pt); \draw (5, 0.8) -- (5, 4.1); \draw (4.9, 0.8) -- (5.1, 0.8); \draw (4.9, 4.1) -- (5.1, 4.1); \node at (5, 0.2) {\(i_1\)};
      \filldraw (8, 1.4) circle (2pt); \draw (8, 0.5) -- (8, 3.8); \draw (7.9, 0.5) -- (8.1, 0.5); \draw (7.9, 3.8) -- (8.1, 3.8); \node at (8, 0.2) {\(i_0\)};
      \filldraw (9, 2.6) circle (2pt); \draw (9, 0.8) -- (9, 4.1); \draw (8.9, 0.8) -- (9.1, 0.8); \draw (8.9, 4.1) -- (9.1, 4.1); \node at (9, 0.2) {\(i_1\)};
      \filldraw (12, 1.4) circle (2pt); \draw (12, 0.5) -- (12, 3.8); \draw (11.9, 0.5) -- (12.1, 0.5); \draw (11.9, 3.8) -- (12.1, 3.8); \node at (12, 0.2) {\(i_0\)};
      \filldraw (13, 2.6) circle (2pt); \draw (13, 0.8) -- (13, 4.1); \draw (12.9, 0.8) -- (13.1, 0.8); \draw (12.9, 4.1) -- (13.1, 4.1); \node at (13, 0.2) {\(i_1\)};

      \node at (0.5, -1) {Case 1};
      \filldraw[red] (0, 2.1) circle (2pt); \filldraw[blue] (1, 2.1) circle (2pt); \draw[->, blue] (0.1, 2.1) -- (0.9, 2.1); \filldraw[green!50!black] (1, 2.4) circle (2pt); \draw[->, green!50!black] (1, 2.2) -- (1, 2.3);
      \filldraw[red] (1, 1.9) circle (2pt); \filldraw[blue] (0, 1.9) circle (2pt); \draw[->, blue] (0.9, 1.9) -- (0.1, 1.9); \filldraw[green!50!black] (0, 1.6) circle (2pt); \draw[->, green!50!black] (0, 1.8) -- (0, 1.7);
      \draw (0, 1.4) -- (-0.43, 1.4); \draw (1, 2.6) -- (1.43, 2.6);
      \draw[line width = 2pt, blue] (-0.4, 1.4) -- (-0.4, 1.9); \draw[line width = 2pt, green!50!black] (-0.2, 1.4) -- (-0.2, 1.6);
      \draw[line width = 2pt, green!50!black] (1.2, 2.4) -- (1.2, 2.6); \draw[line width = 2pt, blue] (1.4, 2.1) -- (1.4, 2.6);

      \node at (4.5, -1) {Case 2};
      \filldraw[red] (4, 3.2) circle (2pt); \filldraw[blue] (5, 3.2) circle (2pt); \draw[->, blue] (4.1, 3.2) -- (4.9, 3.2); \filldraw[green!50!black] (5, 3.5) circle (2pt); \draw[->, green!50!black] (5, 3.3) -- (5, 3.4);
      \filldraw[red] (5, 1.9) circle (2pt); \filldraw[blue] (4, 1.9) circle (2pt); \draw[->, blue] (4.9, 1.9) -- (4.1, 1.9); \filldraw[green!50!black] (4, 1.6) circle (2pt); \draw[->, green!50!black] (4, 1.8) -- (4, 1.7);
      \draw (4, 1.4) -- (3.57, 1.4); \draw (5, 2.6) -- (5.43, 2.6);
      \draw[line width = 2pt, blue] (3.6, 1.4) -- (3.6, 1.9); \draw[line width = 2pt, green!50!black] (3.8, 1.4) -- (3.8, 1.6);
      \draw[line width = 2pt, blue] (5.2, 2.6) -- (5.2, 3.2); \draw[line width = 2pt, green!50!black] (5.4, 2.6) -- (5.4, 3.5);

      \node at (8.5, -1) {Case 3};
      \filldraw[red] (8, 2.1) circle (2pt); \filldraw[blue] (9, 2.1) circle (2pt); \draw[->, blue] (8.1, 2.1) -- (8.9, 2.1); \filldraw[green!50!black] (9, 2.4) circle (2pt); \draw[->, green!50!black] (9, 2.2) -- (9, 2.3);
      \filldraw[red] (9, 1) circle (2pt); \filldraw[blue] (8, 1) circle (2pt); \draw[->, blue] (8.9, 1) -- (8.1, 1); \filldraw[green!50!black] (8, 0.7) circle (2pt); \draw[->, green!50!black] (8, 0.9) -- (8, 0.8);
      \draw (8, 1.4) -- (7.57, 1.4); \draw (9, 2.6) -- (9.43, 2.6);
      \draw[line width = 2pt, green!50!black] (7.6, 1.4) -- (7.6, 0.7); \draw[line width = 2pt, blue] (7.8, 1.4) -- (7.8, 1);
      \draw[line width = 2pt, green!50!black] (9.2, 2.4) -- (9.2, 2.6); \draw[line width = 2pt, blue] (9.4, 2.1) -- (9.4, 2.6);
      
      \draw[dashed] (10.5, -1) -- (10.5, 6);
      
      \node at (12.5, -1) {Case 4};
      \filldraw[red] (12, 3.3) circle (2pt); \filldraw[blue] (13, 3.3) circle (2pt); \draw[->, blue] (12.1, 3.3) -- (12.9, 3.3); \filldraw[green!50!black] (13, 3.6) circle (2pt); \draw[->, green!50!black] (13, 3.4) -- (13, 3.5);
      \filldraw[red] (13, 1) circle (2pt); \filldraw[blue] (12, 1) circle (2pt); \draw[->, blue] (12.9, 1) -- (12.1, 1); \filldraw[green!50!black] (12, 0.7) circle (2pt); \draw[->, green!50!black] (12, 0.9) -- (12, 0.8);
      \draw(12, 1.4) -- (11.57, 1.4); \draw (13, 2.6) -- (13.43, 2.6);
      \draw[red, line width = 2pt] (11.8, 1.4) -- (11.8, 3.3); \draw[blue, line width = 2pt] (11.8, 1) -- (11.8, 1.4); \draw[green!50!black, line width = 2pt] (11.6, 1.4) -- (11.6, 0.7);
      \draw[red, line width = 2pt] (13.2, 1) -- (13.2, 2.6); \draw[blue, line width = 2pt] (13.2, 2.6) -- (13.2, 3.3); \draw[green!50!black, line width = 2pt] (13.4, 2.6) -- (13.4, 3.6);
      
    \end{tikzpicture}
  \end{center}
  \caption{\emph{%
    This diagram shows various possibilities of the relative ordering of
    \(R_{i_0}^*\), \(R_{i_1}^*\), \(\hat R_{i_0}\), and \(\hat R_{i_1}\) in the
    final section of the proof of Lemma~\ref{lem:maximize-simp}. Here, the upper
    and lower bounds are shown by the error bars, the original estimates (\(\hat
    {\mathbf R}\)) are shown in black, the starting risk vector (\(\mathbf
    R^*)\) as red dots, the risk vector \emph{after} indices \(i_0\) and \(i_1\)
    have been exchanged (\(\mathbf R^\dagger\)) as blue dots, and the vector
    with increased objective (\(\mathbf R^\dagger + \delta (\mathbf e_{i_1} -
    \mathbf e_{i_0})\)) as green dots. Here we see that in Case~1, where \(\hat
    R_{i_1} \geq R_{i_0}^* \geq R_{i_1}^* \geq \hat R_{i_0}\), perturbing
    \(\mathbf R^\dagger\) by \(\delta (\mathbf e_{i_1} - \mathbf e_{i_0})\)
    actually \emph{decreases} the \(L_1\) distance to \(\hat {\mathbf R}\); the
    \(L_1\) distance between \(\hat {\mathbf R}\) and \(\mathbf R^\dagger\) is
    shown with the heavy blue lines, and between \(\hat {\mathbf R}\) and
    \(\mathbf R^\dagger + \delta (\mathbf e_{i_1} - \mathbf e_{i_0})\) by the
    heavy green lines. In Case~2, where \(R_{i_0}^* \geq \hat R_{i_1} \geq
    R_{i_1}^* \geq \hat R_{i_0}\), we see that perturbing \(\mathbf R^\dagger\)
    by \(\delta (\mathbf e_{i_1} - \mathbf e_{i_0})\) increases the absolute
    distance from \(\hat {\mathbf R}\) at \(i_1\), but this increase is exactly
    compensated by a decrease at \(i_0\). The situation is identical in Case~3,
    where  \(\hat R_{i_1} \geq R_{i_0}^* \geq \hat R_{i_0} \geq R_{i_1}^*\),
    except with the roles of \(i_0\) and \(i_1\) reversed. Finally, in Case~4,
    which is treated separately, we see that the initial transformation from
    \(\mathbf R^*\) to \(\mathbf R^\dagger\) strictly reduces the \(L_1\)
    distance to \(\hat {\mathbf R}\) when we compare \(\mathbf R^*\) (heavy red
    lines) to \(\mathbf R^\dagger\) (heavy blue lines), so that the \(L_1\)
    distance between \(\hat {\mathbf R}\) and the perturbed risk vector (heavy
    green lines) is still less than the constraint.
  }}%
\label{fig:bad-case}
\end{figure}

Lemma~\ref{lem:maximize-simp} allows us to efficiently search over the space of
risk vectors to find the maximizer. In particular, all that is necessary is to
decrease the risk at the lower pivot and increase it at the upper pivot until
the \(L_1\) budget is expended. The linear-time algorithm for doing so is given
in Algorithm~\ref{algo:max-one}. (As in the case of
Algorithm~\ref{algo:min-one}, there is a potential preprocessing step, given in
Algorithm~\ref{algo:max-adjust}.) As we did in the minimization algorithm, we
capture the maximization algorithm's value at a variety of \(L_1\) budgets,
controlled by some step-size \(\gamma\)---which, we assume, satisfies \(k \gamma
= \epsilon\) for some \(k \in \mathbb N\)---rather than just the full budget
\(\epsilon\). We denote this collection with \(\boldsymbol \Sigma\). The reason
for doing this is to allow us to maximize across strata.

\subsubsection{Optimizing over all strata}

Consider the simplified maximization problem in Eq.~\eqref{eq:simp-optim}. Since
the objective is separable, the problem would be separable in the different
strata, except that the \(L_1\) budget constraint ranges over all of the strata
simultaneously. Much as we reduced the base problem to the parameterized problem
by introducing additional parameters \(\tau_j\), \(j = 1, \ldots, m\), if the
optimal budget allocation to each stratum were known, then we could reduce to
the single stratum case, applying Algorithm~\ref{algo:max-one}. In particular,
we note that the following optimization problem (hereafter \textbf{``the
separable problem''}) parameterized by the stratum-specific budgets
\(\epsilon_{1,0}, \ldots, \epsilon_{m,0}, \epsilon_{1,1}, \ldots,
\epsilon_{m,1}\) is separable:
\begin{equation}
\label{eq:sep-optim}
  \arraycolsep=1.4pt
  \begin{array}{rrclc}
    \mathop\mathrm{Optimize}\limits_{\mathbf R \in \mathbb R^n}
      & \multicolumn{3}{c}{\displaystyle\frac 1 n \displaystyle\sum_{i=1}^n R_i^2} \\
    \mathrm{s.t.} & \frac 1 {|\mathcal S_{j,1}|} \sum_{i \in \mathcal S_{j,a}}^n
      |R_i - \hat R_i|
                        &\leq& \epsilon_{j,a}, & (a = 0, 1;\; j = 1, \ldots, m) \\[4pt]
                  & \sum_{i \in \mathcal S_{j,1}} R_i
                        &=&     \rho_j, & (j = 1, \ldots, m) \\[4pt]
                  & \sum_{i \in \mathcal S_{j,0}} R_i
                        &=&     \tau_j,    & (j = 1, \ldots, m) \\
                  & R_i &\leq&  u_i,    & (i = 1, \ldots, n) \\
                  & R_i &\geq&  \ell_i. & (i = 1, \ldots, n) \\
  \end{array}
\end{equation}
In particular, the separable problem can be solved stratum by stratum, since the
constraints and objective are both separable at the stratum level.

Reducing solving the simplified problem to solving the separable problem,
however, is complex; There is no obvious way to determine the optimal \(L_1\)
budget allocation other than by sweeping over all possible budgets. However, it
is nevertheless possible to efficiently discover an \emph{approximately} optimal
budget allocation using a divide-and-conquer approach. In particular, as shown
in Algorithm~\ref{algo:max-one}, it is only marginally more difficult
computationally to output the maximum value of the objective for the entire
budget than to output the maximum value of the objective for a range of budgets,
i.e., to output \(\boldsymbol \Sigma\), which gives the maximum value of the
objective for the budgets \(0\), \(\gamma\), \(2\gamma\), \ldots, \(\epsilon\).

Given \(\boldsymbol \Sigma^{(0)}\) and \(\boldsymbol \Sigma^{(1)}\)
corresponding to two different strata when the allocated budget is \(0\),
\(\gamma\), \(2\gamma\), \ldots, \(\epsilon\), we can find the approximate
maximum value of the objective when the \emph{total} budget allocated to both
strata is \(0\), \(\gamma\), \(2\gamma\), \ldots, \(\epsilon\) by sweeping over
the two dimensional grid of points \(\{0, \gamma, \ldots, \epsilon\} \times \{0,
\gamma, \ldots, \epsilon\}\), and, for \(k \gamma\), outputting
\[
  \max \left( \Sigma_0^{(0)} + \Sigma_{k+1}^{(1)}, \Sigma_1^{(0)} +
  \Sigma_k^{(1)}, \ldots, \Sigma_{k+1}^{(0)} + \Sigma_0^{(1)} \right).
\]
For completeness, we explicitly give this maximization routine in
Algorithm~\ref{algo:max-all}.

\begin{lemma}
\label{lem:max-runtime}
  If \(\boldsymbol \ell\), \(\hat {\mathbf R}\) and \(\mathbf u\) have been
  sorted, then the approximate solution to the maximization problem in
  Eq.~\eqref{eq:simp-optim} given by applying Algorithms~\ref{algo:max-one}
  and~\ref{algo:max-all} is
  \[
    O \left( m \cdot \left( \frac \epsilon \gamma \right)^2 + n \right).
  \]
\end{lemma}

We note that \(\delta\), as used in Section~\ref{sec:sens} above, equals \(2 m
\gamma\).

\begin{proof}
  We begin by analyzing Algorithm~\ref{algo:max-adjust}. We note that the number
  of iterations of the while loop on lines~7 through~12 is capped by the length
  of the input, since \(i^{\mathrm{upr}}\) is decremented on each input.
  Likewise, Algorithm~\ref{algo:max-one} is almost linear in the size of the
  input: in the while loop on lines~14 though~30, either \(i^{\mathrm{lwr}}\) is
  incremented, or the subsequent multiple of \(\gamma\) is reached, meaning that
  the algorithm as a whole is linear in the size of the input and \(\epsilon /
  \gamma\). Therefore, running Algorithm~\ref{algo:min-one} over all \(2m\)
  strata requires \(O(n + \tfrac {m \epsilon} \gamma)\) time.

  All of the strata can be combined using \(2m-1\) applications of
  Algorithm~\ref{algo:min-all}. Reviewing the for loop on lines~3 through~11 and
  the while loop on lines~5 through~8, we see that exactly
  \[
    \frac {\frac \epsilon \gamma \cdot (\frac \epsilon \gamma + 1)} 2
  \]
  iterations are performed. Adding this to the previous runtime obtained and
  simplifying gives the desired expression.
\end{proof}

\subsection{Controlling approximation error}

There are two sources of approximation error in our algorithm. The first is the
approximation error introduced by solving the maximization problem only
approximately. The second is the approximation error introduced by the fact that
we cannot sweep over all possible values of \(\boldsymbol \tau = (\tau_1,
\ldots, \tau_m)\) when solving the simplified problem; instead, we must sweep
over a grid where. Characterizing how much error is introduced by each of these
approximations is our final task.

We begin with the following simple lemma.

\begin{lemma}%
\label{lem:sq-error}
  Suppose \(\mathbf R^{(0)}\) and \(\mathbf R^{(1)}\) are risk vectors. Then
  \begin{equation}
    \left| \sum_{i=1}^n (R^{(1)}_i)^2 - \sum_{i=1}^n (R^{(0)}_i)^2 \right| \leq
    2 \cdot \|\mathbf R^{(1)} - \mathbf R^{(2)}\|.
  \end{equation}
\end{lemma}

\begin{proof}
  The claim, which is a version of Hölder's inequality, follows
  straightforwardly from considering the following difference:
  \begin{align*}
    \left| \sum_{i=1}^n (R^{(1)}_i)^2 - (R^{(0)}_i)^2 \right|
      &\leq \sum_{i=1}^n |(R^{(1)}_i)^2 - (R^{(0)}_i)^2| \\
      &\leq \sum_{i=1}^n (R^{(1)}_i + R^{(0)}_i) \cdot |R^{(1)}_i - R^{(0)}_i|
      \\
      &\leq \sum_{i=1}^n 2 \cdot |R^{(1)}_i - R^{(0)}_i| \\
      &= 2 \cdot \| \mathbf R^{(1)} - \mathbf R^{(0)}\|_1.
  \end{align*}
  The first inequality follows from the triangle inequality, and the second and
  third from the fact that \(0 \leq R^{(0)}_i, R^{(1)}_i \leq 1\).
\end{proof}

Using Lemma~\ref{lem:sq-error}, we obtain the following characterization of the
potential error in our maximization algorithm, again recalling that \(\delta = 2
m \gamma\), where \(\delta\) is as in Section~\ref{sec:sens}.

\begin{lemma}%
\label{lem:max-error}
  The difference between the true maximum of the simplified problem---i.e.,
  Eq.~\eqref{eq:simp-optim}---and the quantity obtained from applying
  Algorithms~\ref{algo:min-one} and~\ref{algo:min-all} is at most \(2 m
  \gamma\).
\end{lemma}

\begin{proof}
  Let \(\mathbf R^*\) be the true maximizing solution. We will show that
  Algorithm~\ref{algo:max-all} computes the value of the objective at a
  ``close'' point at which we can apply Lemma~\ref{lem:sq-error}.
  
  For each stratum \(\mathcal S_{j,a}\), let \(\epsilon_{j,a}\) denote
  \[
    \frac 1 n \sum_{i \in \mathcal S_{j,a}} |R^*_i - \hat R_i|.
  \]
  Then, there exist \(\tilde \epsilon_{j,a} \in \{0, \gamma, 2 \gamma, \ldots,
  \epsilon\}\) such that \(\tilde \epsilon_{j,a} < \epsilon_{j,a}\) and
  \(\epsilon_{j,a} - \tilde \epsilon_{j,a}\) is less than \(\gamma\).

  Let
  \[
    \Delta = \sum_{a=0}^1 \sum_{j=1}^m \epsilon_{j,a} - \tilde \epsilon_{j,a} =
    \epsilon - k \gamma
  \]
  for some \(k\). Since \(\epsilon\) is a multiple of \(\gamma\), it follows
  that \(\Delta = k' \gamma\) for some \(k \in \mathbb N\), where \(k' < 2 m\),
  the number of strata.

  Set \(\epsilon'_{j,a} = \tilde \epsilon_{j,a}\) except for \(k'\) arbitrarily
  chosen strata, where we set \(\epsilon'_{j,a} = \tilde \epsilon_{j,a} +
  \gamma\) instead. Then, there exists \(\mathbf R\) that is in (maximization)
  normal form such that
  \[
    \frac 1 n \sum_{i=1}^n |R_i - \hat R_i| = \epsilon, \qquad \frac 1 n
    \sum_{\substack{A_i = a\\C_i = c_j}} |R_i - \hat R_i| = \epsilon'_{j,a}.
  \]

  The remainder of the proof follows simply by comparing \(\mathbf R\) and
  \(\mathbf R^*\) stratum by stratum. In particular, consider the restrictions
  of \(\mathbf R\) and \(\mathbf R^*\) to a single stratum \(\mathcal S_{j,a}\).
  By Eq.~\eqref{eq:simp-optim}, we have that
  \[
    \sum_{i \in \mathcal S_{j,a}} R_i = \sum_{i \in \mathcal S_{j,a}} R_i^*,
  \]
  and, moreover, by Lemma~\ref{lem:maximize-simp}, both \(\mathbf R_{j,a}\) and
  \(\mathbf R_{j,a}^*\) can be assumed to be in maximization normal form.
  Assume, without loss of generality, that \(\epsilon'_{j,a} \leq
  \epsilon_{j,a}\). (The proof is virtually identical if the inequality is
  reversed.)

  Let \(i_0\) and \(i_1\) denote the pivots of \(\mathbf R\), and \(i_0^*\) and
  \(i_1^*\) of \(\mathbf R^*\), so that \(i_0 \leq i_0^* \leq i_1^* \leq i_1\).
  (To avoid unnecessary notational complication, we write as if all \(n\)
  individuals belong to the stratum \(\mathcal S_{j,a}\) and the constraints
  were sorted.) Then, we note that
  \begin{equation}
  \label{eq:ordering}
    R_i \geq R_i^* \quad (i_0 \leq i \leq i_0^*), \qquad R_i \leq R_i^* \quad
    (i_1^* \leq i \leq i_1), \qquad R_i = R_i^* \quad (\text{otherwise}).
  \end{equation}
  It follows that
  \begin{equation}
  \label{eq:diff}
    \sum_{i=1}^n (R_i^*)^2 - R_i^2 = \left[ \sum_{i = i_1^*}^{i_1} (R_i^*)^2 -
    R_i^2 \right] - \left[ \sum_{i = i_0}^{i_0^*} R_i^2 - (R_i^*)^2 \right].
  \end{equation}
  Both terms in Eq.~\eqref{eq:diff} are positive by Eq.~\eqref{eq:ordering}.
  Moreover, by Lemma~\ref{lem:sq-error}, both terms are less than or equal to
  \[
    2 \cdot \left[ \sum_{i = i_1}^{i_1^*} R_i^* - R_i \right] = 2 \cdot \left[
    \sum_{i = i_0}^{i_0^*} R_i - R_i^* \right] = 2n \cdot \frac {\epsilon_{a,j}
  - \epsilon_{a,j}'} 2 \leq n \gamma.
  \]
  In particular, their difference must also be less than this quantity, and so,
  summing across strata, we have that
  \[
    \sum_{i=1}^n (R_i^*)^2 - R_i^2 \leq 2 n m \gamma.
  \]
  Dividing through by \(n\) gives the result.
\end{proof}

The second element we need is a bound on the sensitivity of the objective of the
simplified problem to changes in the parameters.

\begin{lemma}%
\label{lem:simp-error}
  Let \(\mathbf R^{(0)}\) and \(\mathbf R^{(1)}\) be solutions to the simplified
  problem in Eq.~\eqref{eq:simp-optim} with parameters
  \[
    \boldsymbol \tau^{(0)} = (\tau^{(0)}_1, \ldots, \tau^{(0)}_m), \qquad
    \boldsymbol \tau^{(1)} = (\tau^{(1)}_1, \ldots, \tau^{(1)}_m),
  \]
  respectively. Then the difference between the objective values of the two
  solutions is at most \(4 \cdot \| \boldsymbol \tau^{(0)} - \boldsymbol
  \tau^{(1)}\|_1\).
\end{lemma}

\begin{proof}
  The strategy is simple: we will transform \(\mathbf R^{(0)}\) into a feasible
  solution of the simplified problem with parameters \(\boldsymbol \tau^{(1)}\)
  and \emph{vice versa} for \(\mathbf R^{(1)}\) and \(\boldsymbol \tau^{(0)}\).
  These new solutions lower or upper bound---depending on whether we are
  considering minimization or maximization---the objective of the new problem;
  however, by Lemma~\ref{lem:sq-error}, we can also bound their difference in
  objective value from the original solutions, giving a bound on the difference
  in objective value of the original solutions.

  In particular, we see that the lemma immediately follows from
  Lemma~\ref{lem:sq-error} and if there exist \(\mathbf R^{(2)}\) and \(\mathbf
  R^{(3)}\) such that
  \[
    \| \mathbf R^{(0)} - \mathbf R^{(2)} \|_1, \| \mathbf R^{(1)} - \mathbf
    R^{(3)} \|_1 \leq 2 \cdot \| \boldsymbol \tau^{(0)} - \boldsymbol \tau^{(1)}
    \|_1
  \]
  and such that \(\mathbf R^{(2)}\) and \(\mathbf R^{(3)}\) are feasible for the
  simplified problem with parameters \(\boldsymbol \tau^{(1)}\) and
  \(\boldsymbol \tau^{(0)}\), respectively. For, in that case, we have that,
  assuming without loss of generality that we are solving the minimization
  problem,
  \[
    \sum_{i=1}^n (R^{(0)}_i)^2 \leq \sum_{i=1}^n (R^{(2)}_i)^2 \leq \sum_{i=1}^n
    (R^{(1)}_i)^2 + 4 \cdot \| \boldsymbol \tau^{(0)} - \boldsymbol \tau^{(1)}
    \|_1,
  \]
  and similarly
  \[
    \sum_{i=1}^n (R^{(1)}_i)^2 \leq \sum_{i=1}^n (R^{(3)}_i)^2 \leq \sum_{i=1}^n
    (R^{(0)}_i)^2 + 4 \cdot \| \boldsymbol \tau^{(0)} - \boldsymbol \tau^{(1)}
    \|_1.
  \]
  Therefore, it suffices to construct \(\mathbf R^{(2)}\) and \(\mathbf
  R^{(3)}\). The construction involves two steps:
  \begin{enumerate}
    \item Ensure that the sum of the risk vector is correct within each stratum;
    \item Ensure that the \(L_1\) budget constraint is satisfied.
  \end{enumerate}
  The \(L_1\) distance needed to achieve each of these steps is bounded by \(\|
  \boldsymbol \tau^{(0)} - \boldsymbol \tau^{(1)} \|_1\), and so \(\mathbf R^{(2)}\)
  and \(\mathbf R^{(3)}\) will have the required property.

  Assume again that we are solving the minimization problem. Consider the
  \(j\)-th unobserved stratum (i.e., \(i \in \mathcal S_{j,1}\)) with associated
  upper and lower thresholds \(t^{\mathrm{lwr}}\) and \(t^{\mathrm{upr}}\) and
  suppose without loss of generality that \(\tau_j^{(0)} \leq \tau_j^{(1)}\).
  Then, to achieve the first step, we simply raise \(t^{\mathrm{upr}}\) until
  the sum of the risk vector is \(\tau_j^{(1)}\) or \(t^{\mathrm{upr}} = 1\). If
  \(t^{\mathrm{upr}} = 1\), then we raise \(t^{\mathrm{lwr}}\) until the sum of
  the risk vector is \(\tau_j^{(1)}\).  We note that since our risk vector has
  strictly increased, the \(L_1\) distance between the original and new risk
  vectors is exactly the difference in their sums, i.e., \(\tau_j^{(1)} -
  \tau_j^{(0)}\). Call the risk vector that results from performing this
  operation across all strata \(\mathbf R'\). Then, it follows that
  \[
    \| \mathbf R' - \hat {\mathbf R} \|_1 \leq \| \mathbf R^{(0)} - \hat
    {\mathbf R} \|_1 + \tau_j^{(1)} - \tau_j^{(0)} \leq \epsilon + \tau_j^{(1)}
    - \tau_j^{(0)}.
  \]

  To achieve the second step, in strata \(\mathcal S_{j,a}\) where
  \(t^{\mathrm{upr}} < 1\) and \(t^{\mathrm{lwr}} > 0\), we simply raise
  \(t^{\mathrm{upr}}\) and lower \(t^{\mathrm{lwr}}\) so as to ensure that the
  sum of the risk vector does not change. Doing so reduces the \(L_1\) distance
  between the risk vector and \(\hat {\mathbf R}\), so we do so until the
  \(L_1\) budget constraint is satisfied. In particular, this process can be
  carried out until \(t^{\mathrm{upr}} = 1\) or \(t^{\mathrm{lwr}} = 0\) in
  every stratum. If the budget constraint is still not satisfied, then the
  simplified problem for \(\boldsymbol \tau^{(1)}\) is not feasible, contrary to
  our assumption that there exists a solution \(\mathbf R^{(1)}\). Therefore,
  the process halts, and, moreover, requires moving at most \(\| \boldsymbol
  \tau^{(0)} - \boldsymbol \tau^{(1)} \|_1\) in \(L_1\) distance. Call the
  resulting risk vector \(\mathbf R^{(2)}\). Then, it follows that
  \[
    \sum_{i \in \mathcal S_{j,1}} R^{(2)}_i = \tau_j^{(1)} \quad (j = 1, \ldots,
    m), \qquad \| \mathbf R^{(2)} - \hat {\mathbf R} \|_1 \leq \epsilon.
  \]
  The other constraints are satisfied trivially. We can construct \(\mathbf
  R^{(3)}\) similarly, and the proof is complete for minimization. For
  maximization, the argument is exactly similar, except that we increase and
  decrease \(R_{i_0}\) and \(R_{i_1}\) instead of \(t^{\mathrm{lwr}}\) and
  \(t^{\mathrm{upr}}\).
\end{proof}

We can extend the previous results to bound the error in the calculation of the
coefficients themselves.

\begin{lemma}%
\label{lem:all-error}
  Let \(V(\epsilon)\) denote the value of
  \begin{align}
  \label{eq:veps}
    \begin{split}
      \arraycolsep=1.4pt
      \begin{array}{rrclc}
          \mathop\mathrm{Minimize}\limits_{\mathbf R \in \mathbb R^n}
            & \multicolumn{3}{c}{\displaystyle \frac 1 n \displaystyle\sum_{j=1}^m
              n_j \cdot \operatorname{\emph{\textsc{Var}}} \left((R_i)_{i \in
              \mathcal G_j}\right)} \\
          \mathrm{s.t.} & \frac 1 n \sum_{i=1}^n |R_i - \hat R_i|
                              &\leq& \epsilon \\
                        & \sum_{i \in \mathcal S_{j,1}} R_i
                              &=&     \rho_j, & (j = 1, \ldots, m) \\[4pt]
                        & R_i &\leq&  u_i,    & (i = 1, \ldots, n) \\
                        & R_i &\geq&  \ell_i. & (i = 1, \ldots, n)
      \end{array}.
    \end{split}
  \end{align}
  To run Algorithm~\ref{algo:main} with parameter \(\epsilon\), let
  \[
    \boldsymbol \eta = (\eta_1, \ldots, \eta_m)
  \]
  denote the step size of our grid search in each of the \(m\) strata on the
  \(\boldsymbol \tau = (\tau_1, \ldots, \tau_m)\) scale. Let \(\delta^*\) denote
  the true optimum of the base problem, and let \(\delta^\dagger\) denote the
  value of the objective returned by applying Algorithm~\ref{algo:main}. Then
  \[
    |\delta^* - \delta^\dagger| \leq \frac {6 \|\boldsymbol \eta\|_1 + 2 m
    \gamma} {V(\epsilon)^2} + \frac {\frac 2 {n_1} \eta_1 + \frac 2 {n_j}
    \eta_j} {V(\epsilon)} + \frac {2 \|\boldsymbol \eta\|_1} {V(\epsilon)}.
  \]
\end{lemma}

\begin{proof}
  The proof is a straightforward exercise in bounding the various terms in
  Eq.~\eqref{eq:est} in Lemma~\ref{lem:est}. In particular, by
  Eq.~\eqref{eq:est}, \(|\delta^* - \delta^\dagger|\) is equivalent to
  \begin{equation}
  \label{eq:err-simp}
    \left| \frac {a_0 \cdot b_0} {c_0 + d_0} - \frac {a_1 \cdot b_1} {c_1 + d_1}
    \right|,
  \end{equation}
  where
  \begin{align*}
    a_0 &= \frac 1 n \sum_{j=1}^m \sigma_j \cdot \tau^*_j - (1 - \sigma_j) \cdot
        \rho_j,
        & a_1 &= \frac 1 n \sum_{j=1}^m \sigma_j \cdot \tau^\dagger_j - (1 -
        \sigma_j) \cdot \rho_j, \\
    b_0 &= \frac {\tau^*_j + \rho_j} {n_j} - \frac {\tau^*_1 - \rho_j} {n_j},
        & b_1 &= \frac {\tau^\dagger_j + \rho_j} {n_j} - \frac {\tau^\dagger_1 -
        \rho_1} {n_1}, \\
    c_0 &= \frac 1 n \sum_{i=1}^n (R_i^*)^2,
        & c_1 &= \frac 1 n \sum_{i=1}^n (R_i^\dagger)^2, \\
    d_0 &= \frac 1 n \sum_{j=1}^m \left( \frac {\rho_j + \tau^*_j} {n_j} \right)^2,
        & d_1 &= \frac 1 n \sum_{j=1}^m \left( \frac {\rho_j + \tau^\dagger_j}
        {n_j} \right)^2.
  \end{align*}
  Here, \(\mathbf R_i^*\) is the optimal risk vector---which, by
  Corollary~\ref{cor:simp}, must correspond to an optimum of the simplified
  problem---and \(\mathbf R_i^\dagger\) is the risk vector found by
  Algorithm~\ref{algo:main}.\footnote{%
    We note that Algorithms~\ref{algo:min-all} and~\ref{algo:max-all}, for
    clarity, do not actually return the risk vector \(\mathbf R\) itself, but
    could easily be modified to do so. This modification is easily carried
    through to Algorithm~\ref{algo:main}.
  }
  Since the gap between the optimal and approximate solutions can only increase,
  we may assume without loss of generality that \(\mathbf R^\dagger\) is the
  risk vector found by Algorithm~\ref{algo:min-all} or
  Algorithm~\ref{algo:max-all} at the gridpoint nearest \((\tau^*_1, \ldots,
  \tau^*_m)\). Finally, since only maximizing the objective involves
  approximation error, we may assume without loss of generality that
  \(\mathbf R^\dagger\) is the risk vector found by
  Algorithm~\ref{algo:max-all}.

  Now, it follows by a straightforward algebraic manipulation that
  \begin{align*}
    \delta^* - \delta^\dagger
      &= a_0 \cdot b_0 \cdot \left( \frac 1 {c_0 + d_0} - \frac 1 {c_1 + d_1}
      \right) + a_0 \cdot (b_0 - b_1) \cdot \frac 1 {c_1 + d_1} +
      (a_0 - a_1) \cdot b_1 \cdot \frac 1 {c_1 + d_1}.
  \end{align*}
  Therefore, we seek bounds \(A\), \(B\), and \(V\) such that
  \[
    |a_0|, |a_1| \leq A, \qquad |b_0|, |b_1| \leq B, \qquad |c_0 + d_0|, |c_1 +
    d_1| \geq V
  \]
  and \(\Delta_a\), \(\Delta_b\), and \(\Delta_v\) such that
  \[
    |a_0 - a_1| \leq \Delta_a, \qquad |b_0 - b_1| \leq \Delta_b, \qquad \left|
    \frac 1 {c_0 + d_0} - \frac 1 {c_1 + d_1} \right| \leq \Delta_v.
  \]
  Then, it will follow that
  \[
    |\delta^* - \delta^\dagger| \leq AB \Delta_v + \frac A V \Delta_b + \frac B
    V \Delta_a.
  \]

  To find \(A\), \(B\), and \(V\), we note that for any possible \(t_j\)
  and \(r_j\), we have that
  \[
    \sum_{j=1}^m \left| \sigma_j \cdot t_j \right| \leq n, \qquad \sum_{j=1}^m
    \left| (1 - \sigma_j) \cdot r_j \right| \leq n,
  \]
  whence \(|a_i| \leq 2\) for \(i = 0, 1\). Therefore we can take \(|A| = 2\).
  In the same way, we have the bound
  \[
    \left| \frac {\tau_j + \rho_j} {n_j} - \frac {\tau_1 - \rho_j} {n_1} \right|
    \leq 2,
  \]
  so we take \(B = 2\) as well. Finally, we have that \(|c_0 + d_0| \geq
  V(\epsilon)\) and \(|c_1 + d_1| \geq V(\epsilon)\) by definition, so we take
  \(V = V(\epsilon)\).

  Now, note that
  \[
    |a_0 - a_1| = \left| \frac 1 n \sum_{j=1}^m \sigma_j \cdot (\tau^*_j -
    \tau^\dagger_j) \right| \leq \frac 1 n \sum_{j=1}^m \left| \tau^*_j -
    \tau^\dagger_j \right| \leq \frac 1 n \sum_{j=1}^m n_j \eta_j = \|
    \boldsymbol \eta \|_1,
  \]
  so we take \(\Delta_a = \|\boldsymbol \eta \|_1\). Next, since \(|\tau_j^* -
  \tau_j^\dagger| \leq \eta_j\) and similarly for \(\tau^*_1\) and
  \(\tau^\dagger_1\), we have that
  \[
    |b_0 - b_1| \leq \frac 1 n \sum_{j=1}^m \left| \frac {\tau^*_j - \tau^\dagger_j}
    {n_j} \right| + \left| \frac {\tau^*_1 - \tau^\dagger_1} {n_j} \right| \leq
    \frac {2 \eta_1} {n_1} + \frac {2 \eta_j} {n_j},
  \]
  so we take \(\Delta_b = \tfrac {2 \eta_1} {n_1} + \tfrac {2 \eta_j} {n_j}\).
  Finally, we have that
  \[
    \left| \frac 1 {c_0 + d_0} - \frac 1 {c_1 + d_1} \right| = \left| \frac {c_1
    - c_0 + d_1 - d_0} {(c_0 + d_0) (c_1 + d_1)} \right| \leq \frac {|c_0 - c_1|
    + |d_0 - d_1|} {V^2}.
  \]
  Therefore, it only remains to bound \(|c_0 - c_1|\) and \(|d_0 - d_1|\). By
  Lemma~\ref{lem:simp-error}, if \(\mathbf R^\dagger\) were the \emph{true}
  solution to the simplified problem, then \(|c_0 - c_1|\) would be bounded by
  \(4 \|\boldsymbol \eta\|_1\), since the \(L_1\) distance to the nearest
  gridpoint is at most \(\|\boldsymbol \eta\|_1\). However, since \(\mathbf
  R^\dagger\) is only an approximate solution, applying
  Lemma~\ref{lem:max-error} gives that \(|c_0 - c_1|\) is bounded by \(2 m
  \gamma + 4 \|\boldsymbol \eta\|_1\). Similarly, using the fact that
  \[
    \left| \frac {\tau^*_j + \rho_j} {n_j} - \frac {\tau^\dagger_j + \rho_j}
    {n_j} \right| \leq \eta_j,
  \]
  we can apply Lemma~\ref{lem:sq-error} to obtain that \(|d_0 - d_1|\) is less
  than or equal to \(2 \|\boldsymbol \eta\|_1\). Combining these bounds gives
  that
  \[
    |\delta^* - \delta^\dagger| \leq \frac {6 \|\boldsymbol \eta\|_1 + 2 m
    \gamma} {V(\epsilon)^2} + \frac {\frac 2 {n_1} \eta_1 + \frac 2 {n_j}
    \eta_j} {V(\epsilon)} + \frac {2 \|\boldsymbol \eta\|_1} {V(\epsilon)},
  \]
  as desired.
\end{proof}

Lemma~\ref{lem:all-error} gives intuition for how to choose the step sizes
appropriately so as to minimize the error in the coefficients for a given amount
of computation. In particular, since solving the parameterized problem requires
roughly the same number of steps at each grid point, the computation scales like
the number of gridpoints, i.e., like
\[
  \frac 1 {\prod_{j=1}^m \eta_j}.
\]
The problem of choosing step sizes \(\boldsymbol \eta\) so as to maximize
accuracy for a given amount of computation is therefore essentially equivalent
to the following optimization problem:
\[
  \arraycolsep=1.4pt
  \begin{array}{rrclc}
    \mathop\mathrm{Minimize}\limits_{\boldsymbol\eta \in \mathbb R^m} &
      \multicolumn{3}{c}{\displaystyle\sum_{j=1} \lambda_j \eta_j} \\
    \text{s.t.} & \eta_j                            &>& 0, & (j = 1, \ldots, m) \\
                & \displaystyle\prod_{j=1}^m \eta_j &=& M.
    \end{array}
\]
Applying the first-order necessary KKT conditions yields that
\[
  \lambda_j = \nu \prod_{k \neq j} \eta_k, \qquad \text{i.e.,} \quad \lambda_j
  \eta_j = \nu M,
\]
whence we have that
\[
  \eta_j = \frac {\sqrt[m]{M \prod_{j=1}^m \lambda_j}} {\lambda_j}.
\]
In particular, the step sizes should be chosen so that they are inversely
proportional to their weights in the error bound in Lemma~\ref{lem:all-error}.

Unfortunately, these weights vary with \(\epsilon\). For large \(\epsilon\),
when \(V(\epsilon)\) is close to zero, the weights are dominated by the
first term, which is optimized when \(\eta_1 = \cdots = \eta_m\). For small
\(\epsilon\), when \(V(\epsilon)\) is large, the weights are mixture of
(somewhat larger) terms that would be optimized by \(\eta_1 = \cdots = \eta_m\)
and (somewhat smaller) terms that would be optimized by \(\tfrac {\eta_1} {n_1}
= \cdots = \tfrac {\eta_m} {n_m}\).

Thus, one reasonable heuristic that is likely to perform well across a range of
\(\epsilon\) is to choose \(\eta_j\) to be equal to some fixed \(\eta\) for all
\(j\). We use this heuristic in our experiments and software implementation.

Alternatively, to simplify the error bounds and eliminate the dependence on the
data through the \(n_j\), one could choose \(\eta_j\) to be proportional to
\(n_j\) for all \(j\). Making this choice gives the error bounds in
Theorem~\ref{thm:main}, while the runtime bounds are given by
Lemmata~\ref{lem:min-runtime} and~\ref{lem:max-runtime}.

Lastly, since it is a convex problem, it is generally practical to compute
\(V(\epsilon)\). However, to obtain a bound on the error entirely in terms of
the input parameters, we can use the following lemma.

\begin{lemma}%
\label{lem:var-bound}
  Let \(V(\epsilon)\) be defined as in Lemma~\ref{lem:all-error}. Then
  \begin{equation}
  \label{eq:var-bound}
    V(\epsilon) \geq \left[ \frac 1 n \sum_{j=1}^m
    {\operatorname{\emph{\textsc{Var}}}} \left( (\hat R_i)_{i \in \mathcal
    S_{j,1}} \right) \right] - 4 \epsilon.
  \end{equation}
\end{lemma}

\begin{proof}
  Let \(\mathbf R^*\) be the solution to the optimization problem defining \(V(\epsilon)\).
  Then, we note that \(\|\mathbf R^* - \hat {\mathbf R}\|_1 \leq \epsilon\),
  and, moreover, that
  \[
    V(\epsilon) - \frac 1 n \sum_{j=1}^m {\operatorname{\textsc{Var}}} \left(
    (\hat R_i)_{i \in \mathcal S_{j,1}} \right) = \frac 1 n \left[ \left(
    \sum_{i=1}^n (R_i^*)^2 - \hat R_i^2 \right) + \left( \sum_{j=1}^m \left[
    \frac {\sum_{i \in \mathcal G_j} R_i^*} {n_j} \right]^2 - \left[ \frac
    {\sum_{i \in \mathcal G_j} \hat R_i} {n_j} \right]^2 \right) \right].
  \]
  Applying Lemma~\ref{lem:sq-error} to the first term yields a bound of \(2
  \epsilon\). For the second term, we note that
  \[
    \sum_{j=1}^m \left| \frac {\sum_{i \in \mathcal G_j} R_i^*} {n_j} - \frac
    {\sum_{i \in \mathcal G_j} \hat R_i} {n_j} \right| \leq \sum_{i=1}^n |R_i^*
    - \hat R_i| \leq \epsilon,
  \]
  and so we can apply Lemma~\ref{lem:sq-error} to obtain a bound of \(2
  \epsilon\) on the second term as well. Combining these bounds gives the
  desired result.
\end{proof}

\begin{algorithm}[p]
\raggedright
  \textbf{Input:} The collections \(\boldsymbol \Delta\) and \(\mathbf K\), as
    well as the \(L_1\) ``budget'' \(\epsilon\). \\
  \textbf{Output:} The value of \(\Sigma(\Delta^*)\), where \(\epsilon(\Delta^*)
    = \epsilon\).
  \begin{algorithmic}[1]
    \STATE Set \(N \gets \texttt{length}(\boldsymbol \delta)\)
    \STATE Initialize \(i \gets 1\) \hfill \COMMENT{\textsl{Pointer to current
      position in \(\boldsymbol \delta\)}}
    \STATE Initialize \(\varepsilon \gets 0\) 
    \hfill \COMMENT{\textsl{Budget used so far}}
    \STATE Initialize \(t \gets 0\) \hfill \COMMENT{\textsl{Gap between current
      and next value of \(\boldsymbol \Delta\)}}
    \STATE Initialize \(\Sigma \gets 0\) \hfill \COMMENT{\textsl{Change in sum
      of squares}}
    \WHILE {\(\varepsilon + K_i (\Delta_{i+1} - \Delta_i) < \epsilon\) \AND \(i
    < N\)}
      \STATE Set \(\varepsilon \gets \varepsilon + K_i (\Delta_{i+1} -
        \Delta_i)\)
      \STATE Set \(\Sigma \gets \Sigma - K_i \cdot \Delta_i \cdot (\Delta_{i+1}
        - \Delta_i) + \tfrac K 2 \cdot (\Delta_{i+1} - \Delta_i)^2\)
      \STATE Set \(i \gets i + 1\)
    \ENDWHILE
    \IF {\(K_i = 0\)}
      \RETURN \(\Sigma\) \hfill \COMMENT{\textsl{\(K_i = 0\) if and only if the
        budget was exactly exhausted on the last iteration of the loop}}
    \ELSIF {\(i = N\)}
      \RETURN \(\Sigma\) \hfill \COMMENT{\textsl{\(i = N\) if and only if we
        have made it to the end of \(\boldsymbol\delta\), i.e., to \(\delta =
        0\)}}
    \ELSE
      \STATE Set \(t \gets (\epsilon - \varepsilon) / K_i\)
      \STATE Set \STATE Set \(\Sigma \gets \Sigma - K_i \cdot \Delta_i \cdot t +
        \tfrac K 2 \cdot t^2\)
      \RETURN \(\Sigma\)
    \ENDIF
  \end{algorithmic}
  \caption{Piecewise quadratic function evaluation}%
\label{algo:K-delta-solve}
\end{algorithm}

\begin{algorithm}[p]
\raggedright
  \textbf{Input:} The bounds \(\boldsymbol \ell\) and \(\mathbf u\), the
    estimates \(\hat {\mathbf R}\), and the sum \(\mu\). \\
  \indent\textbf{Output:} A risk vector \(\mathbf R\) of the form in
    Lemma~\ref{lem:minimize} minimizing \(\|\mathbf R - \hat {\mathbf R}\|_1\) and
    satisfying \(\sum_{i=1}^n R_i = \mu\), along with the ending
    \(k^{\mathrm{lwr}}\), \(k^{\mathrm{upr}}\), \(i^{\mathrm{lwr}}\),
    \(i^{\mathrm{upr}}\), \(t^{\mathrm{upr}}\), and \(t^{\mathrm{lwr}}\).
  \begin{algorithmic}[1]
    \STATE Set \(\texttt{pts}\) to be the concatenation of \(\boldsymbol \ell\),
      \(\hat {\mathbf R}\), and \(\mathbf u\) in ascending order \hfill
      \COMMENT{\textsl{Points at which rates can change}}
    \STATE Set \(\mathbf R \gets \hat {\mathbf R}\) \hfill \COMMENT{\textsl{Risk
      vector}}
    \STATE Set \(n \gets \texttt{length}(\texttt{pts})\)
    \STATE Set \(i^{\mathrm{lwr}} = 1\), \(i^{\mathrm{upr}} = n\) \hfill
      \COMMENT{\textsl{Indices of next values at which rates might change}}
    \STATE Set \(t^{\mathrm{lwr}} \gets \texttt{pts}[i^{\mathrm{lwr}}]\),
      \(t^{\mathrm{upr}} \gets \texttt{pts}[i^{\mathrm{upr}}]\) \hfill
      \COMMENT{\textsl{Thresholds}}
    \STATE Set \(k^{\mathrm{lwr}} \gets 0\), \(k^{\mathrm{upr}} \gets 0\) \hfill
      \COMMENT{\textsl{Number of active indices}}
    \STATE \(D \gets \mu - \sum_{i=1}^n R_i\) \hfill \COMMENT{\textsl{Difference
      between required sum and actual sum}}
    \IF {\(D > 0\)}
      \STATE \(t^{\mathrm{nxt}} \gets \texttt{pts}[i^{\mathrm{lwr}} + 1]\)
      \STATE \(d \gets 0\) \hfill \COMMENT{\textsl{Change in sum from moving
        from \(t^{\mathrm{lwr}}\) to \(t^{\mathrm{nxt}}\)}}
      \WHILE {\(D > d\) \AND \(i^{\mathrm{lwr}} < n\)}
        \STATE \(D \gets D - d\)
        \STATE \(i^{\mathrm{lwr}} \gets i^{\mathrm{lwr}} + 1\)
        \STATE \(t^{\mathrm{lwr}} \gets t^{\mathrm{nxt}}\)
        \STATE \(t^{\mathrm{nxt}} \gets \texttt{pts}[i^{\mathrm{lwr}} + 1]\)
        \IF {\(t^{\mathrm{nxt}}\) corresponds to an element of \(\hat {\mathbf
        R}\)}
          \STATE \(k^{\mathrm{lwr}} \gets k^{\mathrm{lwr}} + 1\)
        \ELSIF {\(t^{\mathrm{nxt}}\) corresponds to an element of \(\mathbf u\)}
          \STATE \(k^{\mathrm{lwr}} \gets k^{\mathrm{lwr}} - 1\)
        \ENDIF
        \STATE \(d \gets k^{\mathrm{lwr}} (t^{\mathrm{nxt}} -
          t^{\mathrm{lwr}})\)
      \ENDWHILE
      \IF {\(i^{\mathrm{lwr}} < n\)}
        \STATE \(t^{\mathrm{lwr}} \gets t^{\mathrm{lwr}} +
          \frac{D}{k^{\mathrm{lwr}}}\)
      \ENDIF
    \ELSIF {\(D < 0\)}
      \STATE \COMMENT{\textsl{Similar steps to the \(D > 0\) case but adapted
        for the upper threshold}}
    \ENDIF
    \STATE \RETURN \(\mathbf R\), \(k^{\mathrm{lwr}}\), \(k^{\mathrm{upr}}\),
      \(i^{\mathrm{lwr}}\), \(i^{\mathrm{upr}}\), \(t^{\mathrm{upr}}\), and
      \(t^{\mathrm{lwr}}\)
  \end{algorithmic}
  \caption{Minimization algorithm (sum adjustment)}%
\label{algo:min-adjust}
\end{algorithm}

\begin{algorithm}[p]
\raggedright
  \textbf{Input:} The bounds \(\boldsymbol \ell\) and \(\mathbf u\), the
    estimates \(\hat {\mathbf R}\), and the sum \(\mu\). \\
  \indent\textbf{Output:} The collections \(\boldsymbol\Delta\) and \(\mathbf
    K\).
  \begin{algorithmic}[1]
    \STATE Initialize \(\texttt{pts}\), \(\mathbf R\), \(t^{\mathrm{lwr}}\),
      \(t^{\mathrm{upr}}\), \(k^{\mathrm{lwr}}\), \(k^{\mathrm{upr}}\),
      \(i^{\mathrm{lwr}}\), and \(i^{\mathrm{upr}}\) as in
      Algorithm~\ref{algo:min-adjust}
    \STATE Set \(n \gets \texttt{length}(\texttt{pts})\)
    \STATE Set \(\varepsilon \gets 0\) \hfill \COMMENT{\textsl{Amount of budget
      expended so far}}
    \STATE Set \(\Delta \gets t^{\mathrm{upr}} - t^{\mathrm{lwr}}\) \hfill
      \COMMENT{\textsl{Gap between thresholds}}
    \STATE Using Algorithm~\ref{algo:min-adjust}, update \(\mathbf R\),
      \(k^{\mathrm{lwr}}\), \(k^{\mathrm{upr}}\), \(i^{\mathrm{lwr}}\),
      \(i^{\mathrm{upr}}\), \(t^{\mathrm{upr}}\), and \(t^{\mathrm{lwr}}\) \hfill
      \COMMENT{\textsl{Ensure that sum is correct}}
    \STATE Set \(\epsilon \gets \epsilon - \|\mathbf R - \hat {\mathbf R}\|_1\),
      returning that the problem is infeasible if the result is negative
    \IF {\(n = 1\)}
      \RETURN \(\boldsymbol\Delta = (\Delta_0)\) and \(\mathbf K = 0\)
    \ENDIF
    \STATE \(D^{\mathrm{lwr}} \gets k^{\mathrm{lwr}}
      (\texttt{pts}[i^{\mathrm{lwr}}+1] - t^{\mathrm{lwr}})\) \hfill
      \COMMENT{\textsl{Cost of moving from lower threshold to next change point}}
    \STATE \(D^{\mathrm{upr}} \gets k^{\mathrm{upr}}(t^{\mathrm{upr}} -
      \texttt{pts}[i^{\mathrm{upr}}-1])\) \hfill \COMMENT{\textsl{Cost of moving
      from upper threshold and next change point}}
    \STATE \(D \gets \min(D^{\mathrm{lwr}}, D^{\mathrm{upr}})\) \hfill
      \COMMENT{\textsl{Smaller of the costs}}
    \STATE \(K \gets 0\)
    \WHILE {\(i^{\mathrm{upr}} - i^{\mathrm{lwr}} > 1\)}
      \STATE Append \(\Delta\) to \(\boldsymbol\Delta\) and \(K\) to \(\mathbf
        K\)
      \IF {\(D^{\mathrm{lwr}} = D\)}
        \STATE \(i^{\mathrm{lwr}} \gets i^{\mathrm{lwr}} + 1\) \hfill
          \COMMENT{\textsl{Increment lower active index}}
        \STATE \(t^{\mathrm{lwr}} \gets \texttt{dps}[i^{\mathrm{lwr}}]\) \hfill
          \COMMENT{\textsl{Update the lower threshold}}
        \IF {\(t^{\mathrm{lwr}}\) came from \(\hat {\mathbf R}\)}
          \STATE \(k^{\mathrm{lwr}} \gets k^{\mathrm{lwr}} + 1\) \hfill
            \COMMENT{\textsl{Increment \(k^{\mathrm{lwr}}\) if an index is
            activated}}
        \ELSIF {\(t^{\mathrm{lwr}}\) came from \(\mathbf u\)}
          \STATE \(k^{\mathrm{lwr}} \gets k^{\mathrm{lwr}} - 1\) \hfill
            \COMMENT{\textsl{Decrement \(k^{\mathrm{lwr}}\) if an index is
            deactivated}}
        \ENDIF
        \STATE \(D^{\mathrm{upr}} \gets D^{\mathrm{upr}} - D^{\mathrm{lwr}}\)
          \hfill \COMMENT{\textsl{Calculate new upper active gap}}
        \STATE \(t^{\mathrm{upr}} \gets t^{\mathrm{upr}} -
          \frac{D^{\mathrm{lwr}}}{k^{\mathrm{upr}}}\) \hfill
          \COMMENT{\textsl{Calculate new upper threshold}}
        \STATE \(D^{\mathrm{lwr}} \gets k^{\mathrm{lwr}}
          (\texttt{dps}[i^{\mathrm{lwr}}+1] - t^{\mathrm{lwr}})\) \hfill
          \COMMENT{\textsl{Calculate new lower active gap}}
      \ELSE
        \STATE \COMMENT{\textsl{Similar steps as in previous branch, adapted to
          the upper threshold}}
      \ENDIF
      \STATE \(D \gets \min(D^{\mathrm{lwr}}, D^{\mathrm{upr}})\) 
      \STATE \(\Delta \gets t^{\mathrm{upr}} - t^{\mathrm{lwr}}\) 
      \STATE \(K \gets \tfrac {k^{\mathrm{lwr}} \cdot k^{\mathrm{upr}}}
        {k^{\mathrm{lwr}} + k^{\mathrm{upr}}}\)
    \ENDWHILE
    \STATE Append \(0\) to \(\boldsymbol\Delta\) and \(\mathbf K\)
    \RETURN \(\boldsymbol \Delta, \mathbf K\)
  \end{algorithmic}
  \caption{Minimization algorithm (single stratum)}%
\label{algo:min-one}
\end{algorithm}

\begin{algorithm}[p]
\raggedright
  \textbf{Input:} The collections \(\boldsymbol \Delta^{(0)}\), \(\mathbf
    K^{(0)}\), \(\boldsymbol \Delta^{(1)}\), and \(\mathbf K^{(1)}\).
  \textbf{Output:} A pair \((\boldsymbol\Delta^{(2)}, \mathbf K^{(2)})\)
    representing the sum of functions \(\epsilon^{(0)}(\Delta) +
    \epsilon^{(1)}(\Delta)\) and \(\Sigma^{(0)}(\Delta) + \Sigma^{(1)}(\Delta)\).
  \begin{algorithmic}[1]
    \STATE Initialize \(i_0 \gets 0\), \(i_1 \gets 1\) \hfill
      \COMMENT{\textsl{Pointers to smallest indices not yet combined}}
    \STATE Initialize \(K_0 \gets 0\), \(K_1 \gets 0\), \(\Delta_0 \gets 1\),
      \(\Delta_1 \gets 1\) \hfill \COMMENT{\textsl{Current values of the
      parameters}}
    \WHILE {\(i_0 \leq \texttt{length}(\boldsymbol\Delta_0)\) \AND \(i_1 \leq
    \texttt{length}(\boldsymbol\Delta_1)\)}
      \IF {\(\Delta^{(0)}_{i_0} > \Delta^{(1)}_{i_1}\)}
        \STATE Set \(i_0 \gets i_0 + 1\), \(K_0 \gets K^{(0)}_{i_0}\), and
          \(\Delta_0 \gets \Delta^{(0)}_{i_0}\)
        \STATE Append \(\Delta_0\) to \(\boldsymbol \Delta^{(2)}\) and \(K_0 +
          K_1\) to \(\mathbf K^{(2)}\)
      \ELSE
        \STATE Set \(i_1 \gets i_1 + 1\), \(K_1 \gets K^{(1)}_{i_1}\), and
          \(\Delta_1 \gets \Delta^{(1)}_{i_1}\)
        \STATE Append \(\Delta_1\) to \(\boldsymbol \Delta^{(2)}\) and \(K_0 +
          K_1\) to \(\mathbf K^{(2)}\)
      \ENDIF
    \ENDWHILE
    \RETURN \(\boldsymbol \Delta^{(2)}\) and \(\mathbf K^{(2)}\)
  \end{algorithmic}
  \caption{Minimization algorithm (combining strata)}%
\label{algo:min-all}
\end{algorithm}

\begin{algorithm}[p]
\raggedright
  \textbf{Input:} The bounds \(\boldsymbol \ell\) and \(\mathbf u\), the
    estimates \(\hat {\mathbf R}\), and the sum \(\mu\). \\
  \indent\textbf{Output:} A risk vector \(\mathbf R\) of the form in
    Lemma~\ref{lem:maximize} minimizing \(\|\mathbf R - \hat {\mathbf R}\|_1\)
    and satisfying \(\sum_{i=1}^n R_i = \mu\), along with \(i^{\mathrm{lwr}}\),
    \(i^{\mathrm{upr}}\).
  \begin{algorithmic}[1]
    \STATE Set \(\mathbf R \gets \hat {\mathbf R}\) \hfill \COMMENT{\textsl{Risk
      vector}}
    \STATE Set \(n \gets \texttt{length}(\mathbf R)\)
    \STATE Set \(i^{\mathrm{lwr}} = 1\), \(i^{\mathrm{upr}} = n\) \hfill
      \COMMENT{\textsl{Pivots}}
    \STATE \(D \gets \mu - \sum_{i=1}^n R_i\) \hfill \COMMENT{\textsl{Difference
      between required sum and actual sum}}
    \IF {\(D > 0\)}
      \STATE \(d \gets u_{i^{\mathrm{upr}}} - R_{i^{\mathrm{upr}}}\)
      \WHILE {\(D > d\) \AND \(i^{\mathrm{upr}} > 0\)}
        \STATE \(R_{i^{\mathrm{upr}}} \gets u_i\)
        \STATE \(D \gets  D - d\)
        \STATE \(i^{\mathrm{upr}} \gets i^{\mathrm{upr}} - 1\)
        \STATE \(d \gets u_{i^{\mathrm{upr}}} - R_{i^{\mathrm{upr}}}\)
      \ENDWHILE
      \STATE \(R_{i^{\mathrm{upr}}} \gets R_{i^{\mathrm{upr}}} + D\)
    \ELSIF {\(D < 0\)}
      \STATE \COMMENT{\textsl{Similar steps to the \(D > 0\) case but adapted
        for the lower pivot}}
    \ENDIF
    \STATE \RETURN \(\mathbf R\), \(i^{\mathrm{lwr}}\), and
      \(i^{\mathrm{upr}}\).
  \end{algorithmic}
  \caption{Maximization algorithm (sum adjustment)}%
\label{algo:max-adjust}
\end{algorithm}

\begin{algorithm}[p]
\raggedright
  \textbf{Input:} The bounds \(\boldsymbol \ell\) and \(\mathbf u\), the
    estimates \(\hat {\mathbf R}\), the sum \(\mu\), and the step size
    \(\gamma\). \\
  \indent\textbf{Output:} The collection \(\boldsymbol \Sigma\).
  \begin{algorithmic}[1]
    \STATE Initialize \(\mathbf R\), \(i^{\mathrm{lwr}}\), and
      \(i^{\mathrm{upr}}\) as in Algorithm~\ref{algo:max-adjust}
    \STATE Set \(n \gets \texttt{length}(R)\)
    \STATE Set \(\varepsilon \gets 0\) \hfill \COMMENT{\textsl{Amount of budget
      expended so far}}
    \STATE Using Algorithm~\ref{algo:max-adjust}, update \(\mathbf R\),
      \(i^{\mathrm{lwr}}\), and \(i^{\mathrm{upr}}\), \(t^{\mathrm{upr}}\)
      \hfill \COMMENT{\textsl{Ensure that sum is correct}}
    \STATE Set \(\varepsilon \gets \|\mathbf R - \hat {\mathbf R}\|_1\) and
      \(\epsilon \gets \epsilon - \varepsilon\), returning that the problem is
      infeasible if \(\varepsilon > \epsilon\)
    \WHILE {\(\varepsilon \geq \gamma\)}
      \STATE Append \(\infty\) to \(\boldsymbol \Sigma\) \hfill
        \COMMENT{\textsl{Sentinel value to indicate infeasibility}}
      \STATE \(\varepsilon \gets \varepsilon - \gamma\)
    \ENDWHILE
    \STATE \(\Delta^{\mathrm{lwr}} \gets R_{i^{\mathrm{lwr}}} -
      \ell_{i^{\mathrm{lwr}}}\)
    \STATE \(\Delta^{\mathrm{upr}} \gets u_{i^{\mathrm{upr}}} -
      R_{i^{\mathrm{upr}}}\)
    \STATE \(\Delta \gets \min(\Delta^{\mathrm{lwr}}, \Delta^{\mathrm{upr}})\)
    \STATE \(\Sigma \gets \sum_{i=1}^n R_i^2\) \hfill \COMMENT{\textsl{Sum of
      squares}}
    \WHILE {\(i^{\mathrm{upr}} < i^{\mathrm{lwr}}\)}
      \IF {\(\Delta \geq \gamma - \varepsilon\)}
        \STATE \(\Sigma \gets \Sigma + 2 \gamma \cdot (\gamma +
          [R_{i^{\mathrm{upr}}} - R_{i^{\mathrm{lwr}}}])\) \hfill
          \COMMENT{\textsl{If next step would exceed step size, record change}}
        \STATE Append \(\Sigma\) to \(\boldsymbol \Sigma\)
        \STATE \(R_{i^{\mathrm{lwr}}} \gets R_{i^{\mathrm{lwr}}} + \gamma\),
          \(R_{i^{\mathrm{upr}}} \gets R_{i^{\mathrm{upr}}} + \gamma\) \hfill
          \COMMENT{\textsl{Update risk vector at pivots}}
        \STATE \(\Delta \gets \Delta - \gamma\), \(\Delta^{\textrm{min}} \gets
          \Delta^{\textrm{min}} - \gamma\), \(\Delta^{\textrm{max}} \gets
          \Delta^{\textrm{max}} - \gamma\) \hfill \COMMENT{\textsl{Update gaps}}
        \STATE \(\varepsilon \gets 0\) \hfill \COMMENT{\textsl{Only adjust step
          size on first iteration}}
      \ELSE
        \IF {\(\Delta^{\mathrm{lwr}} = \Delta\)}
          \STATE \(\Sigma \gets \Sigma + 2 \Delta \cdot (\Delta +
            [R^{i^{\mathrm{upr}}} - R^{i^{\mathrm{lwr}}}])\)
          \STATE \(R_{i^{\mathrm{lwr}}} \gets \ell_{i^{\mathrm{lwr}}}\),
            \(R_{i^{\mathrm{upr}}} \gets R_{i^{\mathrm{upr}}} + \Delta\)
          \STATE \(i^{\mathrm{lwr}} \gets i^{\mathrm{lwr}} + 1\) \hfill
            \COMMENT{\textsl{Increment lower active index}}
        \ELSE
          \STATE \COMMENT{\textsl{Similar steps as in previous branch, adapted
            to the upper pivot}}
        \ENDIF
      \ENDIF
    \ENDWHILE
    \STATE Append \(\Sigma\) to \(\boldsymbol \Sigma\) until it has the
      appropriate length
    \RETURN \(\boldsymbol \Sigma\)
  \end{algorithmic}
  \caption{Maximization algorithm (single stratum)}
\label{algo:max-one}
\end{algorithm}

\begin{algorithm}[p]
\raggedright
  \textbf{Input:} The collections \(\boldsymbol \Sigma^{(0)}\) and \(\boldsymbol
    \Sigma^{(1)}\).
  \textbf{Output:} A collection \(\boldsymbol \Sigma^{(2)}\) representing the
    (approximate) maximum objective across both collections.
  \begin{algorithmic}[1]
    \STATE Initialize \(i_0 \gets 0\), \(i_1 \gets 1\) \hfill
      \COMMENT{\textsl{Pointers to indices currently being combined}}
    \STATE Initialize \(M \gets -\infty\) \hfill \COMMENT{\textsl{Current
      maximum}}
    \FOR {\(i = 0\), \ldots, \(\epsilon / \gamma\)}
      \STATE \(i_0 \gets 0\), \(i_1 \gets i\)
      \WHILE {\(i_0 < i\)}
        \STATE \(M \gets \Sigma^{(0)}_{i_0} + \Sigma^{(1)}_{i_1}\) \hfill
          \COMMENT{\textsl{Find the maximum across gridpoints whose total budget
          is \(i \gamma\)}}
        \STATE \(i_0 \gets i_0 + 1\), \(i_1 \gets i_1 - 1\)
      \ENDWHILE
      \STATE Append \(M\) to \(\boldsymbol\Sigma^{(2)}\)
      \STATE \(M \gets -\infty\)
    \ENDFOR
    \RETURN \(\boldsymbol \Sigma^{(2)}\)
  \end{algorithmic}
  \caption{Maximization algorithm (combining strata)}%
\label{algo:max-all}
\end{algorithm}

\clearpage

\newpage

\section{Estimation of the Non-Parametric Estimand by Linear Regression}%
\label{app:sim}

To assess the quality of risk-adjusted regression as an estimator of the
non-parametric estimand in \eqref{eq:nonpar}, we construct synthetic datasets
based on the NYPD data. On this synthetic dataset, we compare the true value of
the risk adjusted disparities---as defined in \eqref{eq:nonpar}---to estimates
from a risk-adjusted regression.

To construct the synthetic datasets, we first use the real NYPD data to estimate
the probability of frisk, \(\Pr(A = 1 \mid \tilde X)\), conditional on the same
observed pre-frisk covariates \(\tilde X\) used to estimate risk in the main
analysis, excluding race, suspected crime, and precinct.\footnote{%
    We exclude race to demonstrate that disparate impact can occur in the
    absence of disparate treatment. We exclude suspected crime and precinct
    because the large number of strata for these covariates leads to sampling
    difficulties.
}
We similarly estimate the probability a weapon is recovered on a frisked
individual, \(\Pr(W = 1 \mid \tilde X, A = 1)\), again conditional on the same
pre-frisk covariates. In both cases, we use logistic regression to estimate the
probabilities. The frisk model is trained on all data from 2008 and 2009, and
the weapon-recovery model is trained on the subset of stops of the same in which
an individual was frisked.

With the resulting covariate estimates \(\hat {\boldsymbol
\beta}_{\text{weapon}}\) and \(\hat {\boldsymbol \beta}_{\text{frisk}}\), for
the \(k\)-th iteration of the \(m = 100\) simulation instances, we generate a
population of \(n = 1{,}000{,}000\) observations as follows:
\begin{enumerate}
  \item Sample noise terms \(\boldsymbol \epsilon_{\text{weapon}} \sim \mathcal
    N(0, \tfrac 1 {50} \cdot I)\) and \(\boldsymbol \epsilon_{\text{frisk}} \sim
    \mathcal N(0, \tfrac 1 {50} \cdot I)\), resulting in new weapon possession
    and frisk model coefficients \(\boldsymbol \beta_{k,\text{weapon}} = \hat
    {\boldsymbol \beta}_{\text{weapon}} + \boldsymbol \epsilon_{\text{weapon}}\)
    and \(\boldsymbol \beta_{k,\text{frisk}} = \hat {\boldsymbol
    \beta}_{\text{frisk}} + \boldsymbol \epsilon_{\text{frisk}}\).
  \item Sample, with replacement, \(n\) covariate vector and race pairs
    \((\tilde X_i, C_i)_{i = 1, \ldots, n}\) from the original NYPD dataset
    covering the years 2010 and 2011.
  \item For all \(i = 1, \ldots, n\):
  \begin{enumerate}
    \item Define the probability the \(i\)-th individual is frisked: \(p_i =
      \text{logit}^{-1}(\boldsymbol \beta_{k,\text{frisk}}^\top \tilde X_i)\).
      Then sample \(A_i \sim \text{Bernoulli}(p_i)\).
    \item Define the probability the \(i\)-th individual has a weapon: \(R_i =
      \text{logit}^{-1}(\boldsymbol \beta_{k,\text{weapon}}^\top \tilde X_i)\).
      Then, sample \(W_i \sim \text{Bernoulli}(R_i)\).
  \end{enumerate}
\end{enumerate}

The above procedure produces a set of tuples, \(\Omega^j = \{(\tilde X_i, C_i,
A_i, R_i, W_i)\}_{i=1, \ldots, n}\), each of which is a synthetic population. On
this population, we compute the ground-truth risk-adjusted disparities via
Eq.~\eqref{eq:nonpar}.\footnote{%
    The definition in Eq.~\eqref{eq:nonpar} involves conditioning on risk, which
    is continuous. This is problematic on a finite population, as there may be
    only one individual with any given risk. Binning risks avoids this issue,
    and, in this case, we get nearly identical values for any appropriately
    small bin size.
}
These disparities are typically non-zero; note, though, that there is no
disparate \emph{treatment} in this example since, by construction, the
probability \(p_i\) of frisking an individual does not explicitly depend on
race. The disparate impact we measure in this scenario thus arises from
decisions that are simply not appropriately tailored to risk, as in
\emph{Griggs}.

We next evaluate the ability of our own risk-adjusted regression to recover the
ground-truth disparate impact. To do so, we first divide \(\Omega^j\) into two
random sets of equal size, \(\Omega_1^j\) and \(\Omega_2^j\). On \(\Omega_1^j\),
we use logistic regression to estimate the probability that frisked individuals
are found to have a weapon,\footnote{%
  To avoid risk estimates not being well defined when certain feature levels
  appear in \(\Omega_2^j\) but not \(\Omega_2^j\), the models are fit using
  penalized maximum likelihood with the \texttt{glmnet} package and an elastic
  net penalty of \(\alpha = \tfrac 1 {10}\).
}
based on the pre-frisk covariates \(\tilde X\). This model is trained on the
subset of stops in \(\Omega_1^j\) in which a frisk was carried out. We then use
this fitted model to predict \emph{ex ante} risk \(\hat R_i\) for every stop in
\(\Omega_2^j\). Lastly, we fit a linear risk-adjusted regression on
\(\Omega_2^j\) to estimate the Black-white and Hispanic-white disparities.

\begin{figure}[t]
  \begin{center}
    \includegraphics{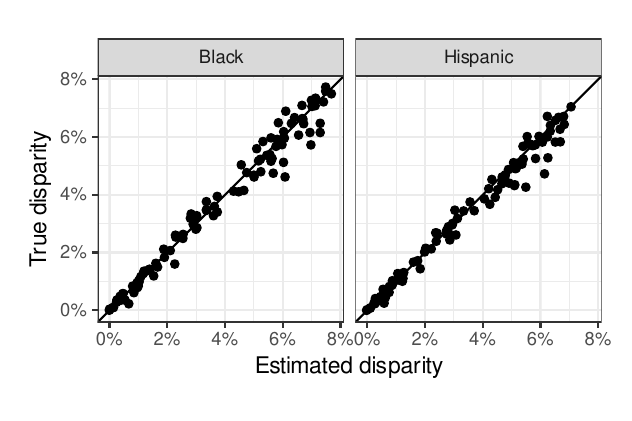}
  \end{center}
  \caption{\emph{%
    Estimates of disparate impact using risk-adjusted regressions on \(m = 100\)
    synthetic datasets, each consisting of \(n = 1{,}000{,}000\) simulated
    stops. Despite model misspecification, the risk-adjusted regressions
    coefficient is a reasonably robust estimator of the true disparity.
  }}%
\label{fig:simulation}
\end{figure}

We note that to simulate real-world settings, we have ensured that, by
construction, the linear probability model used to estimate disparities is
misspecified. Nevertheless, it is a reasonably robust estimator of the true
disparity, as can be seen in Figure~\ref{fig:simulation}. For both Black and
Hispanic pedestrians in the synthetic data, our risk-adjusted estimates of
disparate impact are very close to the ground-truth estimands, with model
misspecification leading to slight overestimates in some instances, and slight
underestimates in others.

\clearpage
\newpage

\section{Accounting for factors beyond risk}%
\label{app:beyond}

Our analysis in the main text implicitly assumed that risk of weapon possession
is the only legitimate consideration for carrying out a frisk---an assumption
motivated by the fact that a frisk is legally allowed only to ensure officer
safety. Officers, however are not required to frisk every individual who they
are legally allowed to frisk (i.e., the law only sets a lower bar), leaving open
the possibility that risk is not the only factor officers consider when making
frisk decisions.\footnote{%
  In \emph{Floyd}, the court found that officers at times frisked individuals
  even in the absence of safety concerns, in violation of the Fourth Amendment
  of the U.S.\ Constitution~\citep{goel_2016c}.
}
For example, in theory, resource-constrained officers might choose to frisk only
the riskiest individuals they stop, effectively setting the bar to frisk
individuals higher than the law demands. If resources differ across
neighborhoods, which in turn correlate with racial composition, then the
risk-adjusted disparities we see may accordingly have a policy-relevant
justification.

\begin{figure}[t]
  \begin{center}
    \includegraphics{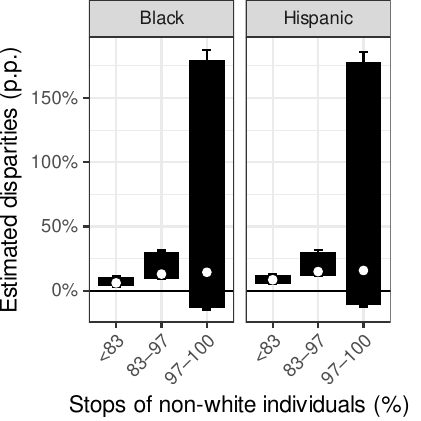}
  \end{center}
  \caption{\emph{%
    Estimates of disparate impact across precincts stratified by racial
    composition of stops, with the black bands showing the range of possible
    estimates if the odds ratio of estimated risk and true risk differ by at
    most \(\epsilon = 0.7\ \mathrm{p.p.}\)
  }}%
\label{fig:sw}
\end{figure}

In Figure~\ref{fig:sw}, we aim to account for this possibility by repeating our
analysis on three subsets of stops stratified by geography. Specifically, we
split the 76 police precincts in our data into three bins based on the racial
composition of stopped individuals. Across strata, we find qualitatively similar
estimates of disparate impact, corroborating our main results. These
within-strata estimates are also reasonably robust to mismeasurement of risk, as
indicated by the black bands, although less so than our main result---a \(0.7\
\textrm{p.p.}\) average absolute difference in risk would potentially be
sufficient to change the sign of the disparity in the most non-white
neighborhoods, or to mask a disparity roughly ten times larger than our
estimate.

That factors beyond risk may justifiably inform frisk decisions can be viewed as
a form of omitted-variable bias, but one distinct in kind from that typically
considered in studies of discrimination. Our disparate impact analysis is
predicated on the understanding that risk, appropriately defined and estimated,
captures nearly all policy-relevant considerations. Although it can be important
in practice to accommodate exceptions when there is a clearly articulated
rationale---as we have done above---care must be taken not to blindly adjust for
every available factor, lest one re-introduce included-variable bias.

\clearpage

\newpage

\section{Additional figures}%
\label{app:figures}

\vfil
\begin{figure}[h]
  \begin{center}
    \includegraphics[width=0.8\textwidth]{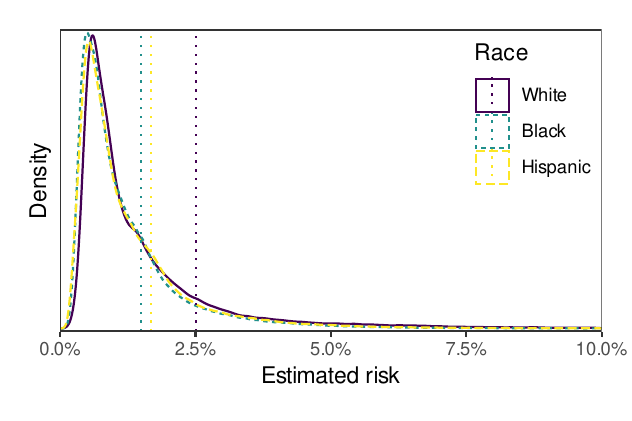}
  \end{center}
  \caption{\emph{%
    Distributions of the estimated \emph{ex ante} risk of probability of
    carrying a weapon for stopped pedestrians, on a log scale. The vertical
    lines indicate each group's average risk (i.e.,~the estimated rate at which
    stopped members of the group carry weapons): 2.7\% for white pedestrians,
    1.5\% for Black pedestrians, and 1.7\% for Hispanic pedestrians.
  }}%
\label{fig:risk}
\end{figure}
\vfil

\begin{figure}[p]
  \begin{center}
    \includegraphics{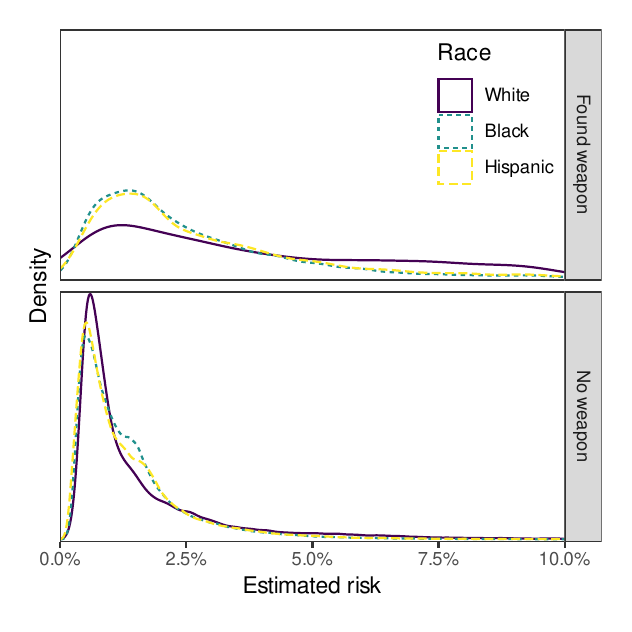}
  \end{center}
  \caption{\emph{%
    Distributions of the estimated \emph{ex ante} risk of probability of
    carrying a weapon for stopped pedestrians, on a log scale among frisked
    individuals, faceted by whether or not a weapon was found.
  }}
\label{fig:risk_by_outcome}
\end{figure}

\begin{figure}[p]
  \begin{center}
    \begin{subfigure}[b]{0.49\textwidth}
      \includegraphics[width=\textwidth]{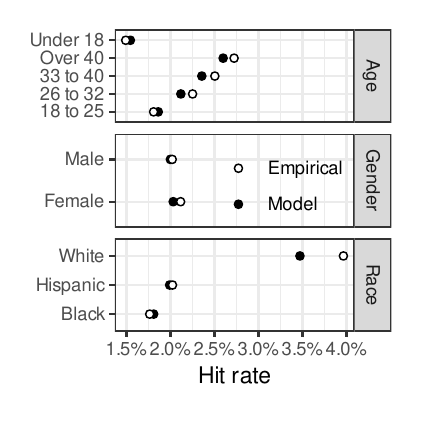}
      \caption{Demographic groups}
    \label{fig:modelchecksdemo}
    \end{subfigure}
    \begin{subfigure}[b]{0.49\textwidth}
      \includegraphics[width=\textwidth]{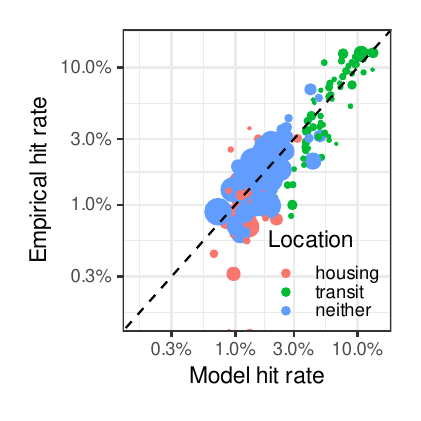}
      \caption{Locations}
    \label{fig:modelcheckslocation}
    \end{subfigure}
  \end{center}
  \caption{\emph{
    Comparison of model-predicted versus empirical weapon recovery rate (``hit
    rate''). Figure~\ref{fig:modelchecksdemo} shows that the model-predicted hit
    rates are close to their empirical counterparts, conditional on values of
    age, race, and gender. In Figure~\ref{fig:modelcheckslocation}, stops are
    binned by precinct and stop location. Points are plotted for each bin with
    more than 100 stops, sized by the number of stops, with colors representing
    the stop location type: transit, housing, or other. The plotted points are
    near the diagonal, suggesting that the outcome model is well-calibrated and
    predicts well over the full range of hit rates and frisk rates,
    respectively. The model itself achieves an AUC of 81\% on the second half of
    the data.
  }}%
\label{fig:modelchecks}
\end{figure}

\begin{figure}[p]
  \begin{center}
    \includegraphics{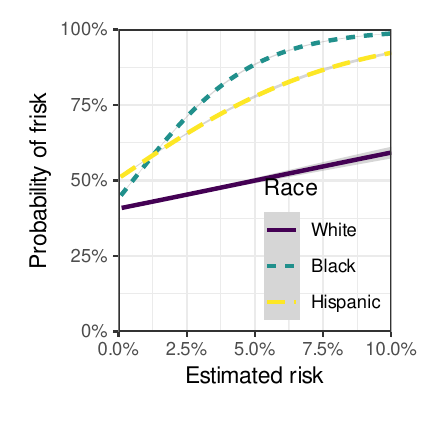}
  \end{center}
  \caption{\emph{%
    Frisk rates vs.\ risk (where risk has been estimated without race as a
    covariate), as estimated via logistic regression curves fit separately for
    each race group. Across risk levels, stopped Black and Hispanic pedestrians
    are frisked substantially more frequently than comparably risk white
    individuals, indicative of disparate impact.
  }}%
\label{fig:policy_rb}
\end{figure}

\begin{figure}[p]
  \begin{center}
    \includegraphics{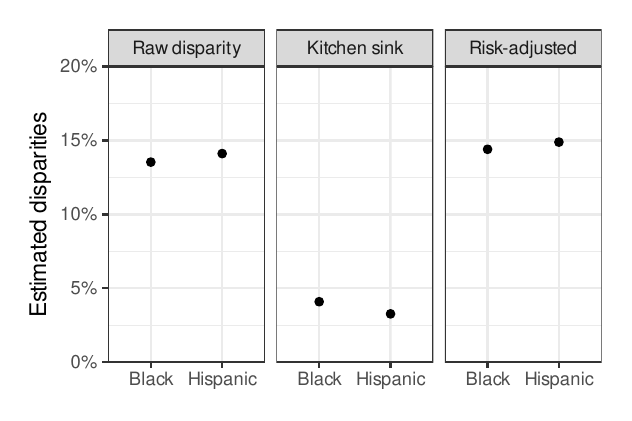}
  \end{center}
  \caption{\emph{%
    Racial gaps in frisk rates adjusting for different sets of covariates, where
    the \(y\)-axis shows the percentage point difference relative to stopped
    white individuals. The left panel shows the raw disparities in frisk rates.
    As a measure of discrimination, raw disparities suffer from omitted-variable
    bias: there may, in theory, be legitimate reasons why Black and Hispanic
    pedestrians are more likely to be frisked. The middle panel shows the
    estimated race effects in a kitchen-sink regression, adjusting for all
    pre-frisk covariates. These estimates suffer from included-variable bias
    because they adjust for features that are correlated with race but unrelated
    to risk. The right panel shows the results of our risk-adjusted
    regression---where risk has been estimated without race as a
    covariate---adjusting exclusively for estimated risk of weapon possession.
    In all cases, estimated standard errors are less than 0.2 percentage points,
    and so are not visible in the plot.
  }}%
\label{fig:comparison_rb}
\end{figure}

\begin{figure}[p]
  \begin{center}
    \includegraphics{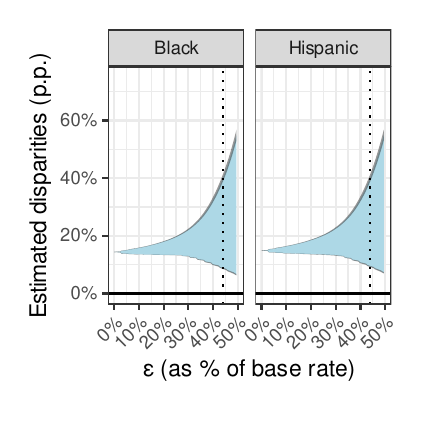}
  \end{center}
  \caption{\emph{%
    Sensitivity of the risk-adjusted disparities in frisk decisions to
    mismeasurement of risk, where risk has been estimated without race as a
    covariate. The blue bands bound our estimates of disparate impact as a
    function of the average absolute difference between the true and estimated
    risks \(\epsilon\), relative to the base rate (1.7\%). The dotted line at
    44\% (\(\epsilon = 0.7\ \mathrm{p.p.}\)) corresponds to a simulated
    situation with severe confounding. The grey bands represent 95\% percentile
    bootstrapped confidence
    intervals~\citep[\(N=1000\);][]{zhao2019sensitivity}. The step size for the
    grid search over group-level total risks was \(0.05\ \mathrm{p.p.}\) for all
    groups, and \(\gamma = 0.01\ \mathrm{p.p.}\) was the approximation parameter
    for the maximization routine.
  }}%
\label{fig:sensitivity_rb}
\end{figure}

\begin{figure}[p]
  \begin{center}
    \includegraphics{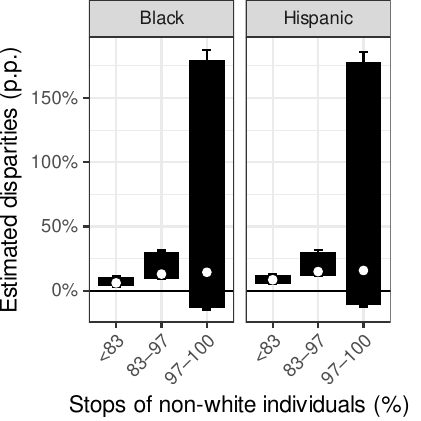}
  \end{center}
  \caption{\emph{%
      Estimates of disparate impact across precincts stratified by racial
      composition of stops, with the black bands showing the range of possible
      estimates if the odds ratio of estimated risk and true risk differ by at
      most \(\epsilon = 0.7\ \mathrm{p.p}\). Here, risk is estimated without race
      as a covariate.
  }}%
\label{fig:sw_rb}
\end{figure}

\begin{figure}[p]
  \begin{center}
    \includegraphics{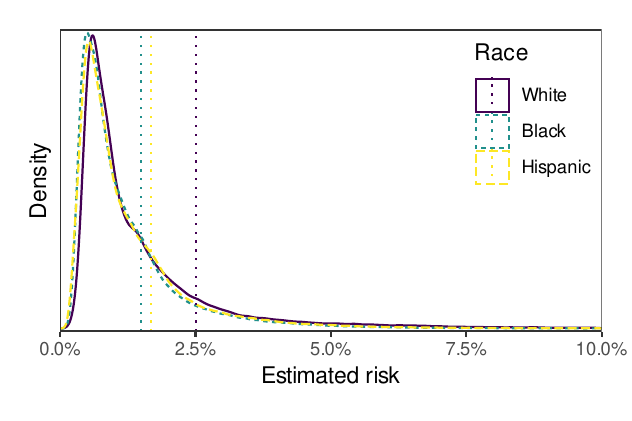}
  \end{center}
  \caption{\emph{%
    Distributions of the estimated \emph{ex ante} risk of carrying a weapon for
    stopped pedestrians, on a log scale and estimated without race as a
    covariate. The vertical lines indicate each group's average risk (i.e.,~the
    estimated rate at which stopped members of the group carry weapons): 2.7\%
    for white pedestrians, 1.5\% for Black pedestrians, and 1.7\% for Hispanic
    pedestrians.
  }}%
\label{fig:risk_rb}
\end{figure}

\begin{figure}[p]
  \begin{center}
    \includegraphics{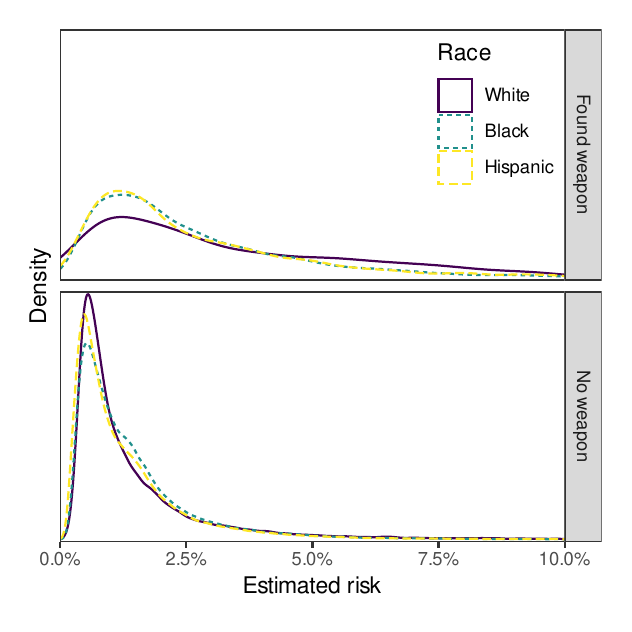}
  \end{center}
  \caption{\emph{%
    Distributions of the estimated \emph{ex ante} risk of probability of
    carrying a weapon---estimated without race as a covariate---for stopped
    pedestrians, on a log scale among frisked individuals, faceted by whether or
    not a weapon was found.
  }}%
\label{fig:risk_by_outcome_rb}
\end{figure}

\begin{figure}[p]
  \begin{center}
    \begin{subfigure}[b]{0.49\textwidth}
      \includegraphics[width=\textwidth]{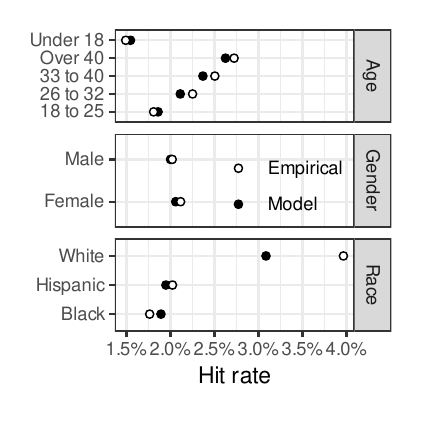}
      \caption{Demographic groups}
    \label{fig:modelchecksdemo_rb}
    \end{subfigure}
    \begin{subfigure}[b]{0.49\textwidth}
      \includegraphics[width=\textwidth]{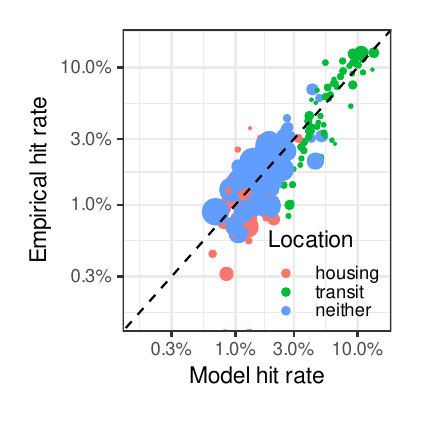}
      \caption{Locations}
    \label{fig:modelcheckslocation_rb}
    \end{subfigure}
  \end{center}
  \caption{\emph{%
    Comparison of model-predicted versus empirical weapon recovery rate (``hit
    rate''), where the risk model does not include race as a covariate.
    Figure~\ref{fig:modelchecksdemo_rb} shows that the model-predicted hit rates
    are close to their empirical counterparts, conditional on values of age,
    race, and gender. In Figure~\ref{fig:modelcheckslocation_rb}, stops are
    binned by precinct and stop location. Points are plotted for each bin with
    more than 100 stops, sized by the number of stops, with colors representing
    the stop location type: transit, housing, or other. The plotted points are
    near the diagonal, suggesting that the outcome model is well-calibrated and
    predicts well over the full range of hit rates and frisk rates,
    respectively. The model itself achieves an AUC of 81\% on the second half of
    the data.
  }}%
\label{fig:modelchecks_rb}
\end{figure}

\clearpage

\newpage

\clearpage

\bibliographystyle{plainnat}
\bibliography{refs.bib}

\begin{thebibliography}{65}
\providecommand{\natexlab}[1]{#1}
\providecommand{\url}[1]{\texttt{#1}}
\expandafter\ifx\csname urlstyle\endcsname\relax
  \providecommand{\doi}[1]{doi: #1}\else
  \providecommand{\doi}{doi: \begingroup \urlstyle{rm}\Url}\fi

\bibitem[Abrams(2014)]{abrams2014law}
David~S. Abrams.
\newblock The law and economics of stop-and-frisk.
\newblock \emph{Loyola University Chicago Law Journal}, 46:\penalty0 369--381, 2014.

\bibitem[Angrist and Pischke(2008)]{angrist2008mostly}
Joshua~D Angrist and J{\"o}rn-Steffen Pischke.
\newblock \emph{Mostly harmless econometrics: An empiricist's companion}.
\newblock Princeton University Press, 2008.

\bibitem[Arnold et~al.(2020)Arnold, Dobbie, and Hull]{arnold2020measuring}
David Arnold, Will~S Dobbie, and Peter Hull.
\newblock Measuring racial discrimination in bail decisions.
\newblock Technical report, National Bureau of Economic Research, 2020.

\bibitem[Arnold et~al.(2021)Arnold, Dobbie, and Hull]{arnold2021measuring}
David Arnold, Will Dobbie, and Peter Hull.
\newblock Measuring racial discrimination in algorithms.
\newblock In \emph{AEA Papers and Proceedings}, volume 111, pages 49--54, 2021.

\bibitem[Arrow(1973)]{arrow1973}
Kenneth Arrow.
\newblock The theory of discrimination.
\newblock \emph{Discrimination in labor markets}, 3\penalty0 (10):\penalty0 3--33, 1973.

\bibitem[Ayres(2002)]{ayres2002}
Ian Ayres.
\newblock Outcome tests of racial disparities in police practices.
\newblock \emph{Justice Research and Policy}, 4\penalty0 (1-2):\penalty0 131--142, 2002.

\bibitem[Ayres(2005)]{ayres2005three}
Ian Ayres.
\newblock Three tests for measuring unjustified disparate impacts in organ transplantation: The problem of ``included variable" bias.
\newblock \emph{Perspectives in Biology and Medicine}, 48\penalty0 (1):\penalty0 68--S87, 2005.

\bibitem[Ayres(2010)]{ayres2010included}
Ian Ayres.
\newblock Testing for discrimination and the problem of ``included variable bias''.
\newblock \emph{Working paper}, 2010.
\newblock Available at \url{http://islandia.law.yale.edu/ayers/ayresincludedvariablebias.pdf}.

\bibitem[Balsa et~al.(2005)Balsa, McGuire, and Meredith]{balsa2005testing}
Ana~I Balsa, Thomas~G McGuire, and Lisa~S Meredith.
\newblock Testing for statistical discrimination in health care.
\newblock \emph{Health Services Research}, 40\penalty0 (1):\penalty0 227--252, 2005.

\bibitem[Becker(1993)]{becker1993}
Gary~S Becker.
\newblock Nobel lecture: The economic way of looking at behavior.
\newblock \emph{Journal of Political Economy}, 101\penalty0 (3):\penalty0 385--409, 1993.

\bibitem[Berkovec et~al.(2018)Berkovec, Canner, Gabriel, and Hannan]{berkovec1996mortgage}
James~A Berkovec, Glenn~B Canner, Stuart~A Gabriel, and Timothy~H Hannan.
\newblock Mortgage discrimination and {FHA} loan performance.
\newblock In \emph{Mortgage Lending, Racial Discrimination, and Federal Policy}, pages 289--305. Routledge, 2018.

\bibitem[Bohren et~al.(2022)Bohren, Hull, and Imas]{bohren2022systemic}
J~Aislinn Bohren, Peter Hull, and Alex Imas.
\newblock Systemic discrimination: Theory and measurement.
\newblock Technical report, National Bureau of Economic Research, 2022.

\bibitem[Boyd and Vandenberghe(2004)]{boyd2004convex}
Stephen~P Boyd and Lieven Vandenberghe.
\newblock \emph{Convex optimization}.
\newblock Cambridge university press, 2004.

\bibitem[Chaudhuri and Sethi(2008)]{chaudhuri2008statistical}
Shubham Chaudhuri and Rajiv Sethi.
\newblock {Statistical Discrimination with Peer Effects: Can Integration Eliminate Negative Stereotypes?}
\newblock \emph{The Review of Economic Studies}, 75\penalty0 (2):\penalty0 579--596, 04 2008.
\newblock ISSN 0034-6527.
\newblock \doi{10.1111/j.1467-937X.2008.00468.x}.
\newblock URL \url{https://doi.org/10.1111/j.1467-937X.2008.00468.x}.

\bibitem[Chouldechova et~al.(2018)Chouldechova, Benavides-Prado, Fialko, and Vaithianathan]{chouldechova2018case}
Alexandra Chouldechova, Diana Benavides-Prado, Oleksandr Fialko, and Rhema Vaithianathan.
\newblock A case study of algorithm-assisted decision making in child maltreatment hotline screening decisions.
\newblock In \emph{Conference on Fairness, Accountability and Transparency}, pages 134--148, 2018.

\bibitem[Cinelli and Hazlett(2020)]{cinelli2020making}
Carlos Cinelli and Chad Hazlett.
\newblock Making sense of sensitivity: Extending omitted variable bias.
\newblock \emph{Journal of the Royal Statistical Society Series B-Statistical Methodology}, 82\penalty0 (1):\penalty0 39--67, 2020.

\bibitem[Corbett-Davies et~al.(2023)Corbett-Davies, Gaebler, Nilforoshan, Shroff, and Goel]{corbett2018measure}
Sam Corbett-Davies, Johann~D. Gaebler, Hamed Nilforoshan, Ravi Shroff, and Sharad Goel.
\newblock The measure and mismeasure of fairness.
\newblock \emph{Journal of Machine Learning Research}, 24\penalty0 (312):\penalty0 1--117, 2023.
\newblock URL \url{http://jmlr.org/papers/v24/22-1511.html}.

\bibitem[Coston et~al.(2020)Coston, Mishler, Kennedy, and Chouldechova]{coston2020counterfactual}
Amanda Coston, Alan Mishler, Edward~H Kennedy, and Alexandra Chouldechova.
\newblock Counterfactual risk assessments, evaluation, and fairness.
\newblock In \emph{Proceedings of the 2020 conference on fairness, accountability, and transparency}, pages 582--593, 2020.

\bibitem[Dobbie et~al.(2018)Dobbie, Goldin, and Yang]{dobbie2018effects}
Will Dobbie, Jacob Goldin, and Crystal~S Yang.
\newblock The effects of pre-trial detention on conviction, future crime, and employment: Evidence from randomly assigned judges.
\newblock \emph{American Economic Review}, 108\penalty0 (2):\penalty0 201--240, 2018.

\bibitem[Dorn and Guo(2022)]{dorn2022sharp}
Jacob Dorn and Kevin Guo.
\newblock Sharp sensitivity analysis for inverse propensity weighting via quantile balancing.
\newblock \emph{Journal of the American Statistical Association}, pages 1--13, 2022.

\bibitem[Elzayn et~al.(2023)Elzayn, Smith, Hertz, Ramesh, Goldin, Ho, and Fisher]{elzayn2023measuring}
Hadi Elzayn, Evelyn Smith, Thomas Hertz, Arun Ramesh, Jacob Goldin, Daniel~E Ho, and Robin Fisher.
\newblock \emph{Measuring and mitigating racial disparities in tax audits}.
\newblock Stanford Institute for Economic Policy Research (SIEPR), 2023.

\bibitem[Espenshade et~al.(2004)Espenshade, Chung, and Walling]{espenshade2004admission}
Thomas~J Espenshade, Chang~Y Chung, and Joan~L Walling.
\newblock Admission preferences for minority students, athletes, and legacies at elite universities.
\newblock \emph{Social Science Quarterly}, 85\penalty0 (5):\penalty0 1422--1446, 2004.

\bibitem[Fagan(2010)]{fagan2010report}
Jeffrey Fagan.
\newblock Report of {Jeffrey Fagan} in the case of {Floyd v.\ the City of New York}, 2010.
\newblock Available at \url{https://www.law.columbia.edu/sites/default/files/microsites/policing-litigation-conference/files/Fagan\%20Report\%20with\%20Technical\%20Appendices.pdf}.

\bibitem[{Floyd v.~City of New York}(2013)]{floyd}
{Floyd v.~City of New York}.
\newblock {959 F. Supp. 2d 540, S.D.N.Y}, 2013.

\bibitem[Gaebler et~al.(2022)Gaebler, Cai, Basse, Shroff, Goel, and Hill]{gaebler2022causal}
Johann Gaebler, William Cai, Guillaume Basse, Ravi Shroff, Sharad Goel, and Jennifer Hill.
\newblock A causal framework for observational studies of discrimination.
\newblock \emph{Statistics and Public Policy}, 9:\penalty0 26--48, 2022.

\bibitem[Garg et~al.(2021)Garg, Li, and Monachou]{garg2021standardized}
Nikhil Garg, Hannah Li, and Faidra Monachou.
\newblock Standardized tests and affirmative action: The role of bias and variance.
\newblock In \emph{Proceedings of the 2021 ACM Conference on Fairness, Accountability, and Transparency}, pages 261--261, 2021.

\bibitem[Gelman et~al.(2007)Gelman, Fagan, and Kiss]{gelman2007analysis}
Andrew Gelman, Jeffrey Fagan, and Alex Kiss.
\newblock An analysis of the {New York City} police department's ``stop-and-frisk'' policy in the context of claims of racial bias.
\newblock \emph{Journal of the American Statistical Association}, 102\penalty0 (479):\penalty0 813--823, 2007.

\bibitem[Goel et~al.(2016{\natexlab{a}})Goel, Rao, and Shroff]{goel_2016a}
Sharad Goel, Justin~M Rao, and Ravi Shroff.
\newblock Personalized risk assessments in the criminal justice system.
\newblock \emph{The American Economic Review}, 106\penalty0 (5):\penalty0 119--123, 2016{\natexlab{a}}.

\bibitem[Goel et~al.(2016{\natexlab{b}})Goel, Rao, and Shroff]{goel_2016c}
Sharad Goel, Justin~M Rao, and Ravi Shroff.
\newblock Precinct or prejudice? {U}nderstanding racial disparities in {New York City's} stop-and-frisk policy.
\newblock \emph{Annals of Applied Statistics}, 10, 2016{\natexlab{b}}.

\bibitem[Goel et~al.(2017)Goel, Perelman, Shroff, and Sklansky]{goel2016b}
Sharad Goel, Maya Perelman, Ravi Shroff, and David Sklansky.
\newblock Combatting police discrimination in the age of big data.
\newblock \emph{New Criminal Law Review}, 20, 2017.

\bibitem[Goel et~al.(2021)Goel, Shroff, Skeem, and Slobogin]{goel2018accuracy}
Sharad Goel, Ravi Shroff, Jennifer Skeem, and Christopher Slobogin.
\newblock The accuracy, equity, and jurisprudence of criminal risk assessment.
\newblock In \emph{Research Handbook on Big Data Law}. Edward Elgar Publishing, 2021.

\bibitem[Greiner and Rubin(2011)]{greiner2011}
D~James Greiner and Donald~B Rubin.
\newblock Causal effects of perceived immutable characteristics.
\newblock \emph{Review of Economics and Statistics}, 93\penalty0 (3):\penalty0 775--785, 2011.

\bibitem[Grossman et~al.(2023{\natexlab{a}})Grossman, Nyarko, and Goel]{grossmanreconciling}
Joshua Grossman, Julian Nyarko, and Sharad Goel.
\newblock Reconciling legal and empirical conceptions of disparate impact, 2023{\natexlab{a}}.
\newblock Available at \url{https://5harad.com/papers/disparate-impact.pdf}.

\bibitem[Grossman et~al.(2023{\natexlab{b}})Grossman, Tomkins, Page, and Goel]{grossman2023disparate}
Joshua Grossman, Sabina Tomkins, Lindsay Page, and Sharad Goel.
\newblock The disparate impacts of college admissions policies on asian american applicants.
\newblock \emph{Working paper}, 2023{\natexlab{b}}.

\bibitem[Hardt et~al.(2016)Hardt, Price, and Srebro]{hardt2016equality}
Moritz Hardt, Eric Price, and Nati Srebro.
\newblock Equality of opportunity in supervised learning.
\newblock \emph{Advances in Neural Information Processing Systems}, 29, 2016.

\bibitem[Huang and Pimentel(2022)]{huang2022variance}
Melody Huang and Samuel~D Pimentel.
\newblock Variance-based sensitivity analysis for weighting estimators result in more informative bounds.
\newblock \emph{arXiv preprint arXiv:2208.01691}, 2022.

\bibitem[Jung et~al.(2020)Jung, Concannon, Shroff, Goel, and Goldstein]{simplerules}
Jongbin Jung, Connor Concannon, Ravi Shroff, Sharad Goel, and Daniel~G Goldstein.
\newblock Simple rules to guide expert classifications.
\newblock \emph{Journal of the Royal Statistical Society: Series A (Statistics in Society)}, 183\penalty0 (3):\penalty0 771--800, 2020.

\bibitem[Kallus et~al.(2022)Kallus, Mao, and Zhou]{kallus2022assessing}
Nathan Kallus, Xiaojie Mao, and Angela Zhou.
\newblock Assessing algorithmic fairness with unobserved protected class using data combination.
\newblock \emph{Management Science}, 68\penalty0 (3):\penalty0 1959--1981, 2022.

\bibitem[Karush(1939)]{karush1939minima}
William Karush.
\newblock Minima of functions of several variables with inequalities as side constraints.
\newblock \emph{M. Sc. Dissertation. Dept. of Mathematics, Univ. of Chicago}, 1939.

\bibitem[Kleinberg et~al.(2018)Kleinberg, Lakkaraju, Leskovec, Ludwig, and Mullainathan]{kleinberg2018human}
Jon Kleinberg, Himabindu Lakkaraju, Jure Leskovec, Jens Ludwig, and Sendhil Mullainathan.
\newblock Human decisions and machine predictions.
\newblock \emph{The quarterly journal of economics}, 133\penalty0 (1):\penalty0 237--293, 2018.

\bibitem[Knowles et~al.(2001)Knowles, Persico, and Todd]{knowles2001}
John Knowles, Nicola Persico, and Petra Todd.
\newblock Racial bias in motor vehicle searches: Theory and evidence.
\newblock \emph{Journal of Political Economy}, 109\penalty0 (1), 2001.

\bibitem[Kuhn and Tucker(1951)]{MR0047303}
H.~W. Kuhn and A.~W. Tucker.
\newblock Nonlinear programming.
\newblock In \emph{Proceedings of the {S}econd {B}erkeley {S}ymposium on {M}athematical {S}tatistics and {P}robability, 1950}, pages pp 481--492. University of California Press, Berkeley-Los Angeles, Calif., 1951.

\bibitem[MacDonald and Raphael(2019)]{macdonald2019effect}
John MacDonald and Steven Raphael.
\newblock The effect of scaling back punishment in racial disparities in criminal case outcomes.
\newblock \emph{University of California Berkeley Working Paper}, 2019.

\bibitem[Mayson(2018)]{mayson2018bias}
Sandra~G Mayson.
\newblock Bias in, bias out.
\newblock \emph{Yale Law Journal}, 128:\penalty0 2218, 2018.

\bibitem[McCartan et~al.(2023)McCartan, Goldin, Ho, and Imai]{mccartan2023estimating}
Cory McCartan, Jacob Goldin, Daniel~E Ho, and Kosuke Imai.
\newblock Estimating racial disparities when race is not observed.
\newblock \emph{arXiv preprint arXiv:2303.02580}, 2023.

\bibitem[Monahan and Skeem(2016)]{monahan2016risk}
John Monahan and Jennifer~L Skeem.
\newblock Risk assessment in criminal sentencing.
\newblock \emph{Annual review of clinical psychology}, 12:\penalty0 489--513, 2016.

\bibitem[Nilforoshan et~al.(2022)Nilforoshan, Gaebler, Shroff, and Goel]{nilforoshan2022causal}
Hamed Nilforoshan, Johann~D Gaebler, Ravi Shroff, and Sharad Goel.
\newblock Causal conceptions of fairness and their consequences.
\newblock In \emph{International Conference on Machine Learning}, pages 16848--16887. PMLR, 2022.

\bibitem[Phelps(1972)]{phelps1972}
Edmund~S Phelps.
\newblock The statistical theory of racism and sexism.
\newblock \emph{The American Economic Review}, 62\penalty0 (4):\penalty0 659--661, 1972.

\bibitem[Pierson et~al.(2018)Pierson, Corbett-Davies, and Goel]{pierson2018fast}
Emma Pierson, Sam Corbett-Davies, and Sharad Goel.
\newblock Fast threshold tests for detecting discrimination.
\newblock In \emph{The 21st International Conference on Artificial Intelligence and Statistics (AISTATS)}, 2018.

\bibitem[Pierson et~al.(2020)Pierson, Simoiu, Overgoor, Corbett-Davies, Jenson, Shoemaker, Ramachandran, Barghouty, Phillips, Shroff, et~al.]{pierson2018large}
Emma Pierson, Camelia Simoiu, Jan Overgoor, Sam Corbett-Davies, Daniel Jenson, Amy Shoemaker, Vignesh Ramachandran, Phoebe Barghouty, Cheryl Phillips, Ravi Shroff, et~al.
\newblock A large-scale analysis of racial disparities in police stops across the {United States}.
\newblock \emph{Nature Human Behaviour}, 4\penalty0 (7):\penalty0 736--745, 2020.

\bibitem[Polachek(2008)]{polachek2008earnings}
Solomon~W Polachek.
\newblock Earnings over the life cycle: The {M}incer earnings function and its applications.
\newblock \emph{Foundations and Trends in Microeconomics}, 4\penalty0 (3):\penalty0 165--272, 2008.

\bibitem[Raghavan(2023)]{raghavan2023should}
Manish Raghavan.
\newblock What should we do when our ideas of fairness conflict?
\newblock \emph{Communications of the ACM}, 67\penalty0 (1):\penalty0 88--97, 2023.

\bibitem[Rehavi and Starr(2012)]{rehavi2012racial}
M~Marit Rehavi and Sonja~B Starr.
\newblock Racial disparity in federal criminal charging and its sentencing consequences.
\newblock \emph{U of Michigan Law \& Econ, Empirical Legal Studies Center Paper}, \penalty0 (12-002), 2012.

\bibitem[Rosenbaum(2002)]{rosenbaum2002sensitivity}
Paul~R Rosenbaum.
\newblock \emph{Observational studies}.
\newblock Springer, 2002.

\bibitem[Rosenbaum and Rubin(1983{\natexlab{a}})]{rosenbaum_1983a}
Paul~R Rosenbaum and Donald~B Rubin.
\newblock Assessing sensitivity to an unobserved binary covariate in an observational study with binary outcome.
\newblock \emph{Journal of the Royal Statistical Society. Series B (Methodological)}, 45:\penalty0 212--218, 1983{\natexlab{a}}.

\bibitem[Rosenbaum and Rubin(1983{\natexlab{b}})]{rosenbaum_1983b}
Paul~R Rosenbaum and Donald~B Rubin.
\newblock The central role of the propensity score in observational studies for causal effects.
\newblock \emph{Biometrika}, 70\penalty0 (1):\penalty0 41--55, 1983{\natexlab{b}}.

\bibitem[Sahni(1974)]{sahni1974computationally}
Sartaj Sahni.
\newblock Computationally related problems.
\newblock \emph{SIAM Journal on computing}, 3\penalty0 (4):\penalty0 262--279, 1974.

\bibitem[{SFFA v. Harvard}(2023)]{2023students}
{SFFA v. Harvard}.
\newblock {Students for Fair Admissions, Inc., Petitioner, v. President and Fellows of Harvard College. Students for Fair Admissions, Inc., Petitioner, v. University of North Carolina, et al.}, 2023.
\newblock \url{https://www.supremecourt.gov/opinions/22pdf/20-1199_l6gn.pdf}.

\bibitem[Shroff(2017)]{shroff2017predictive}
Ravi Shroff.
\newblock Predictive analytics for city agencies: Lessons from children's services.
\newblock \emph{Big Data}, 5\penalty0 (3):\penalty0 189--196, 2017.

\bibitem[Simoiu et~al.(2017)Simoiu, Corbett-Davies, and Goel]{simoiu2017problem}
Camelia Simoiu, Sam Corbett-Davies, and Sharad Goel.
\newblock The problem of infra-marginality in outcome tests for discrimination.
\newblock \emph{The Annals of Applied Statistics}, 11\penalty0 (3):\penalty0 1193--1216, 2017.

\bibitem[Smart and Waldfogel(1996)]{smart1996citation}
Scott Smart and Joel Waldfogel.
\newblock A citation-based test for discrimination at economics and finance journals.
\newblock Technical report, National Bureau of Economic Research, 1996.

\bibitem[VanderWeele and Robinson(2014)]{vanderweele2014}
Tyler~J VanderWeele and Whitney~R Robinson.
\newblock Confounding and mediating variables.
\newblock \emph{Epidemiology}, 25\penalty0 (4):\penalty0 473--484, 2014.

\bibitem[{Washington v. Davis}(1976)]{davis1976}
{Washington v. Davis}.
\newblock {426 U.S. 229}, 1976.

\bibitem[Zhang and Zhao(2022)]{zhang2022bounds}
Yao Zhang and Qingyuan Zhao.
\newblock Bounds and semiparametric inference in $l^\infty$- and $ l^2$-sensitivity analysis for observational studies.
\newblock \emph{arXiv preprint arXiv:2211.04697}, 2022.

\bibitem[Zhao et~al.(2019)Zhao, Small, and Bhattacharya]{zhao2019sensitivity}
Qingyuan Zhao, Dylan~S Small, and Bhaswar~B Bhattacharya.
\newblock Sensitivity analysis for inverse probability weighting estimators via the percentile bootstrap.
\newblock \emph{Journal of the Royal Statistical Society Series B: Statistical Methodology}, 81\penalty0 (4):\penalty0 735--761, 2019.

\end{thebibliography}

\end{document}